\documentclass[12pt]{article}
\usepackage{latexsym}
\usepackage{amsmath}
\usepackage{wasysym}
\usepackage{amsfonts}
\usepackage{dsfont}
\usepackage{pifont}
\usepackage{bm}
\usepackage{epsf}
\usepackage{epsfig,graphicx}
\usepackage{amsmath,amssymb,bm,epsf,epsfig,graphicx}
\usepackage{color}

\usepackage[vcentermath]{youngtab}

\usepackage[colorlinks=true,linkcolor=blue,citecolor=blue,urlcolor=black,filecolor=black,linktocpage=true]{hyperref}

 \csname
@addtoreset\endcsname{equation}{section}

\def\pe{\prime}
\def\3s{{s \choose 3}}
\def\4s{{s \choose 4}}
\def\5s{{s \choose 5}}
\def\6s{{s \choose 6}}

\def\12{\frac{1}{2}}
\def\fr{\frac}
\def\ft{\footnote}

\def\nn{\nonumber}

\def\pr{\partial}

\def\prd{\partial \cdot}

\def\be{\begin{equation}}
\def\ee{\end{equation}}
\def\bea{\begin{eqnarray}}
\def\eea{\end{eqnarray}}
\def\ba{\begin{array}}
\def\ea{\end{array}}
\def\bec{\begin{center}}
\def\ec{\end{center}}

\def\a{\alpha} 
\def\b{\beta}  
\def\g{\gamma} 
\def\G{\Gamma}
\def\d{\delta} 
\def\D{\Delta}
\def\e{\epsilon}

\def\h{\eta}
\def\th{\theta}

\def\l{\lambda}
\def\L{\Lambda}
\def\m{\mu}
\def\n{\nu}
\def\x{\xi}
\def\r{\rho}

\def\vf{\varphi}
\def\c{\chi}

\def\dsll{\not {\! \pr}}
\def\psisl{\not {\! \! \psi}}

\def\cC{{\cal C}}
\def\cD{{\cal D}}

\def\cF{{\cal F}}

\def\cL{{\cal L}}

\def\cO{{\cal O}}

\def\cR{{\cal R}}
\def\cS{{\cal S}}

\begin{document}

\begin{flushright}
AEI-2012-056
\end{flushright}

\vspace{20pt}

\begin{center}

%%%%%%%%%%%%%%%%%%%%%%%%%%%%%%%%%%%%%%%%%%%%%%%%%%%%%%%%%%%%%%%%%%%%

{\Large\sc Maxwell-like Lagrangians for  higher spins}\\

%%%%%%%%%%%%%%%%%%%%%%%%%%%%%%%%%%%%%%%%%%%%%%%%%%%%%%%%%%%%%%%%%%%%

\vspace{30pt} {\sc Andrea Campoleoni${}^{\; a,}$\footnote{Present address: Universit\'e Libre de Bruxelles, ULB-Campus Plaine CP231, B-1050 Brussels, Belgium. Email: \it{andrea.campoleoni@ulb.ac.be}}, Dario Francia${}^{\; b,\,c}$} \\ \vspace{10pt} 
{${}^a$\sl\small
Max-Planck-Institut f\"ur Gravitationsphysik,\\ Albert-Einstein-Institut\\
Am M\"uhlenberg 1, D-14476 Golm, Germany\\
e-mail: {\small \it andrea.campoleoni@aei.mpg.de}}
\vspace{10pt}

${}^b$\sl \small Centro Studi e Ricerche E. Fermi\\
Piazza del Viminale 1, I-00184 Roma, Italy \\
${}^c$Scuola Normale Superiore and INFN \\ Piazza dei Cavalieri 7, I-56126 Pisa, Italy \\
e-mail:
{\small \it dario.francia@sns.it }
\vspace{10pt}

\vspace{10pt}

%%%%%%%%%%%%%%%%%%%%%%%%%%%%%%%%%%%%%%%%%%%%%%%%%%%%%%%%%%%%%%%%%%%%

\vspace{30pt} {\sc\large Abstract}\end{center}
We show how implementing invariance under divergence-free gauge transformations leads to a remarkably simple Lagrangian description of massless bosons of any spin. Our construction covers both flat and (A)dS backgrounds and extends to tensors of arbitrary mixed-symmetry type. Irreducible and traceless fields produce single-particle actions, while whenever trace constraints can be dispensed with the resulting Lagrangians display the same reducible, multi-particle spectra as those emerging from the tensionless limit of free open-string field theory. For all explored options the corresponding kinetic operators take essentially the same form as in the spin-one, Maxwell case.

\vfill
\setcounter{page}{1}

\pagebreak

\tableofcontents

%\newpage

%%%%%%%%%%%%%%%%%%%%%%%%%%%%%%%%%%%%%%%%%%%%%%%%%%%%%%%%%%%%%%%%%%%%%

\section{Introduction}\label{sec:intro}

%%%%%%%%%%%%%%%%%%%%%%%%%%%%%%%%%%%%%%%%%%%%%%%%%%%%%%%%%%%%%%%%%%%%%

The main purpose of this work is to investigate the conditions under which higher-spin free Lagrangians take their simplest possible forms, exploring the case of massless bosons of any spin and symmetry in Minkowski as well as in (Anti)-de Sitter backgrounds. We find that in all these cases it is possible to keep the corresponding kinetic operators essentially as simple as their spin-one, Maxwell counterpart\ft{See \cite{dariokyoto} for a general motivating discussion and \cite{hsp} for reviews on the subject of higher spins.}. For instance, the Lagrangian equations of motion for rank-$s$ symmetric tensors in flat space-time resulting from our approach read
\be \label{Max}
(M\, \vf)_{\, \m_1 \, \cdots \, \m_s} \, \equiv \, \Box \, \vf_{\, \m_1 \, \cdots \, \m_s} \, - \, \left(\, \pr_{\, \m_1} \, 
\pr^{\, \a} \, \vf_{\, \a \, \m_2 \, \cdots \, \m_s} \, + \, \cdots \,\right) \, = \, 0 \, ,
\ee
where the dots stand for symmetrisation of indices while the operator $M$ builds the higher-spin extension of  the Maxwell field equations,
\be \label{spin1-intro}
(M\, A)_{\, \m} \, \equiv \, \Box \,A_{\, \m} \, - \, \pr_{\, \m} \, 
\pr^{\, \a} \, A_{\, \a} \, = \, 0 \, .
\ee
Indeed, considering the on-shell conditions for massless, spin-$s$ propagation \cite{Fierz},
\begin{align} \label{fierzsyst}
&\Box  \, \vf_{\, \m_1 \, \cdots \, \m_s} \, = \, 0\, ,& & &
&\Box  \, \L_{\, \m_1 \, \cdots \, \m_{s - 1}} \, = \, 0\, \, ,\nonumber\\
&\pr^{\, \a}  \vf_{\, \a \,  \m_2\, \cdots \, \m_s} \, = \,  0 \, ,& & &
&\pr^{\, \a}  \L_{\, \a \,  \m_2\, \cdots \, \m_{s-1}} \, = \,  0 \, , \\
&\vf^{\, \a}{}_{\, \a \, \m_3 \, \cdots \, \m_s} \,   = \, 0 \, , & & &
&\L^{\, \a}{}_{\, \a \, \m_3 \, \cdots \, \m_{s-1}} \,   = \, 0 \,  , \nonumber
\end{align}
where $\vf_{\, \m_1 \, \cdots \, \m_s}$ is a rank-$s$ symmetric tensor subject to the abelian gauge transformation 
\be \label{abelian}
\d \vf_{\, \m_1 \, \cdots \, \m_s} \, = \, \pr_{\, \m_1} \, \L_{\, \m_2 \, \cdots \, \m_s} \, + \, \cdots \, ,
\ee
it is already possible to notice that if $\vf_{\, \m_1 \, \cdots \, \m_s}$ and $\L_{\, \m_1 \, \cdots \, \m_{s-1}}$ no longer satisfy the first equations in \eqref{fierzsyst}, as required for the system to be off-shell, then compensating the gauge variation of $\Box  \, \vf_{\, \m_1 \, \cdots \, \m_s}$ leads to forego the condition on the vanishing of its divergence as well and to construct the combination displayed in \eqref{Max}. In this sense one can interpret the Maxwell-like tensor $(M\, \vf)_{\, \m_1 \, \cdots \, \m_s}$ as providing the minimal building block necessary for any off-shell extension of \eqref{fierzsyst}, and our goal in the present paper is to show that the same operator actually also suffices for the same purpose. Differently, the trace conditions in \eqref{fierzsyst} appear at this level as optional possibilities and, as we shall see, keeping or discarding them in the off-shell formulation can affect the spectrum of the resulting theories but not the form of the corresponding Lagrangians.

 The key idea underlying the whole construction is to allow for a restricted form of gauge symmetry with parameters subject to a  suitable set of transversality conditions. For instance, as  we show in section  \ref{sec:flat_symm}, in order to enforce invariance of $(M\, \vf)_{\, \m_1 \, \cdots \, \m_s}$ under \eqref{abelian} the simplest option is indeed to require that the gauge parameter $\L_{\, \m_1 \, \cdots \, \m_{s-1}}$ be \emph{divergence-free},
\be \label{TD}
\pr^{\, \a} \, \L_{\, \a \, \m_2 \, \cdots \, \m_{s-1}} \, = \, 0 \, ,
\ee
allowing to dispense with the introduction of additional terms involving traces of the field, like those appearing for the same class of tensors in Fronsdal's equation \cite{fronsdal}
\be \label{F}
(\cF \, \vf)_{\, \m_1 \, \cdots \, \m_s} \, = \,  (M\, \vf)_{\, \m_1 \, \cdots \, \m_s} \, + \, 
\left(\, \pr_{\, \m_1} \, \pr_{\, \m_2} \, \vf^{\, \a}{}_{\a \, \m_3 \, \cdots \, \m_s} \, + \, \cdots \,\right)\, = \, 0 \, .
\ee
A further distinction concerns the analysis of the spectra: while \eqref{F}, supplemented with the condition that the gauge parameter be traceless, describes the propagation of a single massless particle of spin $s$, the spectrum associated to \eqref{Max} comprises a whole set of particles of spin $s, \, s-2, \, s-4, \, \cdots$ and so on, down to spin $s=1$ or $s = 0$, thus providing a reducible description of higher-spin dynamics. However, without altering the form of the corresponding Lagrangian, easily seen to be given by
\be \label{ML}
\cL \, = \, \12 \, \vf_{\, \m_1 \, \cdots \, \m_s} \, (M\, \vf)^{\, \m_1 \, \cdots \, \m_s} \, ,
\ee
it is also possible to truncate the particle content of divergence-free theories to the single irreducible representation of highest spin $s$ by further restricting both the field $\vf_{\, \m_1 \, \cdots \, \m_s}$ and  the gauge parameter $\L_{\, \m_1 \, \cdots \, \m_{s-1}}$ to be traceless, as originally shown in \cite{SV}. Similar considerations apply to symmetric tensors in (A)dS backgrounds, to which our construction extends with no special difficulties both for reducible and irreducible theories, as discussed in section \ref{sec:ads_symm}.

  Our approach proves especially effective in simplifying the Lagrangian formulation of theories involving tensors with mixed symmetry. In section \ref{sec:flat_mixed} we study the general case of multi-symmetric tensors with $N$ families of indices,
\be 
\vf_{\, \m_1 \, \cdots \, \m_{s_1}, \, \n_1 \, \cdots \, \n_{s_2}, \,  \cdots \,} \, ,
\ee
defining $GL(D)$-reducible representations, showing that a  consistent  Lagrangian for their massless particle content in flat space-time is simply
\be \label{MLmix}
\begin{split}
\cL = \12 \, \vf_{\, \m_1 \, \cdots \, \m_{s_1}, \, \n_1 \, \cdots \, \n_{s_2}, \,  \cdots} \bigg\{ \Box \, \vf^{\, \m_1 \, \cdots \, \m_{s_1}, \, \n_1 \, \cdots \, \n_{s_2}, \,  \cdots}& -  \, (\pr^{\, \m_1} \, \pr_{\a} \, \vf^{\, \a \, \m_2 \, \cdots \, \m_{s_1}, \, \n_1 \, \cdots \, \n_{s_2}, \,  \cdots \,} \, + \, \cdots ) \\ 
&  - \, (\pr^{\, \n_1} \, \pr_{\a} \, \vf^{\, \m_1 \, \cdots \, \m_{s_1}, \, \a\, \n_2 \, \cdots \, \n_{s_2}, \,  \cdots} \, + \, \cdots ) \\
& - \, \cdots \, \bigg\},
\end{split}
\ee
where within parentheses symmetrisations over indices belonging to a single family are understood. Gauge invariance of \eqref{MLmix} under 
\be \label{GT}
\begin{split}
\d \, \vf_{\, \m_1 \, \cdots \, \m_{s_1}, \, \n_1 \, \cdots \, \n_{s_2}, \,  \cdots}\, & =  \, 
(\pr_{\, \m_1} \, \L_{\, \m_2 \, \cdots \, \m_{s_1}, \, \n_1 \, \cdots \, \n_{s_2}, \,  \cdots \,} \, + \, \cdots) \\
& + \, \, ( \pr_{\, \n_1} \, \l_{\, \m_1 \, \cdots \, \m_{s_1}, \, \n_2 \, \cdots \, \n_{s_2}, \,  \cdots \,} \, + \, \cdots )\\
& + \,\,  \cdots \, ,
\end{split}
\ee
is guaranteed provided the $N$ parameters $\L_{\, \m_1 \, \cdots \, \m_{s_1 - 1}, \, \n_1 \, \cdots \, \n_{s_2}, \,  \cdots \,} $, $\l_{\, \m_1 \, \cdots \, \m_{s_1}, \, \n_1 \, \cdots \, \n_{s_2 - 1}, \,  \cdots \,}, \, \cdots $, each missing one index in a given group of symmetric indices, satisfy a set of generalised transversality conditions that, for instance in the case of two-family tensors, take the form
\be \label{TDmix}
\begin{split}
& (\pr_{\m_1} \pr_{\m_2} \, \pr^{\, \a} \, \L_{\, \a \, \m_3 \, \cdots \, \m_{s_1}, \, \n_1 \, \cdots \, \n_{s_2}} \, + \, \cdots )\, + \, 
(\pr_{\n_1} \pr_{\n_2} \, \pr^{\, \a} \, \l_{\, \m_1 \, \cdots \, \m_{s_1}, \, \a \, \n_3 \, \cdots \, \n_{s_2}} \, + \, \cdots)\\
&+ \, \12\, \pr_{\m_1} \pr_{\n_1} \,\pr^{\, \a} \, \left(\L_{\, \m_2 \, \cdots \, \m_{s_1}, \, \a \, \n_2 \, \cdots \, \n_{s_2}\,} \, 
+ \, \l_{\, \a \, \m_2 \, \cdots \, \m_{s_1}, \,\n_2 \, \cdots \, \n_{s_2}}  \right) \, + \, \cdots \, = \, 0 \, ,
\end{split}
\ee
as also synthetically expressed by eq.~\eqref{enh} for the general case of $N$ families, in a more appropriate notation. 

Similarly to the symmetric case, this constrained local symmetry is  indeed sufficient to ensure that the propagating degrees of freedom eventually sit in  the totally transverse reducible tensor of $GL(D-2)$ 
\be
\vf_{\, i_1 \, \cdots \, i_{s_1}, \, j_1 \, \cdots \, j_{s_2}, \,  \cdots}\, , \\
\ee
whose branching in $O(D-2)-$irreps describes the full particle  content associated to \eqref{MLmix}. Comparing \eqref{MLmix} with the constrained Lagrangian of Labastida \cite{labastida} or with its minimal unconstrained versions given in \cite{cfms1,cfms2} (see also \cite{andreaRev,andreaNC} for reviews) allows to appreciate the advantages of our present approach\ft{
Fields of mixed symmetry in Minkowski backgrounds have been subject to intense study since the mid-eighties, following the early progress of string field theory \cite{mixedold}. Here we discuss higher-spin fields as generalisations of the metric tensor for gravity (metric-like approach). Alternative forms of Lagrangians for tensors of mixed-symmetry in flat space have been obtained following various other approaches, including metric-like BRST-inspired formulations \cite{brst-flat} or frame-like ones  \cite{mixednew} generalising Cartan-Weyl gravity.}: while  Lagrangians \eqref{MLmix} always maintain the same form irrespective of the number $N$ of index-families of the tensor $\vf$, in the approach of \cite{labastida,cfms1} the need to implement the more conventional kind of gauge invariance calls for the introduction of a  number of traces of the basic kinetic tensor increasing with $N$. (See eq.~\eqref{l.laba}.) Moreover, the same actions \eqref{MLmix} extend with no modification to the case of tensors in irreducible representations of $GL(D)$ or $O(D)$, so that even in the mixed-symmetry case transverse invariance can accommodate spectra of various degrees of complexity.

 An important remark concerns the analysis of the system \eqref{TDmix} that, when solved in momentum space, displays a different number of solutions according to whether or not the momenta are light-like, and in particular the appearance of additional solutions for the case of light-like momenta allows to exactly compensate a corresponding weakening of the equations of motion in their capability to set to zero some of the unphysical components of the gauge field. In this respect, however, we find some differences in dealing with the $O(D)-$case, where due to the lack of sufficient local symmetry we had to introduce constraints on a subset of double divergences of the field. In this sense, it appears that the reduction to single-particle propagation is not a natural outcome in the framework of transverse-invariant, Maxwell-like theories. However it might be also observed that the corresponding  one-particle description in our transverse-invariant setting appears somehow to mirror the corresponding construction of Labastida \cite{labastida}, where general consistency of the whole setup requires the (symmetric) traces of the gauge parameters  and the (symmetric) double traces of the gauge field to vanish. 

 Remarkably, proceeding along the same  lines it turns out to be possible to extend the Lagrangian formulation for tensors of any symmetry to the case of backgrounds with non-vanishing cosmological constant. Whereas a considerable body of knowledge is by now available for field theories involving symmetric tensors of arbitrary rank, both in flat and in particular in (A)dS backgrounds, where interactions among massless higher-spin particles seem to find a most natural arena \cite{vasiliev4d}, when it comes to tensors with mixed-symmetry the situation is vastly different. Indeed, in cosmological spaces not only are interactions for these types of particles so far little explored \cite{int-ads-mixed}, but even free Lagrangians are available only for special classes of tensors, both in the metric-like approach that we pursue in this paper \cite{mixed-metric} and in the frame-like approach where the higher-spin degrees of freedom are encoded in sets of generalised vielbeins \cite{mixed-frame,zin-fermi,zin-bose} (see also \cite{Skvortsov, AGads, AG}). This essential gap in our knowledge is especially acute since mixed-symmetry states account for the vast majority of the string excitations. In this sense, it seems reasonable to expect that a satisfactory comparison between massless  higher-spins and strings may benefit from a more complete understanding of  the general types of massless particles allowed in a given space-time dimension. 
 
 In this work we propose an action for arbitrary massless fields in (A)dS spaces of any dimension, eq.~\eqref{lag_mix_ads} or \eqref{lag_mix_block}, and in the remainder of this introduction we would like to provide a few additional details on the peculiarities of massless particles in (A)dS spaces, in order to better frame the main lines of our procedure.

 The investigation of theories involving fields in arbitrary representations of the AdS or dS groups in $D$ dimensions, $O(D-1, 2)$ or $O(D, 1)$ respectively, besides the technical complications already present in the flat-space analysis, is fraught with additional subtleties that are absent for the more customary symmetric representations. In particular, as first shown by Metsaev in \cite{metsaev1, metsaev2}, the very notion of single, massless particle does not admit in general a continuous deformation from flat to (A)dS backgrounds and vice-versa, on account of the impossibility of preserving all the gauge symmetries of the flat theory. The analysis of \cite{metsaev1, metsaev2} elucidates the on-shell conditions to be satisfied in order for the wave operator to retain the maximal possible amount of gauge symmetry in (A)dS backgrounds, while also providing the further specifications needed to grant unitarity in Anti-de Sitter space. The general result is that, out of the $p$ gauge parameters in principle available for tensors described by Young diagrams possessing $p$ rectangular blocks of different horizontal lengths, \emph{only one} can be kept in (A)dS. Moreover, while gauge invariance alone does not distinguish among the $p$ options available in principle, for the case of Anti-de Sitter spaces unitarity dictates to preserve the parameter represented by the  diagram missing one box in the upper rectangular block.
 
 As a general consequence, (A)dS massless ``particles'' associated with a given diagram describe the propagation of more degrees of freedom than their flat-space peers. The exact branching of these irreducible (A)dS representations in terms of $O(D-2)$ ones (i.e.\ the structure of the flat-space multiplet corresponding to a single (A)dS particle)   was first conjectured in \cite{bmv} by Brink, Metsaev and Vasiliev and was more recently subject to a detailed group-theoretical analysis in \cite{NCP, AGads}; for instance, the unitary BMV multiplet associated with the massless AdS particle $\vf_{\, \m \, \n, \, \r}$ with the symmetries of the hook tableau $\{2, 1\}$  comprises the degrees of freedom of the flat-space particle described by the same hook diagram together with those of a ``graviton''. However, as already mentioned, while the pattern of flat massless particles branching single (A)dS massless irreps is indeed known in the general case, so far its off-shell realisation has been provided only for special classes of Young diagrams.

 In eq.~\eqref{lag_mix_ads} we propose Lagrangians for general $N$-family, $O(D)-$tensor fields in (A)dS, uniquely determined requiring that they preserve the amount of gauge-symmetry dictated by Metsaev's analysis; in particular for the unitary choice identified in \cite{metsaev1, metsaev2} \eqref{lag_mix_block} reads
\be \label{MLmixAdS}
\begin{split}
\cL = \12  \vf_{\, \m_1  \cdots  \m_{s_1}, \, \n_1  \cdots  \n_{s_2},  \cdots} & \bigg\{ \Box \, \vf^{\, \m_1  \cdots  \m_{s_1}, \, \n_1  \cdots  \n_{s_2}, \,  \cdots} -  (\nabla^{\, \m_1} \, \nabla_{\a} \, \vf^{\, \a \, \m_2  \cdots  \m_{s_1}, \, \n_1  \cdots  \n_{s_2}, \,  \cdots \,} \, + \, \cdots ) \\ 
&  \! \! \!  \! \! \!  \! \! \!  \! \! \! -  \frac{1}{L^2}\, \bigg[(s_1 - t_1 - 1)(D + s_1 - t_1 - 2) \,- \, \sum_{k\,=\,1}^p s_k \, t_k \bigg]\, \vf^{\, \m_1  \cdots  \m_{s_1}, \, \n_1  \cdots  \n_{s_2}, \cdots} \bigg\}\, ,
\end{split}
\ee
where $s_1$ and $t_1$ identify length and height of the first rectangular block, while the sum runs over the products of lengths and heights of all different blocks.  This result provides a relatively simple generalisation of the corresponding Lagrangian for symmetric tensors in (A)dS \eqref{lag_ads_symm}, to which it reduces for  $p = 1$ and  $t_1 = 1$. However, differently from that example and from the flat-space case, where our construction applies both to reducible and to irreducible tensors, the Lagrangians \eqref{MLmixAdS} involve  traceless tensors in irreducible representations of $GL(D)$, and as such define candidate single-particle theories. Consistently with the analysis of \cite{metsaev1}, in this irreducible context the gauge invariance of \eqref{MLmixAdS} is meant under transformations involving a single fully divergenceless and traceless parameter. However, in order to keep the simplicity of our  Lagrangians for this class of field one has to impose severe restrictions on the gauge field too, since, as a consequence of the reduced amount of gauge symmetry available, all divergences of the field but one would be gauge invariant, and cannot be disposed by the equations of motion either. Thus, consistency of the equations obtained from \eqref{MLmixAdS} requires to work with tensors whose gauge-invariant divergences vanish. Due to irreducibility, in the case of \eqref{MLmixAdS} it is sufficient that the fields satisfy
\be
\nabla_{\a} \, \vf^{\, \m_1  \cdots  \m_{s_1}, \, \a\, \n_2  \cdots  \n_{s_2}, \,  \cdots} \, = \, 0 \, ,
\ee
since this condition automatically implements all the constraints we need. Although in our opinion it is still notable that one could encompass the fully general case in such a compact Lagrangian as \eqref{MLmixAdS}, it is clear that in this sense the result is only partially satisfactory. To reiterate, the essential difficulty that characterises massless, mixed-symmetry fields in AdS is that most of the gauge invariance of the corresponding flat theory gets lost and thus new degrees of freedom participate the dynamics. They organise in the BMV patterns, whose structure clearly depends on the relation between the gauge freedom that is lost when moving from flat space and that which is kept, in a way that we shall now try to elucidate.

 A convenient way to get some intuition about the unfamiliar BMV phenomenon is to observe that, in (A)dS backgrounds, the pattern of \emph{reducible} gauge transformations associated in general to a mixed-symmetry gauge potential \emph{gets generically broken by terms proportional to the (A)dS curvature}. With reference to the example of the (traceless) hook tensor $\{2, \, 1\}$ with covariantised gauge variation
\be \label{gauge-intro}
\d \, \vf_{\, \m \, \n, \, \r} \, = \, \nabla_{\m} \, \L_{\, \n, \, \r} \, + \, \nabla_{\n} \, \L_{\, \m, \, \r}\, +
\, \nabla_{\r} \, \l_{\, \m \n}\, - \, \12 \, \big(\nabla_{\m} \, \l_{\, \n \r}\, + \, \nabla_{\n} \, \l_{\, \m \r}\big)\, ,
\ee
where $\L_{\, \m, \, \n}$ is a two-form while $\l_{\, \m \n}$ is a symmetric tensor, it is not hard to verify that, under the ``gauge-for-gauge'' transformations 
\be \label{G2}
\d \, \L_{\, \m, \, \r} \, = \, \nabla_{\r} \, \theta_{\, \m} \, - \, \nabla_{\m} \, \theta_{\,\r} \, , \hspace{1cm} \d \, \l_{\, \m \, \n} \, = \, -\, 2 \,  \big(\nabla_{\m} \, \theta_{\,\n} \, + \, 
\nabla_{\n} \, \theta_{\,\m}\big)\, ,
\ee
that would leave the gauge potential invariant in flat space-time, now $\d \, \vf_{\, \m\n, \, \r}$ acquires a contribution proportional to the (A)dS curvature:
\be
\d \, \vf_{\, \m\n, \, \r} \, = \, - \, 2 \, \bigg([\nabla_{\, \r}, \, \nabla_{\, \m}] \, \theta_{\, \n}\, 
+ \,  [\nabla_{\, \r}, \, \nabla_{\, \n}] \, \theta_{\, \m}\bigg)\,  .
\ee
This observation implies that even if one were able to find a kinetic tensor for $\vf_{\, \m\n, \, \r}$ invariant under the \emph{full} transformation \eqref{gauge-intro} the corresponding theory would possess too much gauge invariance with respect to the flat case, and thus would not describe anymore the degrees of freedom of the $O(D-2)$ hook. As an alternative to the elimination of one of the parameters of the flat theory, suggested by the result of \cite{metsaev1, metsaev2}, one can instead ``neutralise'' the effect of the broken gauge-for-gauge vector $\th_{\, \m}$ ``promoting'' it to play the role of  gauge parameter for a new $O(D)-$field, that at this level could be either a symmetric rank-two tensor or a two-form. One could then proceed to construct the full Stueckelberg Lagrangian for both options, taking care of the fact that not only the ``standard'' gauge invariance but also overall gauge-for-gauge invariance be simultaneously preserved, in such a way that the number of independent gauge components of the resulting (A)dS theory be the same as for the corresponding flat system. The important difference between the two possible options is that, now referring specifically to Anti-de  Sitter backgrounds, only for the rank-two symmetric tensor would the corresponding kinetic operator emerge with the correct sign required by a unitary theory, in accordance with the group-theoretical analysis of \cite{metsaev1, metsaev2}. Proceeding in this fashion for more general cases one can obtain an independent justification of the full BMV-pattern, and construct the corresponding Stueckelberg Lagrangians smoothly deforming the sum of flat-space Lagrangians appropriate for the description of the corresponding multiplet. 

We consider these issues explicitly for the case of tableaux with two rows in section \ref{sec:dof_ads_mixed}, showing how the broken generations of gauge-for-gauge symmetry can be recovered in a Stueckelberg-like scenario where all physical fields of the multiplet get mixed under the full set of gauge transformations. However, in trying to implement this program at the Lagrangian level starting from our $O(D)$ theories, we found that the need to properly deform the double-divergence constraints introduced in that context brings about a number of technical difficulties that prevented so far to extend the construction beyond the case of $\{s, 1\}-$irreducible tensors, for which, at any rate, together with the $O(D)$ construction we also provide the Stueckelberg deformation of the Labastida Lagrangian. Let us also mention that the Stueckelberg construction, also discussed in \cite{bmv} for the AdS particle with the symmetries of the hook tableau, has been explored in particular by Zinoviev in \cite{zin-fermi,zin-bose}, where frame-like Lagrangians for massive two-family tensors in (A)dS were obtained and their massless and partially massless limits were also discussed.

  Whenever a system is found to be invariant under constrained gauge transformations it is natural in our opinion to try and interpret it as resulting from the partial gauge-fixing of a more general theory whose gauge symmetry is not constrained. Indeed, for the irreducible case involving traceless fields and transverse-traceless parameters investigated in \cite{SV}, the Lagrangian \eqref{ML} can be seen to arise from a partial gauge fixing of Fronsdal's Lagrangian itself, whose formulation requires traceless parameters (and doubly traceless fields). In its turn, the Fronsdal-Labastida theory admits minimal unconstrained extensions given in \cite{fs3, fms1, cfms1, cfms2}, building on previous formulations \cite{fs1, fs2, dario07} where the removal of constraints was linked to the possibility of assigning a dynamical role to the higher-spin curvatures of \cite{dwf}\ft{For alternative formulations of the free theory of massless higher spins see e.g.\ \cite{brst-symm}.}. For the transverse-invariant Lagrangians that we propose in this work the most natural unconstrained extensions should be identified with the ``triplets'' associated to the tensionless limit of free open string field theory \cite{triplets, fs2, st, dariotripl} (see also \cite{Bonelli,FoTsu,SorVas}), whenever the corresponding actions are available. However, for the case of (A)dS tensors with mixed symmetry the corresponding unconstrained Lagrangians are not yet known, and exploring the possibility of constructing them and their possible relation with string systems in (A)dS is an interesting question that we leave for future investigation.
 
 For the case of spin $2$ the idea of considering transverse-diffeomorphism invariance is indeed quite old and was explored from a number of perspectives, mainly in connection with so-called unimodular gravity and its relation to the cosmological constant problem \cite{uni-old,uni-ham,uni-new}. The observation is that, while the variation of the Einstein-Hilbert action performed keeping the determinant of the metric fixed provides only the traceless part of Einstein's equations, the (contracted) Bianchi identity allows to recover the relation between the Ricci scalar and the trace of the stress-energy tensor, up to an arbitrary integration constant appearing in the resulting equation as a cosmological term. The connection with our approach is established observing that, at the linearised level, demanding that the determinant of the metric be gauge invariant requires \emph{transverse} vector parameters, thus providing the first non-trivial example of \eqref{TD}.  Let us also mention that conditions of transversality on gauge parameters were recently considered in the context of quantum-mechanical models on K\"ahler manifolds in \cite{BB}.

We present the main results of this work in section \ref{sec:lagr}, where we perform the construction of transverse-invariant Lagrangians in increasing degree of generality, from symmetric tensors in Minkowski space to mixed-symmetry fields in (A)dS. The spectrum of particles propagating in the corresponding equations of motion is then analysed in section \ref{sec:spec} exploiting various approaches. As already mentioned, when built out of traceful fields our Lagrangians propagate reducible spectra of free particles. This is the situation where the highest simplification is obtained (taking into account the structure of the Lagrangians, the form of the equations of motion and the analysis of the spectrum) and the closest contact with the tensionless open string is achieved. In section \ref{sec:diag} we perform an additional step for the case of symmetric tensors, providing the solution to the problem of dissecting the field $\vf$ so as to explicitly identify its lower-spin components; as a result the action gets decomposed into a sum of decoupled terms, one for each particle present in the spectrum of the theory. We conclude summarising our findings while also putting them in a more general perspective, while in the appendices we collect our notations and conventions together with a number of additional technical results.

%%%%%%%%%%%%%%%%%%%%%%%%%%%%%%%%%%%%%%%%%%%%%%%%%%%%%%%%%%%%%%%%%%%%%

\section{Lagrangians}\label{sec:lagr}

%%%%%%%%%%%%%%%%%%%%%%%%%%%%%%%%%%%%%%%%%%%%%%%%%%%%%%%%%%%%%%%%%%%%%

%%%%%%
\subsection{Flat backgrounds}\label{sec:flat}
%%%%%%

%%
\subsubsection{Symmetric tensors}\label{sec:flat_symm}

 Let us consider the Lagrangian\ft{Our notation and conventions are spelled out in appendix \ref{app:formulae}.
Symmetrised indices are always implicit and symmetrisation is understood with no weight factors. In particular ``$\pr$'' stands
for a  gradient,  while the symbol ``$\prd$'' denotes a divergence. Thus, in the Maxwell-like equations for a rank-$s$  tensor, 
$(\Box \, - \, \pr \, \prd)\vf = 0$, the second term actually contains $s$ contributions:
$\pr \, \prd \vf = \pr_{\m_1} \, \pr^{\a} \vf_{\a \m_2 \cdots \m_s} \, + \,  \pr_{\m_2} \, \pr^{\a} \vf_{\a \m_1 \cdots \m_s} \, + \, \cdots$.} 
\be \label{tlagr}
\cL \, = \, \12 \, \vf \, M\, \vf \,  ,
\ee   
where $M$ is the Maxwell operator
\be \label{M}
M \, = \, \Box \, - \, \pr \, \prd \, \, .
\ee
Up to total derivatives, its gauge variation under $\d \, \vf \, = \, \pr \, \L$ is
\be
\d \, \cL \, = \, - \, 2 \, {s \choose 2} \, \prd \prd \vf  \, \prd \L \, ,
\ee 
and thus vanishes assuming the condition of transversality for the gauge parameter:
\be \label{tdiff}
\prd \L \, = \, 0 \, .
\ee
One could alternatively impose a differential constraint on the gauge field of the form $\prd \prd \vf = 0$, that would also guarantee gauge invariance of \eqref{tlagr}. However, in order for this latter condition to be itself gauge invariant, the parameter should satisfy in principle a more involved transversality condition of the form $2 \, \Box \, \prd \L \, + \, \pr \, \prd \prd \L = 0$, so that it does not seem especially convenient to proceed in this direction. We shall reconsider the possibility to assume auxiliary conditions on the double divergences of the field in section \ref{sec:flat_mixed}, when discussing fields with mixed symmetry.

Our interest in these kind of systems has several motivations, the first clearly being the appeal of simplicity. As we will show in this work,  they provide an alternative route to the description of massless higher spins in their full generality, finding their original inspiration in the so-called TDiff-invariant spin-$2$ theories originally considered in \cite{uni-old,uni-ham} and more recently in \cite{uni-new} in connection with unimodular gravity and with the cosmological constant problem. Moreover, in a number of cases one can relate transverse-invariant theories to the triplet Lagrangians emerging from the tensionless limit of the free open string \cite{triplets, fs2, st}, of which they effectively provide a simplified version retaining the same particle content. 

A detailed analysis of the spectrum of  Lagrangian \eqref{tlagr} and of its generalisations is presented in section \ref{sec:spec}. In the specific case of interest in this section one can also connect the corresponding equations of motion,
\be \label{eomsymm}
(\Box  \, - \, \pr \, \prd) \, \vf \,  = \, 0 \, ,
\ee
to the reduced Fierz system \cite{Fierz},
\be \label{FS}
\begin{split}
&\Box \, \vf \, = \, 0 \, , \\
&\prd \vf \, = \, 0\, .
\end{split}
\ee
Indeed computing a divergence of \eqref{eomsymm} gives $\pr \, \prd \prd \vf \,  = 0$, and thereby effectively 
\be \label{doublediv}
\prd \prd \vf \, = \, 0 \, ,
\ee
up to discrete degrees of freedom that we will systematically neglect, since they do not affect the counting of local degrees of freedom which is our main object of interest in the present framework. The remaining, transverse part of the divergence of $\vf$ can be gauged away using the divergence-free parameter $\L$ on account of
\be
\d \, \prd \vf \, = \, \Box \, \L \, ,
\ee
thus showing the equivalence of \eqref{eomsymm} with \eqref{FS} supplemented by the appropriate residual gauge invariance with parameter satisfying
\be \label{FSP}
\begin{split}
& \Box \, \L \, = \, 0 \, , \\
& \prd \L \, = \, 0\, .
\end{split}
\ee
A standard analysis of \eqref{FS} and \eqref{FSP} shows that the propagating polarisations are those associated with the components $\vf_{\, i_1 \cdots \, i_s}$, where the indices $i_k$ refer to directions transverse to the light-cone. Thus, together with the spin-$s$ degrees of freedom contained in  the traceless part of $\vf_{\, i_1 \cdots \, i_s}$ in $D-2$ Euclidean dimensions, lower-spin representations of spin $s - 2k$, with $k = 1, \ldots, [\frac{s}{2}]$, also propagate and sit in the traces of $\vf_{\, i_1 \cdots \, i_s}$.  For the  irreducible case describing a single particle of spin $s$, already studied in \cite{SV}, it will suffice to observe that, up to a restriction of the space of fields to traceless tensors subject to gauge variations with transverse and traceless parameters, one does not need to modify the form \eqref{tlagr} of the Lagrangian, that in this sense applies to both reducible and irreducible descriptions. The corresponding equations of motion obtain  taking the traceless projection of \eqref{eomsymm} and read
\be
\left(\, \Box  \, - \, \pr \, \prd \, + \, \fr{1}{D \, + \, 2\, s \, - \, 4}\, \h \, \prd \prd \,\right) \vf \,  = \, 0 \, .
\ee

 Let us also mention that in our reducible context, with unconstrained fields subject to transverse and \emph{traceful} gauge variations, multiple divergences of $\vf$ of order higher than one and, for even spins, the highest order trace $\vf^{\, [\fr{s}{2}]}$, provide independent gauge-invariant quantities that could possibly enter modified forms of the Lagrangian. This is in particular true for the spin-$2$ case, where $\vf^{\, \pe}$ and $\prd \prd \vf$ could be combined in various forms providing gauge-invariant modifications of \eqref{tlagr}. For arbitrary spins, even limiting the attention to kinetic operators containing no more than two derivatives, we observe that uniqueness of \eqref{tlagr} is always meant up to the scalar sector of the even-spin case admitting indeed possible deformations, both in the forms of mass terms 
\be
\D \, \cL_m \, = \, \12 \, m^{\, 2} \, (\vf^{\, [\frac{s}{2}]})^{\, 2}\, ,
\ee
or as additional kinetic operators for the scalar member of the multiplet,
\be
\D \, \cL_K \, = \, \fr{a}{2}\, \vf^{\, [\fr{s}{2}]} \, \Box \, \vf^{\, [\fr{s}{2}]}\, ,
\ee
amenable in principle to change the propagating nature of the latter, and possibly to eliminate it altogether from the spectrum by a suitable choice of the coefficient $a$\ft{See \cite{SorVas} for a discussion of reducible triplets in the frame-like approach, where in particular a similar arbitrariness at the level of the scalar component of the multiplet was also noticed.}. 

 The reducible particle content associated to eq.~\eqref{eomsymm} in the absence of trace constraints corresponds to that of the Lagrangians obtained from the BRST action for the free open string, after taking the tensionless limit $\a^{\, \pe} \rightarrow \infty$ in the sense described in \cite{triplets, fs2, st}. The spin-$s$ block-diagonal term obtained in that approach (after solving for an additional auxiliary field with algebraic equations of motion) reads in fact
\be \label{tripletnoC}
\cL \, = \, \12 \, \vf \, M \, \vf \, + \, 2 \, \vf \, \pr^{\, 2} \, \cD \, - \, 2 \binom{s}{2} \, \cD \hat{M} \, \cD \, ,
\ee
where $\vf$ and $\cD$ are symmetric  tensors of ranks $s$ and $s-2$ respectively, subject to the unconstrained gauge transformations $\d \, \vf \, = \, \pr \, \L$ and $\d \, \cD \, = \, \prd \L$, while
\be \label{Mdeform}
\hat{M} \, = \, \Box \, + \, \12 \, \pr \, \prd \, ,
\ee
is a sort of deformed Maxwell operator for the field $\cD$. In this respect our analysis shows that  performing the off-shell gauge-fixing $\cD = 0$ does not alter the spectrum of the theory. Conversely, one can generate the Lagrangian \eqref{tripletnoC} from \eqref{tlagr} in two steps: first introducing a  Stueckelberg field $\th$ via the redefinition $\vf \, \rightarrow \, \vf - \pr \, \th$, with $\d \, \vf = \pr \, \L$ and $\d \, \th = \L$, and then identifying the divergence of $\th$ with the  field $\cD$ of \eqref{tripletnoC}.  In this sense the relation between the transverse-invariant Lagrangian \eqref{tlagr} and the unconstrained triplet Lagrangian \eqref{tripletnoC} is analogous to the relative role played by Fronsdal's constrained theory, with traceless gauge parameter and doubly-traceless field, and its minimal unconstrained extension proposed in \cite{fs3} for the description of irreducible massless particles of spin $s$.

Modifying the constraints on the field  and on the gauge parameter in various ways (which include the option of relaxing them altogether) several possible completions of \eqref{tlagr} can be found, local and non-local (see also our comments in the Discussion). Concerning the latter, a rationale for the introduction of non-local terms in unconstrained Lagrangians is found whenever it is possible to interpret them as the result of the integration over unphysical fields \cite{dariocrete}. The corresponding analysis for the triplets was performed in \cite{dariotripl} where it was shown that the elimination of the field $\cD$ in \eqref{tripletnoC} produces  indeed a gauge-invariant completion of \eqref{tlagr} given by the Lagrangian
\be \label{triplet}
\begin{split}
\cL \, &=  \, \12 \, \vf \, M \, \vf \, + \sum_{m \,=\, 2}^{s} \, {s \choose m}\, 
\prd^{\, m} \vf \, \fr{1}{\Box^{\, m - 1 }} \, \prd^m \vf \, \\
& = \, \12 \, \sum_{m \,=\, 0}^{s} \, {s \choose m}\, 
\prd^{\, m} \vf \, \fr{1}{\Box^{\, m - 1 }} \, \prd^m \vf \, ,
\end{split}
\ee
where $\prd^{\, m}$ denotes the $m-$th power of the divergence. In the same context it was also shown  how to combine the various terms in \eqref{triplet} in order  to reproduce the square of the corresponding higher-spin curvatures $\cR^{\, (s)}_{\, \m_1 \cdots\, \m_s, \, \n_1 \cdots\,  \n_s}$ \cite{dwf}, leading to the compact expression\ft{Lagrangians for symmetric bosons and fermions of arbitrary spin were first formulated in terms of metric-like curvatures in \cite{fs1} while an approach similar in spirit was also proposed for mixed-symmetry bosons in \cite{dmh, bb}. Out of the infinitely many options available in principle,  the unique Lagrangians leading to the correct propagators were given for symmetric tensors in \cite{fms1}, while their massive deformations were discussed in \cite{dario07}, together with a more detailed analysis of the role of curvatures for fermionic theories. Their connection with the minimal local Lagrangians of \cite{fs3} was given in \cite{dariocrete}. More recently, first-order non-linear deformation of the curvatures were also found  in \cite{mmrt}.} 
\be \label{Maxwellbose}
\cL \, =  \, \fr{(-1)^{\, s}}{2\, (s + 1)} \, 
\cR^{\, (s)}_{\, \m_1 \cdots\, \m_s, \, \n_1 \cdots\,  \n_s} \, \fr{1}{\Box^{s - 1}} \, \cR^{\, (s)\, \m_1 \cdots\, \m_s, \, \n_1 \cdots\,  \n_s} \, .
\ee 
In \eqref{Maxwellbose} the spin$-1$ case corresponds to the only local option, while the non-local Lagrangians obtained for spin $s \geq 2$ build a metric-like generalisation of Maxwell's Lagrangian in its geometric form bearing the same particle content as the triplet system \eqref{tripletnoC}.

 Let us finally observe that, in analogy with the spin$-1$ case, where the divergence of the field strength defines the equations of motion, in our present setting  the kinetic tensor $M$ can be easily related to the first connexion in the de Wit-Freedman hierarchy\ft{To manipulate the generalised connexions of  \cite{dwf} we resort to a notation \cite{dario07}, where symmetrised indices are denoted with the same symbol, while the subscript denotes the number of indices in a given group. For instance $ \pr_{\, \m} \, \vf_{\, \m_{s-1} \r}$ is a shortcut for $ \pr_{\, \m_1} \, \vf_{\, \m_2 \, \cdots\, \m_{s} \r} \, + \, \pr_{\, \m_2} \, \vf_{\, \m_1 \, \m_3\, \cdots\, \m_{s} \r} \, + \, \cdots \, $, with the index $\r$ excluded from the symmetrisation. For the manipulations required in this section the rules of symmetric calculus given in Appendix \ref{app:form_sym} apply separately for each group of indices.  \label{Notation}} \cite{dwf}
\be \label{G1}
\G^{(1)}{}_{\, \r, \, \m_s} \, =\, \pr_{\,\r} \, \vf_{\, \m_s} \, - \, \pr_{\, \m} \, \vf_{\, \m_{s-1} \r} \, ,
\ee
which clearly reduces to Maxwell's field strength for $s =1$. Indeed, it is simple to check that the divergence of $\G_{\, \r, \, \m_s}$ in the $\r$ index builds the Maxwell operator \eqref{M}, 
\be \label{gamma1}
\pr^{\, \r} \, \G^{(1)}{}_{\, \r, \, \m_s} \, = \, (M \, \vf)_{\, \m_s} \, ,
\ee
while the Lagrangian \eqref{tlagr} can be written as a square of those connexions as follows:
\be \label{Glagr}
\cL \, = \, \fr{1}{4\, (s  -1)} \, \left\{\, \G_{\, \m, \, \r \, \m_{s-1}} \, - \, (s - 2) \, \G_{\, \r, \, \m_s} \,\right\}  \G^{\, \r, \, \m_s} \, .
\ee
This observation suggests a clear parallel with the Fronsdal formulation, where the basic kinetic tensor $\cF$ given in \eqref{F} obtains from the trace of  the second connexion in the hierarchy of \cite{dwf}, 
\be  \label{gamma2}
\G^{(2)}{}_{\, \r \r, \, \m_s} \, = \,  \pr^{\, 2}_{\,\r} \, \vf_{\, \m_s} \, - \, \12 \, \pr_{\, \r} \, \pr_{\, \m} \, \vf_{\, \m_{s-1} \r} \, + \,  \pr^{\, 2}_{\,\m} \, \vf_{\,\m_{s-2} \r \r},
\ee
to be computed in the $\r$-indices, providing a spin$-s$ generalisation of  the linearised spin$-2$ Ricci tensor.

\subsubsection{Mixed-symmetry tensors}\label{sec:flat_mixed}

In this section we would like to extend our analysis to tensor fields of mixed symmetry.  The basic objects under scrutiny will  be \emph{multi-symmetric} tensors,
\be \label{multi}
\vf_{\m_1 \cdots\, \m_{s_1},  \, \n_1 \cdots\, \n_{s_2}, \, \cdots} \, \equiv \, \vf \, ,
\ee
possessing an arbitrary number $N$ of independent sets (``families'') of symmetrised indices and thus defining \emph{reducible} $GL(D)$ tensors, of the kind appearing as coefficients in the expansion of the bosonic string field. This choice provides indeed the highest degree of overall simplification, although our results can  be easily adapted to the case where $\vf$ defines an irreducible representation of $GL(D)$, as we shall see. For irreducible tensors of $O(D)$, on the other hand, while the construction of a gauge-invariant Lagrangian involves no additional difficulties the analysis of the corresponding equations of motion brings about a few novel complications that require a separate discussion.

The proper generalisation of the index-free notation used in the symmetric case was introduced  in \cite{cfms1,cfms2, andreaRev, andreaNC} and  is reviewed in appendix \ref{app:form_mix}. The basic idea is to employ ``family'' labels to denote operations adding or subtracting space-time indices belonging to a given group. More specifically upper family indices are reserved for operators, like gradients, which add space-time indices, while lower family indices are used for operators, like divergences, which remove them. As a result gradients and divergences acting on the $i-$th family are denoted concisely by $\pr^{\, i} \, \vf$ and $\pr_{\, i}\, \vf$ respectively, while  $T_{ij} \, \vf$ refers to a trace contracting one index in the $i-$th family with one index in the $j-$th family. It is also useful to introduce operators, denoted by $S^{\,i}{}_j$, whose effect is to displace indices from one family to another while also implementing the corresponding symmetrisation; namely
\be
S^{\, i}{}_j\, \vf \,\equiv\, \vf_{\,\cdots\,,\,(\m^i{}_1 \cdots\, \m^i{}_{s_i}| ,\, \cdots \,,\,|\m^i{}_{s_i+1})\, \m^j{}_1 \cdots\, \m^j{}_{s_j-1} ,\, \cdots}\, , \qquad \textrm{for}\ \ i \neq j \, ,
\ee
while their diagonal members $S^{\, i}{}_i$ essentially count the number of indices in the $i-$th family:
\be
S^{\, i}{}_i\, \vf \,\equiv\, s_i \ \vf_{\,\cdots\,,\,\m^i{}_1 \cdots\, \m^i{}_{s_i} ,\, \cdots} \, . %\label{def_S1}
\ee
Thus, given the multi-symmetric tensor $\vf$ in \eqref{multi}, the  corresponding $GL(D)$-diagram with the same index structure is characterised by the additional condition
\be \label{irredcond}
S^{\,i}{}_j \, \vf \, = \, 0\, ,  \qquad \mbox{for} \ i<j \, ,
\ee
while supplementing \eqref{irredcond}  with the tracelessness constraint
\be \label{trace}
T_{\, i j} \, \vf \, = \, 0 \, , \qquad \,\forall\, \, i, j \, ,
\ee
allows to deal directly with irreducible tensors of $O(D)$.

We first consider the case of multi-symmetric tensors \eqref{multi} and postulate Maxwell-like equations of motion
\be \label{mixedeom}
M\, \vf \, \equiv \, (\Box  \, - \, \pr^{\, i} \, \pr_{\, i}) \, \vf \,  = \, 0 \, ,
\ee   
together with the gauge transformations
\be \label{mixed_gauge}
\d \, \vf \, = \, \pr^{\, i} \, \L_{\, i} 
\ee
involving a set of multi-symmetric gauge parameters each of them lacking one space-time index in the appropriate group, and thus denoted by $\L_{\, i}$. Under  \eqref{mixed_gauge} the equations of motion transform according to
\be \label{var_mixed}
\d \, (\Box  \, - \, \pr^{\, i} \, \pr_{\, i}) \, \vf \,  = \, - \, \12 \, \pr^{\, i} \, \pr^{\, j} \, \pr_{\, (i} \, \L_{\, j)} \, ,
\ee 
and the main issue to be discussed is the analysis of the constraints that \eqref{var_mixed} forces on the gauge parameters, so as to grant invariance of the corresponding Lagrangians
\be \label{Mlagr}
\cL \, = \, \12 \, \vf \, M \, \vf \, .
\ee
In the following we summarise our findings and comment on their meaning, while postponing to section \ref{sec:dof_flat_mixed} the proofs of our statements and the corresponding detailed analysis of the spectrum described by \eqref{mixedeom}.

Our conclusions are that, for the case of $GL(D)-$tensors, reducible or irreducible, the amount of gauge invariance left after imposing the vanishing of \eqref{var_mixed} is indeed sufficient to avoid the propagation of all unwanted components of the tensor $\vf$. (I.e. those polarisations longitudinal to the light-cone in a given frame.) However,  the way this elimination is achieved is quite a nontrivial one  and exploits the peculiar fact that the system resulting from the condition of gauge invariance of $M$,
\be \label{enh}
p^{\, i} \, p^{\, j} \, p_{\, (i} \, \L_{\, j)} \, = \, 0 \, ,
\ee
here written in momentum space, displays a different number of solutions for momenta satisfying the mass-shell condition $p^{\, 2} = 0$ with respect to the case of off-shell momenta, in neat contrast with the case of symmetric tensors where all symmetrised gradients only possess trivial kernels, for all values of $p$.
 
Actually, when $p^{\, 2} \neq 0$,  computing successive divergences of \eqref{mixedeom} it is possible to iteratively prove that all the double-divergences of $\vf$ vanish  
\be
M \vf \, = \, 0 \, \hspace{1cm} \stackrel{\stackrel{p^{2} \neq 0}{}}{\, \longmapsto} \hspace{1cm} p_{\,i}\, p_{\, j} \, \vf \, = \, 0 \, \qquad \forall \, i, \, j.
\ee
This implies that $p_{\, i} \, \vf$ is transverse and thus, solving for $\vf$ one finds
\be
\vf \, = \, \fr{p^{\, i}}{p^{\, 2}} \, p_{\, i} \, \vf \, ,
\ee
allowing to conclude that only pure gauge solutions are available. In this situation the effective parameter $\L_{\, i} \, = \, \fr{p_{\, i}}{p^{\, 2}} \, \vf$ naturally satisfies \emph{strong} constraints of the form
\be \label{mixtdiff}
p_{\, (i} \, \L_{\, j)} \, = \, 0 \, ,
\ee
thus making manifest that for $p^{\, 2} \neq 0$ there is no room for additional gauge transformations to be effective on the field. 

For $p^{\, 2} = 0$ on the other hand it is convenient to go to the frame where $p_{\, \m} \, = \, p_{\, +}$, where all components transverse to the light-cone (i.e.~the physical ones) are manifestly gauge invariant, regardless of the conditions to be imposed on the parameters. The issue is then to discuss whether or not the available gauge symmetry actually suffices, in conjunction with the equations of motion, to remove all components longitudinal to the light-cone.  As we anticipated, the positive answer to this issue crucially relies on the observation that the set of solutions to the \emph{weak} constraints \eqref{enh} for $p^{\, 2} = 0$  is indeed bigger than those admitted by the apparently natural strong conditions  \eqref{mixtdiff}\ft{Also the amount of gauge-for-gauge symmetry associated to \eqref{mixed_gauge} is different in the two cases. In particular the strong constraints \eqref{mixtdiff} do impose some restrictions on the transformations $ \d \, \L_{\, i}~=~\pr^{\, k} \, \L_{\, [i k]}\, , $ (where square brackets denote antisymmetrisations and indicate that the two missing indices in $\L_{\, [i k]}$ cannot belong to the same family) while the conditions \eqref{enh} clearly do not. However, the point we want to stress is that in the wider set of solutions to \eqref{enh} we also find components of the gauge parameters capable of having an effect on the gauge potential, that would not be available if one had to impose \eqref{mixtdiff}.}, and the surplus of gauge symmetry is instrumental in reaching on-shell the condition of vanishing of all double divergences
\be \label{ddiv}
p_{\, i} \, p_{\, j} \, \vf \, = \, 0 \, .
\ee
In order to avoid possible sources of confusion let us stress that here the issue is not making a choice between two possible sets of constraints equally admissible for the gauge parameters of our theory; of course the Lagrangian dictates the amount of local symmetry it possesses, and we are asking whether or not the manifest solutions \eqref{mixtdiff} to the conditions \eqref{enh} allow to characterise the set of available parameters. Put differently, we would like to obtain a complete characterisation of the kernel of the double-gradient operator in \eqref{enh}. In these terms the issue should be posed for the case of the Labastida theory as well \cite{labastida}, and we shall comment about it at the end of this section.

Once \eqref{ddiv} is reached the remaining gauge invariance involves parameters subject to the strong constraints \eqref{mixtdiff}  still allowing to gauge fix to zero the \emph{transverse} parts of the divergences of $\vf$, on account of the relations
\be \label{divf}
\d \, \pr_{\, i} \, \vf \, = \, \Box \, \L_{\, i} \, + \, \pr^{\, j} \, \pr_{\, i} \, \L_{\, j} \, ,
\ee
given  that $\pr_{\, i} \, \vf$ and the parameters $\L_{\, i}$ subject to \eqref{mixtdiff} possess the same tensorial structure, while the ``tail'' on the r.h.s of  \eqref{divf} does not spoil the independence of the transverse projections of the parameters $\L_{\, i}$ and only accounts for the consistency of \eqref{divf} under gauge-for-gauge transformations. Performing the gauge fixing in \eqref{divf} allows to complete the reduction of \eqref{mixedeom} to the Fierz system
\be \label{FSmix}
\begin{split}
&\Box \, \vf \, = \, 0 \, , \\
& \pr_{\, i} \, \vf \, = \, 0 \, ,
\end{split}
\ee 
while at this stage the parameters $\L_{\, i}$ also satisfy a system analogous to \eqref{FSmix}. On the other hand, if the system did not possess additional gauge symmetry other than that defined by \eqref{mixtdiff} then all double divergences would be gauge invariant\ft{Their gauge variation in general would be $\d \, \pr_{\, i} \, \pr_{\, j} \, \vf \, = \, \Box \, \pr_{\, (i} \, \L_{\, j)} \, + \, \pr^{\, k} \, \pr_{\, i} \, \pr_{\, j} \, \L_{\, k}$, where the last term can be rewritten as follows
$$
\pr_{\, i} \, \pr_{\, j} \, \L_{\, k} \, = \, \pr_{\, (i} \, \pr_{\, j} \, \L_{\, k)} \, - \, \pr_{\, k} \, \pr_{\, (i} \, \L_{\, j)} \, ,
$$
thus showing that under the strong transversality constraints \eqref{mixtdiff} all double divergences are gauge invariant.}, while it is possible to prove that the equations of motion \eqref{mixedeom} are not strong enough to set all of them to zero on the mass-shell.

 Eventually, as we shall detail in section \ref{sec:dof_flat_mixed}, starting from reducible $GL(D)-$tensors satisfying \eqref{mixedeom} and subject to the gauge invariance constrained by the weak conditions \eqref{enh} one finds that the propagating polarisations are of the form
\be \label{polar}
\vf_{i^{1}{}_1 \cdots\, i^{1}{}_{s_1},\, \cdots , \,  i^{N}{}_1 \cdots\, i^{N}{}_{s_N}}\, ,
\ee
with indices $i^{j}{}_{l} = 1, \ldots , D-2\,$, here displayed for additional clarity, taking values along the directions transverse to the light-cone. The resulting  expression can be first decomposed in diagrams of $GL(D-2)$, each carrying in its turn a reducible particle content described by the corresponding branching in irreps of \mbox{$O(D-2)$}, so that working with multi-symmetric tensors leads to a spectrum that is two-fold reducible, in a sense, and which can be directly compared with the one emerging from the component expansion of string field theory.

However, it is also possible to stay closer in spirit to the more customary examples of low spin and choose $\vf$ in an irreducible representation of $GL(D)$, enforcing \eqref{irredcond}, without otherwise spoiling the general scheme of the construction. Indeed, the Maxwell operator defined in \eqref{mixedeom} commutes with the operators $S^{\,i}{}_j$,
\be
[\, M, \, S^{\,i}{}_j \,] \, = \, 0 \, ,
\ee
and thus also with the operator projecting a multi-symmetric tensor to an irrep of $GL(D),$\ft{The projectors take the schematic form $\Pi = \mathbb{I} + S^n$, where $S^n$ denotes a product of $S^i{}_j$ operators with all indices contracted; for instance, the multi-symmetric tensor $\vf_{\m_1 \m_2, \, \n}$ can be projected to the hook diagram of $GL(D)$ by the following operator:
\be \nn
Y_{\{2,1\}} \, \vf \, = \, \fr{4}{3} \left(\, \mathbb{I} \, - \, \fr{1}{4} \, S^{\,1}{}_2 \, S^{\,2}{}_1 \,\right) \vf \, .
\ee
Let us observe that this also implies that the $S^i{}_j$ commute with the projectors themselves.  From a different perspective,  one can observe that acting with $S^i{}_j$ on a Young diagram one obtains a sum of vectors in the same irreducible space.} thus ensuring invariance of the form of the Lagrangian  and of  the equations of motion even in the  irreducible case. Similar considerations are also valid for the Labastida Lagrangian \cite{labastida}, which retains indeed the same form in both cases, as discussed for instance in section $4$ of \cite{cfms1}.  

 The situation appears instead to be different if one wishes to start with fields in irreps of $O(D)$ which should provide, as we saw for the case of symmetric tensors, candidate one-particle theories. In particular two novel aspects are specific of this class of tensors:
\begin{itemize}
 \item following the instance of $GL(D)$ tensors we would like to identify the weakest conditions to be imposed on the gauge parameters granting invariance of the Lagrangian \eqref{Mlagr}; they are defined by the following system
\be
\begin{split}
&\d \, T_{\, i j} \, \vf \, = \, 0 \, , \\
&\d \, \{ M \, \vf \, \}^{\, t} \, = \, 0 \, ,
\end{split}
\ee
where $\{ M \vf\}^{\, t}$ denotes the traceless projection of $M \vf$, providing in this case the Lagrangian equations of motion. The difficulty at this level is that in the gauge variation of $\{ M \, \vf \, \}^{\, t}$ there appear terms involving $\Box \, \pr_{\, i} \, \L_{\, j}$ and it is not clear a priori how one might implement invariance of the Lagrangian equations in its weakest form without imposing a wave equation on the parameters. 
 \item On the other hand, if one tries to avoid this issue by asking $\d \, \{M \, \vf\} \, = \, 0$ then it is possible to show that combining the latter condition with the request of gauge invariance of the traces forces the strong conditions \eqref{mixtdiff} on traceless parameters, thus apparently only allowing an amount of gauge symmetry that is not sufficient to eliminate all unphysical polarisations.

\end{itemize} 
 
All the previous observations strongly depend on the condition that no constraints be imposed on the gauge field itself. As we stressed, the main difference between strong and weak conditions is that the latter allow the elimination of those few components of the double divergences that the equations of motion would not set to zero for momenta on the mass-shell. This indicates that an alternative consistent possibility would be to declare those components to be zero from the very beginning and to formulate an equivalent theory with gauge parameters subject to the strong constraints \eqref{mixtdiff} for any $p$. In order to find a covariant expression for these conditions it is useful to observe that the divergences of the equations of motion,
\be \label{bianchimix}
\pr_{\, i} \, M \, \vf \, = \, - \, \pr^{\, j} \, \pr_{\, i} \, \pr_{\, j} \, \vf \, = \, 0 \, ,
\ee
allow to select a minimal covariant subset of double divergences to be effectively constrained to vanish. Indeed, writing \eqref{bianchimix} explicitly one obtains
\be \label{ddiver}
\begin{split}
& \pr^{\, 1} \, \pr_{\, 1} \, \pr_{\, 1} \, \vf \, + \, \pr^{\, 2} \, \pr_{\, 2} \, \pr_{\, 1} \, \vf \, + \,  \pr^{\, 3} \, \pr_{\, 3} \, \pr_{\, 1} \, \vf \,  + \, \cdots \, = \, 0 \, ,\\
& \pr^{\, 1} \, \pr_{\, 1} \, \pr_{\, 2} \, \vf \, + \, \pr^{\, 2} \, \pr_{\, 2} \, \pr_{\, 2} \, \vf \, + \,  \pr^{\, 3} \, \pr_{\, 3} \, \pr_{\, 2} \, \vf \, + \, \cdots \, = \, 0 \, ,\\
& \cdots \, 
\end{split}
\ee
from which one can appreciate that it suffices to impose\ft{Taking a successive divergence of \eqref{bianchimix} it is immediate to verify that the equations of motion would be automatically projected onto the constraint.}
\be \label{mixtdiv}
\pr_{\, i} \, \pr_{\, j} \, \vf \, = \, 0 \, \qquad \, 1\, < \, i \, \leq \, j \, , 
\ee
in order to eliminate all double-divergences on-shell, since  after substituting \eqref{mixtdiv} in \eqref{ddiver} the equations for the remaining terms only contain a single symmetrised gradient whose kernel is trivial. Let us also observe that for the case of irreducible tensors the conditions \eqref{mixtdiv} are effectively redundant and can be more economically expressed in the form
\be
\pr_{\, 2} \, \pr_{\, 2} \, \vf \, = \, 0 \, ,
\ee
given that the double divergence in the second family contains all the irreducible components of  $\pr_{\, i} \, \pr_{\, j} \, \vf$  for  $1 <  i  \leq  j $. Clearly \eqref{mixtdiv} would not affect symmetric tensors and $(s, 1)-$tensors,  while for the irreducible $GL(D)-$case absence of constraints would extend to tableaux consisting of a single rectangular block of any size, together with possibly one additional single box in the first column. 

Imposing \eqref{mixtdiv} provides an alternative route to the formulation of a consistent single-particle theory for the case of  $O(D)-$tensors, avoiding in particular the aforementioned difficulties specific of this case. It is indeed non hard to show that the elimination of all double divergences from the equations of motion proceeds essentially in the same way as for the $GL(D)$ case, once the conditions \eqref{mixtdiv} are assumed\ft{One can appreciate this point in general by observing that the divergence of the equations of motion for the $O(D)$ case produces the same system \eqref{ddiver} as for the $GL(D)$ case, together with a tail with dimension-dependent coefficients, that as such cannot alter its rank in general.}. To summarise, the Lagrangian \eqref{Mlagr} can be adapted to provide a consistent description of single-particle propagation, once the following conditions are assumed: the fields are taken as tensors in irreducible representations of $O(D)$, subject to the double-divergence constraints \eqref{mixtdiv}, while the parameters are themselves $O(D)$ tensors (in particular they are traceless), subject to the strong form of the transversality constraints \eqref{mixtdiff}. 

However, this formulation of the $O(D)$ theory does not appear fully satisfactory in its present form, both because it is not obvious that restrictions as strong as \eqref{mixtdiv} are really mandatory and due to a few disadvantages that they bring along with them. To begin with, one can check that the conditions imposed by \eqref{mixtdiv} are still redundant, in the sense that the equations of motion would be capable themselves of setting to zero many of the components that we are assuming to vanish a priori, if one were able to write the minimal set of conditions in covariant form. In addition, as we shall see in section \ref{sec:dof_ads_mixed}, we were not able so far to find a deformation to (A)dS of the constraints \eqref{ddiv}, thus preventing the implementation of the Stueckelberg deformation of our one-particle theory.

It might be interesting at this point to draw a parallel with the Labastida construction \cite{labastida} for single-particle propagation. In that framework the fields are irreducible tensors of $GL(D)$, while the basic kinetic tensor takes the form 
\be \label{FL}
\cF \, = \, M \, \vf + \12 \, \pr^{\,i} \pr^{\,j} \, T_{ij} \, \vf \, ;
\ee
consistency of the whole set-up is achieved asking for the gauge parameters and the field to satisfy the tracelessness conditions: 
\begin{subequations}  \label{conslaba}
\begin{align} 
&T_{\, (i j} \, \L_{\, k)} \, = \, 0 \, ,  \label{conslabaA}\\
&T_{\, (ij} \, T_{\, kl)} \, \vf \, = \, 0 \, ,  \label{conslabaB}
\end{align}
\end{subequations}
providing somehow the counterparts of \eqref{mixtdiff} and \eqref{mixtdiv}. In particular, let us observe that also in the Labastida case, as far as consistency of the non-Lagrangian equations $\cF = 0$ is concerned, it would be possible to adopt a double-tracelessness condition weaker than \eqref{conslabaB}, of the form
\be
T_{\, (ij} \, T_{\, kl)} \, \vf \, = \, 0 \, \hskip .5cm \mbox{for} \quad 1 \, \leq \, i \, \leq j < \, k \, \leq \, l \, ,
\ee
since anyway the Bianchi identity 
\be
\pr_{\, i} \, \cF \, - \, \12 \, \pr^{\, j} \, T_{\, ij} \, \cF \, = \, - \, \fr{1}{12} \,  \pr^{\, j} \,  \pr^{\, k} \,  \pr^{\, l} \,  T_{\, (ij} \, T_{\, kl)} \, \vf \, 
\ee
would eventually imply the vanishing of all remaining gauge-invariant double traces. However,  the need for \eqref{conslabaB} arises when building a Lagrangian whose equations reduce to $\cF = 0$, as already manifest for the symmetric case. This is different in our present setting, where because of the self-adjointness of the kinetic operator $M$ no additional conditions are required to the goal of building a gauge invariant Lagrangian.  Let us also recall that the gauge variation of the Labastida kinetic tensor is proportional to a triple gradient of the traces of the field,
\be
\d \, \cF \, = \, \fr{1}{6} \, \pr^{\, i} \, \pr^{\, j} \, \pr^{\, k} \, T_{(\,ij} \, \L_{\,k\,)} \, ,
\ee
so that, according to our previous discussion, the natural condition granting gauge invariance of the equations of motion would look
\be \label{labaweak}
\pr^{\, i} \, \pr^{\, j} \, \pr^{\, k} \, T_{(\,ij} \, \L_{\,k\,)} \, = \, 0 \, ,
\ee
rather than its strong form, where all symmetrised traces are directly required to vanish:
\be \label{labacgaugeb}
T_{(\,ij} \, \L_{\,k\,)} \, = \, 0 \, .
\ee
However, we checked explicitly for the case of an irreducible $\{3, 2\}-$tensor that no enhancement of the symmetry is observed for momenta on the mass-shell. It is to be noted that in the $\{3, 2\}-$case the double-trace constraints \eqref{conslabaB} are not required; this is relevant to our point, since they would eventually force the strong form of the trace constraints \eqref{conslabaA}, thus making it harder to discern the role of a possible non-trivial kernel in \eqref{labaweak}. This indicates that the phenomenon that we observed when solving \eqref{enh} represents a  peculiarity of the Maxwell-like theory presented in this section. Let us remark that the choice of implementing gauge invariance subject to the transversality condition \eqref{enh} leads indeed to a sizeable simplification, as one can appreciate comparing \eqref{Mlagr} with  the Labastida Lagrangian for the $N-$family case \cite{labastida}, that in our notation reads\ft{As already recalled, this form of the Lagrangian applies both to multi-tensors and to irreps of $GL(D)$. We refer the reader to section 3 of \cite{cfms1}  for more details.}
\be
\cL \, = \, \frac{1}{2} \, \vf \, \left\{\cF \, + \, \sum_{p\,=\,1}^{N} \, \fr{(-1)^p}{p! \, (p+1)!} \, \h^{i_1 j_1} \, \ldots \, \h^{i_{\,p} j_{\,p}} \,
Y_{\{2^p \}} \, T_{i_1 j_1} \, \ldots \,
T_{i_p j_p} \, \cF \, \right\} \, ,
\label{l.laba}
\ee
where $Y_{\{2^p\}} \, \equiv \, Y_{\stackrel{\{2,\ldots,2\}}{\leftarrow \, p \, \rightarrow}}$ is the projector onto the two-column, $p$-rows diagram.

 As for what concerns tensionless strings in the sector of mixed-symmetry fields, generalised ``triplets'' were introduced in \cite{st}  and there shown to describe the full spectrum of the free open string collapsed to zero mass. To draw a parallel with our formulation we refer to the form of corresponding equations of motion for the multi-tensor $\vf$ transforming as in \eqref{mixed_gauge}, rephrased in our present notation as
\begin{subequations} \label{triplet_mixed}
\begin{align}
&M \, \vf \, + \, \12 \, \pr^{\, i} \, \pr^{\, j} \, \cD_{\, (i j) } \, = \, 0 \, ,  \label{triplet_mixedA} \\
&\Box \, \cD_{\, (i j)}  \, + \, \pr^{\, k} \, \pr_{\, i} \, \cD_{\, j k} \, = \, \pr_{\, i} \, \pr_{\, j} \, \vf  \, ,\label{triplet_mixedB}
\end{align}
\end{subequations}
where the set of fields $\cD_{\, i j}$ transform according to 
\be
\d \, \cD_{\, i j} \, = \, \pr_{\, i} \, \L_{\, j} \, ,
\ee
while also being  subject to the constraints
\be \label{Ddiv}
\pr_{\, k} \, \cD_{\, i j} \, = \, \pr_{\, i} \, \cD_{\, k j} 
\ee
so as to prevent their gauge-invariant combinations to propagate unwanted degrees of freedom. The condition \eqref{Ddiv} shows how  differential conditions on the fields tend to manifest themselves even in the unconstrained description of Maxwell-like theories too, while \eqref{triplet_mixedB} indicates that even after eliminating the symmetric part of $\cD_{\, i j}$, and thus forcing the parameter to satisfy \eqref{mixtdiff}, still the double divergences of $\vf$ do not necessarily vanish. While in \cite{st} the road to avoid those constraints at the Lagrangian level went through the implementation of the BRST construction, what we presented in this section indicates that an alternative, more economical, option exists in the form of Lagrangian \eqref{Mlagr}, once it is recognised that the full amount of gauge symmetry available on the mass-shell of light-like momenta is sufficient to guarantee the complete reduction of the equations of motion to the Fierz form \eqref{FSmix}.

 Finally, let us also mention that, similarly to the symmetric case, an alternative presentation of \eqref{mixedeom} obtains also in this case introducing the following generalisation of the first connexion of \cite{dwf} 
\be \label{G1mix}
\G_{\, \r, \, \m^{1}{}_{s_1},\, \cdots,\,  \m^{N}{}_{s_N}} \, = \, \pr_{\,\r} \, \vf_{\m^{1}{}_{s_1},\, \cdots,\,  \m^{N}{}_{s_N}} \, - \, \sum_{i\,=\,1}^{N} \, \pr_{\, \m^{i}} \, \vf_{\, \m^{1}{}_{s_1},\, \cdots,\, \m^{i}{}_{s_i-1} \, \r,\, \cdots, \, \m^{N}{}_{s_N} }, \,  
\ee
where the index $\m^{i}{}_{s_i}$ is a shortcut for a group of $s_i$ indices in the $i-$th family, consistently with the notation used for \eqref{G1}. (See footnote \ref{Notation}.) Making use of \eqref{G1mix} it is then possible to recast the equations of motion \eqref{mixedeom} in the form
\be
(M \, \vf)_{\, \m^{1}{}_{s_1},\, \cdots,\,  \m^{N}{}_{s_N}} = \, \pr^{\, \a} \, \G_{\, \a, \, \m^{1}{}_{s_1},\, \cdots,\,  \m^{N}{}_{s_N}} \, ,
\ee
along the lines of their counterparts \eqref{gamma1} for the case of symmetric tensors.

%%%%%%
\subsection{(A)dS backgrounds} \label{sec:ads}
%%%%%%

In this section we show how to build transverse-invariant Lagrangians in (A)dS backgrounds. For symmetric tensors the construction mirrors the procedure exploited in Minkowski space-time, for both reducible and irreducible cases. The case of mixed-symmetry tensors is technically more involved and conceptually more subtle, due to the unconventional branching of the corresponding irreducible representations in terms of $O(D-2)$ ones \cite{metsaev1, metsaev2, bmv}. For this class of fields we focus on single-particle Lagrangians, while still allowing the corresponding tensors to be of arbitrary symmetry type.

\subsubsection{Symmetric tensors}\label{sec:ads_symm}

We would like to construct the simplest deformation of Lagrangian \eqref{tlagr} to the case of maximally symmetric backgrounds; the corresponding spectra are discussed in section \ref{sec:dof_ads_symm}. To this end, starting with the covariantised version of the operator \eqref{M}, we compute the gauge variation of the corresponding tensor
\be
M\, \vf \,\equiv\, \left(\,\Box \, - \, \nabla \, \nabla \cdot\,\right) \vf\, ,
\ee
under the divergence-free gauge transformations
\be \label{gauge_ads}
\d \vf \,=\, \nabla\L \, , \qquad\qquad \nabla\cdot\L \,=\, 0 \, ,
\ee
obtaining\ft{The basic technical device needed is the commutator of two covariant derivatives 
acting on a vector 
$$  %\label{comm_der}
[\, \nabla_\m \,, \nabla_\n \,]\, V_\r \, = \, \frac{1}{L^2}\, \left(\, g_{\n\r}\, V_\m \,-\, g_{\m\r}\, V_\n \,\right) \, ,
$$
where for definiteness we refer to the Anti-de Sitter case, with $L$ denoting the radius and $g$ the metric of a $D$-dimensional AdS space. With the substitution $L \to i L$ one recovers the commutator on dS; this modification would not affect our manipulations so that our results formally apply to the dS case as well. }
\be \label{varads}
M \d \vf \,  = \, \fr{1}{L^{\, 2}} \, 
\left\{\, \left[\,(s-2)(D + s - 3) - s\,\right] \, \nabla \L - \, 2 \, g \, \nabla \L^{\pe} \,\right\} \, ,
\ee
where $g$ is the AdS metric. Its gauge-invariant completion is then easily found to be 
\be \label{M_L_symm}
M_L\, \vf \,\equiv\, M\, \vf \, - \, \fr{1}{L^{\, 2}} \,\left\{\, \left[\,(s - 2)(D + s - 3) \, - \, s\,\right] \, \vf \, - \, 2 \, g \, \vf^{\, \pe} \,\right\},
\ee
displaying the same spin-dependent ``mass term'' as the covariantised Fronsdal theory \cite{fronads}, up to a sign-flip in the trace part due to the different roles played by the variation of $\vf^{\, \pe}$ in the two cases\ft{In the Fronsdal case with traceless parameter one has $\d \, \vf^{\, \pe} \, = \, 2 \, \nabla \cdot  \L$; for divergence-free parameters we have in general $\d \, \vf^{\, \pe} \, = \, \nabla \, \L^{\, \pe}$.}. The corresponding equations of motion, 
\be \label{M_L=0}
M_L\, \vf \, = \, 0 \, ,
\ee
are obtained from the Lagrangian
\be \label{lag_ads_symm}
\cL \,=\, \12 \ \vf \,M_L\, \vf \, ,
\ee
which provides a smooth deformation of \eqref{tlagr} to (A)dS space. In the absence of additional assumptions eq.~\eqref{M_L=0} propagates a reducible spectrum of (A)dS massless particles of the same kind as its flat counterpart  \eqref{tlagr}. 

However, it can be also interesting to further restrict the relevant tensors in \eqref{lag_ads_symm} to be traceless:
\be
\vf^{\, \pe} \, = \, 0 \,, \, \hspace{3cm} \, \L^{\, \pe} \, = \, 0 \,,\, 
\ee
thus providing the (A)dS extension of the irreducible system in flat backgrounds. The proper Lagrangian under these assumptions is still given by \eqref{lag_ads_symm}, with the proviso that now the kinetic tensor $M_L$ does not contain contributions involving the trace of $\vf$, while the corresponding equations of motion
\be \label{ads_irr_eom}
M\, \vf \, - \, \fr{1}{L^{\, 2}} \, \left[\,(s - 2)(D + s - 3) \, - \, s\,\right] \, \vf \, + \, 
\fr{2}{D + 2 \, (s - 2)} \, g \, \nabla \cdot \nabla \cdot \, \vf \, =  \, 0 \, ,
\ee
can be shown to propagate only the massless polarisations of spin $s$. In this sense, eqs. \eqref{lag_ads_symm} (with $\vf^{\, \pe}\, = \, 0$)  and \eqref{ads_irr_eom} build an alternative to Fronsdal's theory in (A)dS \cite{fronads}, involving a minimal number of off-shell field components.

\subsubsection{Mixed-symmetry tensors}\label{sec:ads_mixed}

As a starting point for our analysis we compute the gauge transformation of the covariantised form of \eqref{mixedeom},
\be \label{maxwell_ads}
M \, \vf \, \equiv \, \left(\, \Box - \nabla^i \nabla_i \,\right) \vf \, ,
\ee
where $\vf$ is a multi-symmetric tensor with covariantised gauge variation
\be \label{gauge_ads_mix}
\delta \vf \,=\, \nabla^i \Lambda_{\,i}\, ,
\ee
trying to identify the compensating terms needed to make \eqref{maxwell_ads} gauge invariant. With the help of the commutators collected in appendix \ref{app:form_mix} one can obtain
\be \label{var_mixed_ads}
\begin{split} 
& M\, \d \vf\, = \, - \, \frac{1}{L^2} \left\{\, (D-1)\, \nabla^i\Lambda_i \,-\, (D-N-3)\, \nabla^i S^j{}_i\, \Lambda_j \,-\, \nabla^i S^j{}_k S^k{}_i\, \Lambda_j \,\right\} \\
& - \, \12\, \nabla^i \nabla^j \nabla_{(\,i\,}\Lambda_{\,j)} + \frac{1}{L^2} \left\{\, 2\, g^{ij}\, \nabla_{(\,i\,}\Lambda_{\,j)} \,+\, g^{ij} S^k{}_i \nabla_{[\,j\,}\Lambda_{\,k]} \,-\, 2\, \nabla^i g^{jk}\, T_{ij}\, \Lambda_k  \,\right\} , 
\end{split}
\ee
where $N$ denotes the number of families. From the resulting expression, still rather involved even after imposing  the transversality conditions 
\be \label{constrmix}
\nabla_{(i}\,\Lambda_{\,j)} \,=\, 0 \, ,
\ee
it is possible to appreciate the difficulties met in extending the flat gauge invariance to the (A)dS case, already visible for the case of tensors with two families of indices. Indeed,  rewriting \eqref{var_mixed_ads} for these fields in a more explicit notation as
\be \label{2family}
\begin{split}
(M\, \d \, \vf)_{\,  \m_s, \, \n_r} \, = \, \fr{1}{L^2} & \left\{ [(s-1)(D + s - 3) - (D + 2s - 3)]\, \nabla_{\!\m}\, \L_{\, \m_{s-1},\, \n_r} \right.  \\
+& \, [(r-1)(D + r - 3) - (D + 2r - 3)]\, \nabla_{\!\n}\, \l_{\, \m_{s},\, \n_{r-1}} \\
+& \, \nabla_{\!\m}\, \L_{\, \n \, \m_{s-2}, \, \m \, \n_{r - 1}} \, + \,  \nabla_{\!\n}\, \l_{\, \m_{s-1}\, \n,\, \, \m\, \n_{r-2}} \\
+& \left. \, (D + s + r - 5) \left[\, \nabla_{\!\n}\, \L_{\, \m_{s-1}, \, \m \, \n_{r - 1}} \, + \,  \nabla_{\!\m}\, \l_{\, \n \, \m_{s-1} ,\, \n_{\, r-1}\,} \,\right]
 + \, \cdots \, \right\}\, ,
\end{split}
\ee
where the dots stand for terms involving traces or divergences of the parameters while like indices are understood to be symmetrised, one can recognise that, for $s \neq r$, there is no way of compensating the first two terms in \eqref{2family} with contributions linear in $\vf_{\,  \m_s, \, \n_r}$ of any sort\ft{Considering counterterms involving exchanges of indices would not help. Indeed the variation of the generic term
\be \nn
\begin{split}
\d \vf_{\m_{s-n}\, \n_n \,,\, \m_n\, \n_{r-n}} = & \ \nabla_{\!\m}\, \L_{\m_{s-n-1}\, \n_n ,\, \m_n\, \n_{r-n}} \,+\, \nabla_{\!\n}\, \L_{\m_{s-n}\, \n_{n-1} ,\, \m_n\, \n_{r-n}}\\
+ & \ \nabla_{\!\n}\, \l_{\m_{s-n}\, \n_{n} ,\, \m_n\, \n_{r-n-1}} \,+\, \nabla_{\!\m}\, \l_{\m_{s-n}\, \n_{n} ,\, \m_{n-1}\, \n_{r-n}} \, ,
\end{split}
\ee
makes it manifest that no simultaneous compensation of the first two terms in \eqref{2family} is possible in general.} so that mixed-symmetry tensors in (A)dS are bound to possess a smaller gauge symmetry than their flat-space counterparts. In fact, with hindsight, this phenomenon is maybe not so surprising, given that already for the one-family case, involving symmetric tensors only, the gauge invariant completion of the (A)dS operator  \eqref{M_L_symm} depends on the length of the corresponding row.

This observation does not imply that multi-symmetric tensors cannot be given any Lagrangian formulation in (A)dS spaces, however it renders those maximally reducible objects  less palatable, in the absence of simple criteria allowing to identify the proper gauge symmetry to be implemented off-shell. Indeed, we found it simpler to exploit tensors transforming irreducibly under permutations of their space-time indices, also in order to deal more efficiently with the complications introduced by the operators $S^i{}_j$, and in the remainder of this section we shall focus on this latter option. This means that in the following
\be \label{def_irr}
\vf_{\m^1{}_1 \cdots\, \m^1{}_{s_1} ,\,\cdots\,,\,\m^N{}_{1} \cdots\, \m^N{}_{s_N}} \equiv\, Y_{\{s_1,\ldots,\,s_N\}} \, \vf_{\m^1{}_1 \cdots\, \m^1{}_{s_1} ,\,\cdots\,,\,\m^N{}_{1} \cdots\, \m^N{}_{s_N}} \, ,
\ee
where $Y_{\{s_1,\ldots,\,s_N\}}$ denotes the projector onto the $GL(D)$ representation labelled by the Young diagram\footnote{We identify Young diagrams by ordered lists of the lengths of their rows enclosed between braces. See \cite{permutations} and references therein for some introductory material on the representations of linear and orthogonal groups.} $\{s_1,\ldots,s_N\}$, with $s_1 \geq s_2 \geq \cdots \geq s_N$, a condition that can be expressed in terms of the $S^i{}_j$ operators as
\be \label{irr}
S^i{}_j\, \vf \,=\, 0\, , \hskip .5cm \textrm{for}\ \ i < j \, .
\ee
Eventually, we shall show that for \emph{traceless} fields satisfying \eqref{irr} part of the gauge symmetry of the Maxwell-like Lagrangian \eqref{Mlagr} can be restored in (A)dS  choosing  in
\be
\cL \,=\, \12\, \vf \left\{\, \Box - \nabla^i \nabla_i \,-\, m^2 \,\right\} \vf \, 
\ee
a suitable ``mass-term'', leading to the formulation of candidate single-particle Lagrangians. However, the reduced gauge symmetry available on (A)dS also requires to impose constraints on almost all divergences of the field, as we shall detail more in the following.

Irreducible gauge fields in Minkowski backgrounds transform with irreducible parameters obtained stripping one box from the corresponding tableau, in all admissible ways \cite{labastida-morris}. However, even in this case one can conveniently study the gauge variation of the Maxwell-like Lagrangians in (A)dS starting from \eqref{var_mixed_ads}: one has only to take into account that the multi-symmetric $\L_i$ are no longer independent due to \eqref{irr}. In our formalism we can recover the structure of the irreducible parameters analysing the solutions of the variation of \eqref{irr} given by the set of relations
\be \label{irr_par}
S^i{}_j\, \Lambda_k \,+\, \delta^i{}_k\, \Lambda_j \,=\, 0\, ,  \hskip .5cm \textrm{for}\ \ i < j \, .
\ee
As we discuss more in detail in appendix \ref{app:proofs}, the conditions \eqref{irr_par} select the irreducible components carried by each $\L_k$, that can be decomposed as
\be \label{expansion}
\L_k \,=\, \sum_{n\,=\,k}^N \left(1-\d_{s_n,\,s_{n+1}}\right) Y_{\{s_1,\ldots,\,s_n-1,\ldots,\,s_N\}}\, \L_k \,\equiv\, \sum_{n\,=\,k}^N \left(1-\d_{s_n,\,s_{n+1}}\right) \L^{(n)}_{\,k}\, ,
\ee
where, in particular, no components labelled by $n < k$ are present in $\L_k$, while the factor between parentheses makes it manifest that if $s_n = s_{n+1}$ then $\{\ldots,s_n-1,s_{n+1},\ldots\}$ is not an admissible Young diagram. Moreover, eqs.~\eqref{irr_par} also imply that all $\L^{(n)}_k$ with the same label $(n)$ are proportional, as one can realise setting $i=k$ and acting with the proper Young projector so as to obtain
\be \label{nonhom}
\L^{(n)}_{\,j} \,=\, -\ S^k{}_j\, \L^{(n)}_{\,k}  \, ,\hskip .5cm \mbox{for fixed $k < j$} \, .
\ee
This result (where no summation over $k$ is implicit) also rests on the fact that the operators $S^i{}_j$ commute with Young projectors, as discussed in section \ref{sec:flat_mixed}. Therefore, one could identify the irreducible parameters with proper linear combinations of the $\L^{(n)}_k$ associated to the same Young diagram. However, in the following it will be more convenient to preserve the redundancy of \eqref{gauge_ads_mix}, that in the irreducible case one can rewrite more explicitly as
\be \label{gauge_expl}
\d \vf \,=\, \sum_{n\,=\,1}^N \left(1-\d_{s_n,\,s_{n+1}}\right) \sum_{i\,=\,1}^n \, \nabla^i \L^{(n)}_{\,i} \, ,
\ee  
with the proviso that one can treat separately the various irreducible components labelled by $(n)$, but not the different parameters labelled by $i$.

The key to analyse the gauge variation \eqref{var_mixed_ads} of the Maxwell operator is then that $M$ commutes with all $S^i{}_j$. Therefore, for any fixed irreducible component carried by the parameters the structure of the gradient terms in $M \d\vf$ should agree with \eqref{gauge_expl} in order to be compatible with \eqref{irr}. On the other hand, the irreducibility condition cannot fix the relative coefficients in the sum over $n$ because any addendum is annihilated independently by all $S^i{}_j$ with $i<j$. As a result, using the relations \eqref{irr_par} it should be possible to recast \eqref{var_mixed_ads} in the form
\be \label{var_proj}
M\, \d\vf \,=\, \sum_{n\,=\,1}^N \, k_{\,n}\left(1-\d_{s_n,\,s_{n+1}}\right) \sum_{i\,=\,1}^n \, \nabla^i \L^{(n)}_{\,i} \,+\, \textrm{divergences and traces}.
\ee
This argument is supported by an explicit computation in appendix \ref{app:proofs}, where we also fix the coefficients $k_{\,n}$ obtaining
\be \label{var_proj_k}
k_{\,n} \,=\, \frac{1}{L^2}\, \bigg[\, (s_n-n-1)(D+s_n-n-2) - \sum_{k\,=\,1}^N s_k \,\bigg] \, .
\ee

 Let us now mention that -- even if one works with a traceful $\vf$ -- the terms displayed explicitly in \eqref{var_proj} clearly cannot receive any correction from the gauge variation of traces of the field. Therefore, one can only cancel them with a counterterm involving $\vf$, so that, for Young-projected fields, the only possibility is to define 
\be \label{M_L_mix}
M_L \, \vf \,\equiv\, \left(\, \Box - \nabla^i \nabla_i \,\right) \vf \, - \, m^2\, \vf \, ,
\ee
since all alternative counterterms must be of the type
\be \label{mass_wrong}
\Delta\, \vf \,\equiv \left(\, a_1\, S^i{}_j S^j{}_i \,+\, \sum_{k}\, a_k\, S^i{}_{j_1} S^{j_1}{}_{j_2} \cdots\, S^{j_k}{}_i \,\right) \vf 
\ee
in order to preserve the index structure of $M\vf$. However, $\Delta$ commutes with all $S^i{}_j$ and, as a result, it acts as a multiple of the identity on any irreducible representation of the $gl(N)$ algebra generated by them (see \eqref{[S,S]}). On the other hand, eq.~\eqref{irr} implies that $\vf$ is a highest-weight state, that as such uniquely specifies an irreducible representation of $gl(N)$. Therefore, $\Delta$ acts diagonally on any $\vf$ satisfying \eqref{irr}, and in our present setup can only shift the coefficient $m^2$ in \eqref{M_L_mix}.  One can make this property manifest by casting, for instance, the first addendum of \eqref{mass_wrong} (corresponding to the quadratic Casimir of $gl(N)$) in the form\ft{We can also illustrate this fact displaying explicitly the space-time indices in a simple example. First of all, the action of $S^i{}_j S^j{}_i$ preserves the lengths of the groups of symmetrised indices, but displaces their position. For a field $\vf_{\m\n,\,\r}$ the only alternative is $\vf_{\r\,(\m\,,\,\n)}$, which results from the action of $(S^1{}_2 S^2{}_1 - 2\cdot\mathds{1})$ on $\vf_{\m\n,\,\r}$. However, with a simple direct calculation one can show that
\be \nn
Y_{\{2,1\}} \vf_{\m\n,\,\r} \,=\, \12 \left(\, 2\,\vf_{\m\n,\,\r} - \vf_{\r\,(\m\,,\,\n)} \right) \quad \Rightarrow \quad Y_{\{2,1\}} \vf_{\r\,(\m\,,\,\n)} \,=\, -\ Y_{\{2,1\}} \vf_{\m\n,\,\r} \, .
\ee }
\be \label{casimir}
\cC \, =\, \sum_{i\,=\,1}^N\, S^i{}_i \left(\, S^i{}_i + N - 2i +1 \,\right) \,+\, 2\, \sum_{i\,=\,1}^{N-1} \sum_{j\,=\,i+1}^N S^j{}_i S^i{}_j  
\ee
where in particular the second term vanishes on account of \eqref{irr}. As a consequence, one can only tune a single parameter in $M_L$, whereas in general all $k_{\,n}$ in $M\d\vf$ are different. Therefore, one can cancel at most the gradient terms corresponding to a single irreducible component by suitably tuning $m^2$ in \eqref{M_L_mix}, while it remains  to be verified whether the leftover terms in \eqref{var_mixed_ads} induce extra constraints. 
 
 Let us start from the divergence terms in \eqref{var_mixed_ads}, 
\be \label{var_div}
M\, \d\vf \,= \, \cdots \, - \, \12\, \nabla^i \nabla^j \nabla_{(\,i\,}\Lambda_{\,j)} + \frac{1}{L^2} \left\{\, 2\, g^{ij}\, \nabla_{(\,i\,}\Lambda_{\,j)} \,+\, g^{ij} S^k{}_i \nabla_{[\,j\,}\Lambda_{\,k]} \,\right\} \, + \, \cdots \, ,
\ee
since also for this class of contributions the discussion applies to both traceless and traceful fields.
The novelty with respect to the symmetric case is the term containing the antisymmetric combination $\nabla_{[\,j\,}\Lambda_{\,k]}$, that does not vanish manifestly even after forcing the constraint \eqref{constrmix}. Indeed, the vanishing of the divergence terms in \eqref{var_div} requires that the surviving irreducible parameter be fully divergenceless,
\be \label{adsdiv}
\nabla_{i}\, \Lambda^{(n)}_j \,=\, 0 \, , \hskip .5cm \textrm{for $n$ fixed and}\ \forall\ i, j \, ,
\ee
although, as we show in appendix \ref{app:proofs}, when a single irreducible gauge parameter is present this condition is already implied by the constraints \eqref{constrmix}. Therefore, in the gauge variation of the deformed Maxwell-like equation \eqref{M_L_mix} only the term
\be
M_L\,\d\vf \,=\, -\, \frac{2}{L^2}\, \nabla^i g^{jk}\, T_{ij}\, \Lambda_k
\ee
remains to be discussed, and at this stage working with or without trace constraints makes a notable difference. The simplest possibility is to impose
\be \label{tracemix}
T_{ij}\, \vf \,=\, 0 \, .
\ee
At the level of field equations this extra condition would require to project \eqref{M_L_mix} on its traceless component, but we can discuss gauge invariance directly at the level of the Lagrangian. There the contraction with another traceless field avoids the need for a projection and the self-adjointness of $M_L$ implies
\be
\d\, \cL \,=\, \vf\, M_L\, \d \vf \,=\, \frac{s_i s_j s_k}{L^2} \left(\, \nabla_i\, T_{jk} \, \vf \,\right)T_{ij}\,\L_k \,=\, 0 \, .
\ee 

 In conclusion, if $\vf$ satisfies \eqref{irr} and \eqref{tracemix} then the Maxwell-like Lagrangian
\be \label{lag_mix_ads}
\cL \,=\, \12 \ \vf \left\{\, \Box \,-\, \nabla^i\, \nabla_i\, - \, \frac{1}{L^2}\, \bigg[\, (s_n-n-1)(D+s_n-n-2) \,-\, \sum_{k\,=\,1}^N s_k \,\bigg] \right\} \vf
\ee
is invariant under the gauge transformation generated by a \emph{single} fully divergenceless 
$\{\ldots,s_n-1,\ldots\}$-projected  parameter. Let us observe that the ``masses'' that we found coincide with those appearing in the on-shell system presented in \cite{metsaev2}, while for the particular case $N=1$ \eqref{lag_mix_ads} reproduces our result for symmetric tensors discussed in section \ref{sec:ads_symm}. 

As manifest in eq.~\eqref{expansion}, in the presence of blocks of rows of equal length one cannot choose $n$ arbitrarily in the interval from $1$ to $N$. The allowed values correspond to the rows at the end of each block: it could then be convenient to denote a general Young diagram by $\{(s_1,t_1),\ldots,(s_p,t_p)\}$ where the pair $(s_k,t_k)$ denotes the dimensions of the $k-$th block, so that
\be
\sum_{i\,=\,1}^p\, t_i \,=\, N \, .
\ee
A field transforming in the $\{(s_1,t_1),\ldots,(s_p,t_p)\}$ representation of $GL(D)$ thus admits $p$ independent gauge parameters on Minkowski backgrounds, while in (A)dS backgrounds one can at most keep the invariance under the gauge transformation
\be \label{gauge_expl_bis}
\d \vf \,= \sum_{i\,=\,1}^{t_1+\,\cdots\,+\,t_k} \nabla^i \L^{(t_1+\,\cdots\,+\,t_k)}_{\,i} \, ,
\ee    
for a given value of $k$. Stressing the existence of blocks of rows with equal length leads to rewrite the Lagrangian \eqref{lag_mix_ads} in the form
\be \label{lag_mix_block}
\cL \,=\, \12 \ \vf \left\{\, \Box \,-\, \nabla^i\, \nabla_i\, - \, \frac{1}{L^2}\, \bigg[\, (s_{k}-\sum_{j\,=\,1}^k t_j-1)(D+s_k-\sum_{j\,=\,1}^k t_j-2) \,-\, \sum_{j\,=\,1}^p t_j s_j \,\bigg] \right\} \vf \, .
\ee

 The Lagrangian \eqref{lag_mix_ads} is invariant under the transformation
\be \label{gauge_expl_2}
\d \vf \,=\, \sum_{i\,=\,1}^n \, \nabla^i \L^{(n)}_{\,i} \, ,
\ee  
generated by a single independent irreducible gauge parameter, for any choice of $(n)$ admitted by the block structure of the field. However, in the mixed-symmetry case implementing gauge invariance of the Lagrangian does not suffice to ensure the on-shell propagation of a unitary irreducible representation of the (A)dS group. For instance, as we shall discuss more in detail in section \ref{sec:dof_ads_mixed}, in Anti de Sitter backgrounds in order to deal with unitary representations one must preserve the gauge symmetry associated to the parameter $\L^{(t_1)}$, labelled by the Young diagram missing one box in the first block with respect to the field \cite{metsaev2,bmv}. Moreover, in this case one also has to impose on the field the constraint
\be \label{ultra-constr}
\nabla_{t_1+t_2}\, \vf \,=\, 0
\ee
in order to eliminate all non-physical polarisations. For irreducible fields satisfying \eqref{irr} the constraint \eqref{ultra-constr} implies that all divergences aside from those computed in the first block of rows vanish. In section \ref{sec:dof_ads_mixed} we also show that, when \eqref{ultra-constr} holds, the traceless projection of $M\vf$ satisfies the same constraint as well. Therefore, the field equations does not require any additional projection with respect to the one imposed by the tracelessness condition \eqref{tracemix}. On (A)dS it is thus still possible to describe the dynamics of particles of arbitrary symmetry type with Maxwell-like Lagrangians, although at the price of imposing additional constraints on the field. 
%In section \ref{sec:dof_ads_mixed} we comment on the possibility of forcing \eqref{ultra-constr} as a consequence of the equations of motion introducing a proper set of auxiliary fields.

%%%%%%%%%%%%%%%%%%%%%%%%%%%%%%%%%%%%%%%%%%%%%%%%%%%%%%%%%%%%%%%%%%%%%

\section{Spectra}\label{sec:spec}

%%%%%%%%%%%%%%%%%%%%%%%%%%%%%%%%%%%%%%%%%%%%%%%%%%%%%%%%%%%%%%%%%%%%%

%%%%%%
\subsection{Flat backgrounds}\label{sec:dof_flat}
%%%%%%

 In this section we investigate the spectra described by the equations \eqref{eomsymm} and \eqref{mixedeom}. We already showed in sections \ref{sec:flat_symm} and \ref{sec:flat_mixed} that the transverse-invariant equations of motion reduce to the (traceful) Fierz systems \eqref{FS} and \eqref{FSmix}, respectively. For symmetric tensors here we provide an independent counting of the degrees of freedom evaluating  the role of each component in light-cone coordinates. In addition, we also discuss some aspects of the Hamiltonian analysis and  present in particular a simple argument to count the number of first-class constraints associated to gauge symmetries constrained as in \eqref{tdiff}.  The reduction of the mixed-symmetry equations \eqref{mixedeom} to the Fierz system \eqref{FSmix}, on the other hand, was discussed under the assumption that all double divergences could be set to zero. The light-cone analysis that we present in this section provides in particular a proof of this statement.   
   
 In the ensuing discussion we shall work in momentum space in light-cone coordinates, denoting indices transverse to the light-cone directions with small Latin letters $i, j, k, \dots$, or even omitting them altogether whenever it might be done without ambiguities; for $p^2 \neq 0$ it is easy to prove that only pure gauge solutions exist: indeed, solving for $\vf$ in \eqref{eomsymm} one obtains
\be
\vf \, = \, \fr{p}{p^{\, 2}} \, p \cdot \vf\, ,
\ee
where the combination $\fr{1}{p^{\, 2}} \, p \cdot \vf$ can play the role of a proper gauge parameter in the present framework, due to the condition \eqref{doublediv} ensuring transversality of $p \cdot \vf$. While this observation would allow one to restrict the analysis to null momenta, we prefer anyway to keep it slightly more general at this stage and show how the elimination of all components longitudinal to the light-cone works for the case of arbitrary momenta. Thus in our discussion of symmetric tensors we will only assume 
\be
p_{\, +} \, \neq \, 0\,,
\ee
which is always admissible for physical particles.

 The simplest example of the ensuing analysis is given by the spin$-2$ case that we review here for pedagogical reasons. The condition of transversality \eqref{tdiff} on the vector parameter $\L_{\, \m}$,  
\be
p \cdot \L \, = \, - \, p_+ \, \L_- - \, p_- \, \L_+ \, + \, p_i\, \L_i \, = \, 0 \, ,
\ee
implies that $ \L_-$ is effectively determined in terms of the remaining $D-1$ components, $\L_+$ and  $\L_i$, $i =  1, \dots, D-2$. This implies that fixing the gauge completely (modulo singular gauge transformations) one can eliminate at most $h_{\, + +}$  and $h_{\, + i}$; from the corresponding equations of motion evaluated in this gauge one finds however $(p \cdot h)_+ = 0$ and  $(p \cdot h)_i = 0$, which imply in their turn $h_{- +} = 0$ and $h_{- i } = \fr{p_j}{p_+} \, h_{i j }$. Finally, from the equation for   $h_{- +}$ one finds $h_{- - } = \fr{1}{p_+^2} \, p_i \, p_j \, h_{ij}$, so that the only independent components of $h_{\, \m \n}$ are indeed the transverse ones, subject to the equation  $p^{\, 2} \, h_{\, i j}\, = \, 0$ and thus arbitrary on the light-cone  $p^{\, 2} \, = \, 0$. These components describe an irreducible tensor of $GL\, (D - 2)$, whose branching in terms of irreps of $O (D-2)$ identifies its particle content, as expected,  with that of a massless spin$-2$ particle together with a massless scalar.

\subsubsection{Symmetric tensors}\label{sec:dof_flat_symm}

In this section we shall use the following notation\ft{In practice, we insert an exponent to indicate the number of times a specific ``$+$'' or ``$-$'' component appears, while we denote with a numerical label the total number of indices for components transverse to the light-cone.}:
\be
\vf_{\underbrace{- \, \cdots \, -}_{l} \,\underbrace{+ \, \cdots \, +}_{s - k - l} \, i_1 \, \cdots \, i_k} \, \equiv
\, \vf_{-^l \, +^{s - k - l} \, i_k}\, . 
\ee
The condition of transversality on the gauge parameter
\be \label{symmtdiff}
p\cdot \L_{\, \m_{ s-2}} \, = \, - p_+ \, \L_{\,-\, \m_{ s-2}}  - \, p_- \, \L_{\,+\, \m_{ s-2}}  \, + \, 
p_{i} \, \L_{\,i \, \m_{ s-2}} \, = \, 0 \,
\ee
fixes all components of $\L_{\m_{ s-1}}$ with at least one ``$-$'' index in terms of components of the form
$ \L_{\, +^{s - k - 1} \, i_k}$. Thus a complete gauge-fixing is reached setting 
\be \label{gaugefixs}
 \vf_{\, +^{s - k} \, i_k} \, = \, 0 \, ,
\ee
with $k$ ranging from $0$ to $s - 1$, while in order to obtain conditions on components involving ``$-$'' indices we have to resort to the equations of motion. From \eqref{gaugefixs} we obtain, recursively,
\be \label{dofsymm1}
(M\, \vf)_{\, +^{s - k} \, i_k} \, = \, 0 \, \hspace{1cm} \, \Rightarrow \, \hspace{1cm} \, (p \cdot \vf)_{+^{s - k - 1}\, i_k} \, = \, 0 \, ,
\ee
whose expansion allows to iteratively set to zero all components of $\vf$ with one index along the ``$-$'' direction and at least one index along the ``$+$'' direction:
\be \label{recursive2}
\begin{split}
& \vf_{\, -\, +^{s - k - 1} \, i_k} \, = \, 0\, , \\
& k \, = \, 0 ,\ldots , s\, - \, 2 \, ,
\end{split}
\ee
while also providing the relations
\be \label{no+1}
\vf_{\, -\, i_{s - 1}} \, = \, \fr{p_j}{p_+} \, \vf_{\, j \, i_{s - 1}}\, .
\ee
One can now repeat the procedure, exploiting the consequences of the equations of motion for the components of $\vf$  set to zero in \eqref{recursive2}. In analogy with the previous steps one obtains
\be \label{dofsymm2}
(M\, \vf)_{\,-\, +^{s - k -1} \, i_k} \, = \, 0 \, \hspace{1cm} \, \Rightarrow \, \hspace{1cm} \, (p \cdot \vf)_{-\, +^{s - k - 2}\, i_k} \, = \, 0 \, ,
\ee
with $k = 0, \ldots s-2$. As a consequence one finds that all components with two ``$-$'' indices and at least one ``$+$'' index vanish
\be \label{recursive3}
\begin{split}
& \vf_{\, -^2\, +^{s - k - 2} \, i_k} \, = \, 0\, , \\
& k \, = \, 0 , \ldots , s\, - \, 3 \, ,
\end{split}
\ee
together with an additional relation for the component with no ``$+$'' indices, to be combined with \eqref{no+1}
\be
\vf_{\, -^2\, i_{s - 2}} \, = \, \fr{p_j}{p_+} \, \vf_{\, - \, j  \,  i_{s - 2}}\, = \, \fr{p_j p_k}{p^2_+} \, \vf_{\, j \, k \,  i_{s - 2}}\, .
\ee
The corresponding iterative pattern can be proven by induction and leads to 
\be \label{dofsymm3}
(M\, \vf)_{\,-^l\, +^{s - k -l} \, i_k} \, = \, 0 \, \hspace{1cm} \, \Rightarrow \, \hspace{1cm} \, (p \cdot \vf)_{-^l\, +^{s - k - l -1}\, i_k} \, = \, 0 \, ,
\ee
from which it is possible to deduce the following relations:
\be \label{recursivel}
\begin{split}
& \vf_{\, -^{l}\, +^{s - k - l} \, i_k} \, = \, 0\, , \\
& \vf_{\, -^{l + 1} \, i_{s - l - 1}} \, = \, \fr{1}{(p_+)^{l+1}} \, p_{j_1}  \cdots  p_{j_{l+1}} \, \vf_{\, j_1 \, \cdots \, j_{l+1}   \, i_{s - l - 1}}, \\
& k \, = \, 0 , \ldots , s\, - \, l \, - \, 1 \, , \\
& l \, = \, 0 , \ldots ,\, s \, - \, 1 \, ,
\end{split}
\ee
essentially stating that the only independent  components of $\vf$ are those containing just indices transverse to the light-cone, $\vf_{\, i_s} \equiv \vf_{\, i_1 \cdots \, i_s}\, $, which satisfy the equations
\be \label{independent}
p^{\, 2} \, \vf_{\, i_1 \cdots\, i_s} \, = \, 0 \, ,
\ee
and thus describe a set of massless particles carrying spin $s, s-2, s-4, \ldots, $ down to $1$ or $0$.

 From the perspective of the Hamiltonian analysis \cite{gitman, HT} the peculiarity of transverse-invariant systems is found in the unusual counting of the corresponding first-class constraints, associated to the presence of higher generations of constraints besides the primary and secondary ones present in more conventional situations. (See \cite{uni-ham} for a discussion of the spin$-2$ case and \cite{SV} for the case of symmetric and traceless tensors.) In general, on a Cauchy surface, one has to assign independently the values of a given component of the gauge parameter and of its time derivatives, up to the highest order appearing in the variation of the gauge field, thus implying that they have to be counted as independent constraints; thus, for instance, for conventional  theories with parameters entering with one derivative in $\d \vf$, and in the absence of additional constraints, each gauge component has to be counted twice, since its first time derivative provide an additional independent condition to be imposed on the system. 
  
 Our observation is that for transverse-invariant theories there is a simple procedure allowing to compute the number of components of the parameters, including their time derivatives, that have to be counted as independent on a given Cauchy surface. Indeed, solving the transversality constraint \eqref{tdiff} with respect to the time derivative one finds
\be \label{temporal}
\pr^{\, \a} \, \L_{\, \a \, \m_2 \, \cdots \, \m_{s-1}} = \, 0 \, \hspace{1cm} \, \Rightarrow \, \hspace{1cm} \, \dot{\L}_{\, 0 \, \m_2 \, \cdots \, \m_{s-1}} = \, \vec{\nabla} \, 
\cdot \L_{\,  \m_2 \, \cdots \, \m_{s-1}} \, ,
\ee
where in the r.h.s.\ the divergence is computed along the spatial directions. One can thus appreciate that for all components of $\L_{\, \m_1 \, \m_2 \, \cdots \, \m_{s-1}}$ carrying at least one temporal index the time derivatives are not to be regarded as independent, in view of  the condition \eqref{temporal}.
This means that the total number of first-class constraints is twice the number of components of the parameters with only spatial indices, $\L_{\, a_1 \, a_2 \, \cdots \, a_{s-1}}$, \,  $a_k = 1 , \ldots , D-1$, since for the latter their time derivatives are really independent, and only once the number of components possessing at least one temporal index, in view of the previous observation. Thus, for the case of rank$-(s-1)$ symmetric parameters discussed in this section the total number of first class constraints is given by the formula
\be \label{1stclass}
\#\, \mbox{1st class} \, = \, 2 \, \underbrace{{D + s -3 \choose s-1}}_{\L_{\, a_1 \, a_2 \, \cdots \, a_{s-1}}} \, + \, 
\underbrace{{D + s - 3 \choose s-2}}_{\L_{\, 0\, \m_2 \, \cdots \, \m_{s-1}}}\, .
\ee
In the absence of second-class constraints one can use \eqref{1stclass} to directly compute the propagating degrees of freedom of the transverse-invariant system using the formula \cite{HT}
\be \label{count-1stclass}
\#\, \mbox{d.o.f.} \, = \, \#\, \mbox{(components in $\vf$)} \, - \, \#\, \mbox{1st class}\, ,
\ee
finding agreement with our result \eqref{independent}. The light-cone analysis in its turn implicitly provides a proof of the absence of second-class constraints, thus dispensing the need to study  the full Hamiltonian system of constraints associated with \eqref{tlagr}.

\subsubsection{Mixed-symmetry tensors}\label{sec:dof_flat_mixed}

Having discussed in some detail the counting of degrees of freedom for the case of symmetric tensors we are now in the position to extend our proof to the more general case of $GL(D)-$tensors  subject to \eqref{mixedeom}, \eqref{mixed_gauge} and \eqref{enh}. As already stressed, the main point under scrutiny is to investigate the role of the double divergences of $\vf$ for light-like momenta; however, for the case of $O(D)-$tensors we saw in section \ref{sec:flat_mixed} that our present solution requires to deal with this issue by imposing the constraints \eqref{mixtdiv}. Once this is assumed the reduction of the equations of motion to the Fierz conditions proceeds as we described in section \ref{sec:flat_mixed} and need not to be repeated here. For this reason in the following we focus on the case of $GL(D)-$tensors. Our treatment applies to both reducible and irreducible cases.

 We discuss separately the two cases $p^{\, 2} \neq 0$ and $p^{\, 2} = 0$. In particular, as we saw in section \ref{sec:flat_mixed}, in the former case one can solve for $\vf$ in  \eqref{mixedeom} obtaining
\be \label{gaugefix}
\vf \, = \, \fr{p^{\, i}}{p^{\, 2}} \, p_{\, i} \, \vf
\ee
thus implying that $\vf$ only contains pure gauge components, provided one also shows that under the same conditions all double divergences vanish as well, 
\be
p_{\,i}\, p_{\, j} \, \vf \, = \, 0 \, ,
\ee
which, in its turn, can be iteratively proven to hold taking successive divergences of the equations of motion in this kinematical regime. For $p^{\, 2} = 0$ on the other hand it is no longer true that the divergences of the equations of motion imply the vanishing of all double divergences of the field:
\be \label{kerddiv}
p^{\, i} \, p_{\,i}\, p_{\, j} \, \vf \, = \, 0 \, \hspace{1cm} \stackrel{p^{2} = 0}{\, \, \not \!  \! \longmapsto} \hspace{1cm} p_{\,i}\, p_{\, j} \, \vf \, = \, 0 \, .
\ee
In other words, for light-like momenta the gradient operator on the l.h.s. of \eqref{kerddiv} possesses a non-trivial kernel. For instance, in the frame where $p_{\, \m} \, = \, p_{\, +}$, for an irreducible tensor with the symmetries of the diagram $\yng(3,2)$ the component of the double divergence given by $p^{\, \a} p^{\, \b} \vf_{+ + +, \, \a \b}$ can never appear in the first of \eqref{kerddiv}, as it should arise from the divergence of components of the equations of motion with four ``$+$'' indices altogether, which are identically zero due to irreducibility. However, it might still be possible to overcome this difficulty without introducing constraints on the gauge fields if the system \eqref{enh} providing the conditions for gauge invariance
\be \label{enh2}
p^{\, i} \, p^{\, j} \, p_{\, i} \, \L_{\, j} \, = \, 0 \, ,
\ee
admits more solutions for $p^{\, 2} = 0$ than for $p^{\, 2} \, \neq \, 0$, and those additional solutions be capable to gauge fix to zero just those components of the double divergences that live in the kernel of the gradient in $p^{\, i} \, p_{\,i}\, p_{\, j} \, \vf$.

 The direct evaluation of the rank of the algebraic system \eqref{enh2} in a few non-trivial examples, e.g. tensors of the form $\{3, 2\}$ and $\{4, 2\}$, shows explicitly that the number of independent solutions changes in the two kinematical regimes, and that in particular for $p^{\, 2} = 0$ there is an enhancement of the gauge symmetry.  In order to avoid possible sources of ambiguities let us stress that the two systems \eqref{kerddiv} (first equation) and \eqref{enh2} provide the exact algebraic outcome,  for all $p$'s, of the computation of the divergences and of the gauge variation of $M \vf$, in the sense that in both expressions there are no D'Alembertian operators involved, that we might be possibly discarding in the regime when we consider $p^{\, 2} = 0$. The rest of this section is essentially devoted to the systematic evaluation of the gauge symmetry available for light-like momenta and to its implementation in the elimination of unphysical components from our equations. We shall start analysing in detail the case of tensors with two families of indices, to then move to the description of the general case. 
 
 For light-like momenta it is simpler to work in a frame where $p = p_{\, +}$. To make things more explicit we introduce a notation highlighting the number of ``$+$'' and ``$-$'' indices in a given family; for instance for the gauge variation of a generic component of the gauge field $\vf_{\m_s,  \, \n_r}$ we shall write
\be \label{deltaphi}
\d \, \vf_{+^m -^k,  \, +^n -^l} \, =  \, m \, \L_{+^{m-1} -^k,  \, +^n -^l} \, + \, n \, \l_{+^m -^k,  \, +^{n-1} -^l} \, ,
\ee
meaning that $\vf$ carries $m$ indices along the ``$+$'' direction and $k$ indices along the ``$-$'' direction  in the first family, and similarly $n$ indices along the ``$+$'' direction and $l$ indices along the ``$-$'' direction  in the second family. Moreover, in order to streamline the presentation we shall systematically omit overall coefficients proportional to  $p_+$, together with indices transverse to the light-cone, e.g. the complete form for $\vf$ in this case should read
\be
\vf_{+^m -^k,  \, +^n -^l} \, \equiv \, \vf_{+^m -^k i_{s-m-k},  \, +^n -^l j_{r-n-l}}\, .
\ee
In this notation the variation of the equations of motion looks\ft{To further simplify our formulas in this section we shall denote simply with $M$ the tensor providing the equations of motion: $M = (\Box \, - \, \pr^{\, i} \, \pr_{\, i}) \, \vf$.}
\be \label{deltaM}
\begin{split}
\d \, M_{+^m -^k,  \, +^n -^l} \, = & \, 2 \, {m \choose 2} \,  \L_{+^{m-2} -^{k+1},  \, +^n -^l} \, + \,  
2 \, {n \choose 2} \, \l_{+^m -^k,  \, +^{n-2} -^{l+1}} \, \\
& + m \cdot n \, (\L_{+^{m-1} -^k,  \, +^{n-1} -^{l+1}} \, + \, n \, \l_{+^{m-1} -^{k+1},  \, +^{n-1} -^l}) \, ,
\end{split}
\ee
up to an overall factor of $- p_{+}^{\, 2}$ that once again we shall always neglect. One should observe at this point that, due to \eqref{deltaphi}, it is possible to rewrite \eqref{deltaM} as follows
\be \label{deltaMphi}
\d \, M_{+^m -^k,  \, +^n -^l} \, = \, m \,\d \, \vf_{+^{m-1} -^{k+1},  \, +^n -^l} \,  + \, n \, \d \, \vf_{+^m -^k,  \, +^{n-1} -^{l+1}}  \, ,
\ee
where $m + n \geq 2$, otherwise the variation of $M$ would vanish identically. Let us also observe that on the r.h.s. of \eqref{deltaMphi} there is always at least one ``$-$'' index, thus implying that the components of the parameters $\L$ and $\l$ carrying only ``$+$'' and transverse indices are not constrained, and can thus be freely used to gauge fix all components of $\vf$ with at least one ``$+$'' index and no ``$-$'' indices at all.

The form \eqref{deltaMphi} of $\d M$ will play an important role in our discussion. In order to systematically analyse its consequences we proceed ordering the equations for $\d M = 0$ according to increasing number of ``$-$'' indices, starting from those components possessing none of them:
\be \label{deltaM+}
\d \, M_{+^m,  \, +^n} \, = \, m \,\d \, \vf_{+^{m-1} -,  \, +^n} \,  + \, n \, \d \, \vf_{+^m,  \, +^{n-1} -}  \, = \, 0 \, .
\ee
Generally speaking, in order to uncover the full set of conditions imposed by \eqref{enh2} on the parameters one has to consider the possibility that the two terms on the r.h.s. might appear in the variation of some other components of $M$, and then solve the full resulting system; in the case of \eqref{deltaM+} however one can see than both terms can only appear in the variation of the l.h.s.; in this sense \eqref{deltaM+} provides a rectangular system of one equation in two variables\ft{In terms of gauge parameters \eqref{deltaM+} formally involves four variables; however, the effective variables capable of acting on the components of $\vf$, barring gauge-for-gauge transformations, are those identified by the combinations $\d \, \vf_{+^{m-1} -,  \, +^n}$ and $ \d \, \vf_{+^m,  \, +^{n-1} -}$.} indicating the possibility of performing free gauge fixings exploiting in general the combination of  $\d \, \vf_{+^{m-1} -,  \, +^n}$  and  $\d \, \vf_{+^m,  \, +^{n-1} -}$ orthogonal to \eqref{deltaM+}. For instance one could choose to eliminate with a gauge fixing all components of the form $\vf_{+^m,  \, +^{n-1} -}$, i.e. with only one ``$-$'' index, placed in the second family in this choice, while also freezing because of \eqref{deltaM+} the parameters in  the combination $\d \, \vf_{+^{m-1} -,  \, +^n}$. Let us observe that this gauge fixing and the previous one mentioned above would be still allowed by the strong constraints \eqref{mixtdiff}.

 In the second step we look at those components of $\d M$ containing one ``$-$'' index, first placing it in the second family:
\be \label{deltaM-}
\d \, M_{+^m,  \, +^n - } \, = \, m \,\d \, \vf_{+^{m-1} -,  \, +^n - } \,  + \, n \, \d \, \vf_{+^m,  \, +^{n-1} - -}  \, = \, 0 \, .
\ee
The difference with the previous case is that, while the second term  on the r.h.s. can only appear in the variation of this component of $M$, the first one also contributes to another term with the same overall number of ``$+$'' and ``$-$'' indices:
\be \label{deltaM-bis}
\d \, M_{+^{m-1} -,  \, +^{n+1}} \, = \, (m-1) \,\d \, \vf_{+^{m-2} - -,  \, +^{n+1}} \,  + \, (n+1) \, \d \, \vf_{+^{m-1} - ,  \, +^{n} -}  \, = \, 0 \, ,
\ee
where in the first place we consider $m \geq 2$, i.e. we assume that both terms in the variation make sense. Under these conditions
an additional new term appears at this level, which however cannot contribute to other variations of $M$, so that \eqref{deltaM-} and \eqref{deltaM-bis} form a rectangular $2 \times 3$ system whose matrix can be arranged as 
$$
\left(\begin{array}{ccc}
n & m & 0 \\
0 & n+1 & m-1
\end{array}\right)
$$
thus showing that, out of the three combinations of parameters involved, one is left available for gauge fixing even after imposing the condition that the variation of $M$ be zero. The cases $m = 0, 1$ are instead representative of a different type of possible systems: for $m = 0$ and 
$n \geq 2$ the first equation simply sets to zero the combination of parameters in $\d \, \vf_{ ,  \, +^{n-1} - -}$, while for $m = 1$ and $n \geq 1$ we can still perform the second step but \eqref{deltaM-bis} would contain in this case only one term, and the system would degenerate to a square system with no eigensolutions. The same kind of situation holds reversing the roles of $m$ and $n$.

The outcome of this first piece of analysis is that, although the conditions \eqref{enh2} effectively obstruct many of the possible gauge fixings, there appears to be a class of constraints (that we shall characterise soon) leaving room for some free parameters to be used to eliminate components of $\vf$. Before showing the details of how to systematically implement the latter, here we would already like to observe that the options emerging at this level are of a type that would be not permitted if we were to  impose the strong conditions \eqref{mixtdiff}.

 To this end, let us consider for a moment the general case of $N$ families and $N$ parameters $\L_{\, i}$, and let us first expand the ``diagonal'' sector of \eqref{mixtdiff}, given by $i = j$, obtaining
%:
\be \label{diagtdiff}
\L_{\, i\, (-)_i} \, = \, 0 \, ,
\ee
where for instance with the notation 
\be
 \L_{\, i \, (-)_i} 
\ee
we denoted a component of the gauge parameter $\L_{\, i}$ with one ``$-$'' index in the $i-$th family. (Here latin letters label family numbers.) It is not hard, then, to recognise that \eqref{diagtdiff} imposes on each parameter $\L_{\, i}$ a condition analogous to \eqref{symmtdiff} for the symmetric case, essentially stating that the components with ``$-$'' indices in the $i-$th family are not independent, and thus cannot be used to gauge fix some components of $\vf$, once all components of $\L_{\, i}$ with ``$+$'' and transverse indices in the $i-$th family have been used. Now let us consider the role of the ``off-diagonal'' constraints
\be
\pr_{\, i} \, \L_{\, j} \, + \, \pr_{\, j} \, \L_{\, i} \, = \, 0 \, , \hspace{3cm} i \, < \, j\, , 
\ee
which can be expanded as
\be \label{offdiagtdiff}
\L_{\, i \, (-)_j} \, + \, \L_{\, j \, (-)_i}  \, = \, 0 \, ,
\ee
where for instance the notation
\be
\L_{\, i \, (-)_j}
\ee
identifies a gauge parameter with one index less in the $i-$th family, possessing at least one ``$-$'' index in the $j-$th family. 
In particular, for the case of two families  it is possible to recognise that \eqref{diagtdiff} and \eqref{offdiagtdiff} essentially set to zero all components of the parameters with two ``$-$'' indices, in whatever position, thus obstructing any of the gauge fixings still allowed after imposing \eqref{deltaM-} and \eqref{deltaM-bis}, and therefore showing that the strong constraints in force for generic values of the momenta get relaxed to some weaker conditions when $p^{\, 2} = 0$.

 Coming back to the analysis of \eqref{enh2}, it turns out that the possibilities explored for the variations of the components of $M$ with one ``$-$'' index provide instances of all possible outcomes in the general case: 
\begin{itemize}
 \item the system \eqref{enh2} decomposes in diagonal blocks, each block identified by the overall number of ``$+$'' and ``$-$'' indices, together with the disposition of those indices in the two families for any one of the $\d M$'s giving rise to the equations in the block.
 \item Once a representative is chosen among the equations of the system, say $\d \, M_{+^m -^k,  \, +^n -^l }$, all other equations of the block are uniquely fixed by considering the variables $\d \vf$ appearing in each equation and adding to the system the possible other equation where the same variable appears. Any equation can host at most two variables and each $\d \vf$ can appear at most in two different equations, according to whether or not it possesses ``$-$'' in both families; with reference to the case discussed above we see that while  $\d \, \vf_{+^{m-1} -,  \, +^n - }$ appears in  two equations $\d \, \vf_{+^m,  \, +^{n-1} - -}$ is present only in \eqref{deltaM-}, since its two ``$-$'' indices are both in the second family and thus it can only emerge as the result of computing a divergence in this group of indices.
 \item The possible blocks can be either ``square'' blocks or rectangular ones: 
 \begin{itemize}
  \item the ``square'' blocks are those containing at least one equation in which all ``$+$'' indices belong to a single family; the number of independent equations in the systems containing these kind of representatives is always the same as the number of unknowns, so that this class of equations implies that the corresponding gauge fixings are all obstructed. Another way to see this point is that for the representative whose ``$+$'' indices all lie within a single family the equation is always of the form $\d \vf = 0$, and thus, due to the banded structure of the matrices (with bands of width $2$) all the variables in the same systems are set to zero as well.
  \item rectangular blocks arise whenever there is a representative of the two possible forms\ft{The two options go together and are not alternative, in the sense that given a representative of one of the two kinds the other kind will be also present in the same block. Although for the explicit construction of the systems it might be useful to start with one of the two and then generate the other equations iteratively, this presentation better displays the symmetry between the two families.}
 \be
  \begin{split}
   & \d \, M_{+^{m_1},  \, +^{n_1} -^k} \, = \, 0 \hspace{1cm} m_1 >  k, \, n_1 >  0 \\   
   & \d \, M_{+^{m_2} -^k, \, +^{n_2}} \, = \, 0  \hspace{1cm} m_2 >  0, \, n_2 >  k \, ,
  \end{split} 
 \ee
 with $m_1 + n_1 = m_2 + n_2 \geq 2$, while $k > 0$ (there is always at least one ``$-$'' index). These blocks have dimensions $(k + 1) \times (k + 2)$ and are described by banded matrices of the form
$$
{\mathcal A} = \left(
\begin{array}{ccccc}
n & m & \cdots & 0 & 0\\
0 & n + 1 & m-1 & \cdots & 0\\
\vdots & \vdots & \ddots & \ddots & \vdots\\
0 & 0 & \cdots & n + k + 1 & m - k - 1\\
\end{array}
\right)
$$
These are the systems where additional gauge symmetry sits, of a kind that would not be allowed by the strong constraints \eqref{mixtdiff}\ft{This applies to $k > 1$. For $k = 1$ the corresponding gauge symmetry would still be available assuming strong constraints.}: for each of these blocks one can save for the gauge fixing a combination of the parameters at will among the $k + 2$ present; for definiteness we shall choose to preserve the combinations appearing in $\d \, \vf_{+^{m},  \, +^{n} -^k}$.
 \end{itemize}
\end{itemize}
This completes the analysis of the weak constraints \eqref{enh2}. 

Our remaining task is to show that the resulting gauge symmetry in combination with the equations of motion allow to conclude that all components of $\vf$ with indices along directions longitudinal to the light cone can be set to zero. We shall prove it by induction in the number of ``$-$'' indices. As already noticed, the parameters with no  ``$-$'' indices are not constrained; this means that it is possible to choose 
$\d \, \vf_{+^m,  \, +^n}$  s.t.
\be
\begin{split}
&\vf_{+^m,  \, +^n} \, + \, \d \, \vf_{+^m,  \, +^n} \, = \, 0 \, , \\
& m + n \geq 1 \, .
\end{split}
\ee
From the equations of motion for these components we obtain
\be \label{M-}
M_{+^m,  \, +^n} \, = \, 0 \qquad \rightarrow \qquad m\, \vf_{+^{m-1} -,  \, +^n} \,  + \, n \, \vf_{+^m,  \, +^{n-1} -}  \, = \, 0 \, .
\ee
We see that for $m \geq 1, \, n = 0$ or $m = 0, \, n \geq 1$ \eqref{M-} sets to zero those components of the form (restoring transverse indices)
$\vf_{+^{m-1} - \, i_{s - m}, \, j_r}$ or $\vf_{i_s,\, +^{n-1} - \, j_{r - n}}$, respectively.  Whenever both $m$ and $n$ are greater than zero, however, \eqref{M-} would not imply that both components vanish. On the other hand, as noticed above when commenting \eqref{deltaM+}, the constraints allow to gauge away e.g. $\vf_{+^m,  \, +^{n-1} -}$ so that putting everything together we eventually get, for all possible $m$ and 
$n$
\be
\begin{split}
&\vf_{+^{m-1} -,  \, +^n} \, = \, 0 \, ,\\
&\vf_{+^m,  \, +^{n-1} -} \, = \, 0 \, .
\end{split}
\ee
Let us also observe that equation \eqref{M-} retains the same form as \eqref{deltaM+}, although the latter only holds for $m + n \geq 2$. This observation will be the key to the next step and to the discussion of the general, $N-$family case. 

Now that we showed how the combined action of gauge fixing and equations of motion leads to the elimination of all components with one ``$-$'' index let us assume that the same holds for all components with $t$ ``$-$'' indices overall, i.e.
\be
\vf_{+^m -^k,  \, +^n -^{t-k}} \, = \, 0 \qquad \forall\, m, \, n \qquad k = 0, \, \ldots, t
\ee
and let us compute the corresponding equations of motion 
\be \label{M-t}
M_{+^m -^k,  \, +^n -^{t-k}} \, = \, 0 \qquad \rightarrow \qquad m\, \vf_{+^{m-1} -^{k + 1},  \, +^n -^{t-k}} \,  
+ \, n \, \vf_{+^m -^{k},  \, +^{n-1} -^{t-k +1}}  \, = \, 0 \, .
\ee
As anticipated, the key observation is that \eqref{M-t} bears the same structure as \eqref{deltaMphi} (with $l = t - k$), with the implication that the analysis of the corresponding system is the same as that of the constraints, provided that $m + n \geq 2$, as required for the very existence of \eqref{deltaMphi}, and thus it proceeds as follows:
\begin{itemize}
 \item for $m + n= 1$  \eqref{M-t} directly implies 
 $$
 \vf_{-^{k},  \, -^{t-k +1}}  \, = \, 0 \qquad k = 0, \ldots, t
 $$
i.e. all components with no  ``$+$'' indices and $t + 1$ ``$-$'' indices vanish due to the equations of motion;
 \item for $m + n \geq 2$  \eqref{M-t} can be discussed in the same way as \eqref{deltaMphi}; in particular the overall system organises in blocks s.t.
  \begin{itemize}
  \item the group of equations containing representatives whose ``$+$'' indices all belong to a single family organise in  ``square'' blocks of maximal rank: all components of $\vf$ appearing in these system vanish due to the equations of motion. 
  \item Equations of the form
  \be
  \begin{split}
   & M_{+^{m_1},  \, +^{n_1} -^t} \, = \, 0 \hspace{1cm} m_1 >  t, \, n_1 >  0\, , \\   
   & M_{+^{m_2} -^t, \, +^{n_2}} \, = \, 0  \hspace{1cm} m_2 >  0, \, n_2 >  t \, ,
  \end{split} 
 \ee
with $m_1 + n_1 = m_2 + n_2$ and $t > 1$ do not set to zero all components, as they ``generate''  rectangular systems of $t+1$ equations and $t+2$ unknowns, of rank $t + 1$.  However, as the analysis of the constraints shows, for each choice of component of $\vf$ that is not set to zero by these system there is a corresponding gauge fixing available allowing to eliminate it from the equations. For instance, we might decide for definiteness to solve all these systems in terms of the components of the form $\vf_{+^{m},  \, +^{n} -^{t + 1}}$, and then to exploit the parameters in $\d \, \vf_{+^{m},  \, +^{n} -^{t + 1}}$, in terms of which we can also decide to solve the corresponding block of constraint equations, to eventually eliminate this last set of components by means of a gauge fixing.
 \end{itemize}
\end{itemize}
One can appreciate that the two steps allow to set to zero all components of $\vf$ with $t + 1$ ``$-$'' indices altogether, thus completing the proof.

We are now in the position to discuss the general case in relatively simple terms, as the steps we need are essentially the same as those we just discussed.

Let us consider a tensor with $N$ families of indices and compute its gauge variation for $p = p_+$
\be \label{deltaphiN}
\d \, \vf_{+^{m_1} -^{k_1}, \,\ldots\, , +^{m_i} -^{k_i}, \,\ldots\, , +^{m_N} -^{k_N}} \, = \, \sum_{i = 1}^N\, m_i  \, \L^{(i)}_{+^{m_1} -^{k_1}, \,\ldots\, , +^{m_i - 1} -^{k_i}, \,\ldots\, , +^{m_N} -^{k_N}}\, .
\ee
Now let us impose that the gauge variation of $M$ (written up to overall signs and factors of $p_+^{\, 2}$) be zero and let us make use of \eqref{deltaphiN} to rewrite it in terms of the variation of $\vf$:
\be \label{deltaMN}
\begin{split}
\d \, M_{\ldots\,, +^{m_i} -^{k_i}, \, \ldots } \, = \,  & \sum_{i = 1}^N\, 2 \, {m_i  \choose 2} \L^{(i)}_{\ldots\,, +^{m_i - 2} -^{k_i + 1}, \, \ldots } \\
& + \,  
\sum_{i < 1}^N\, 2 \, m_i \cdot m_j \left\{\L^{(i)}_{\ldots\,, +^{m_j - 1} -^{k_j + 1}, \, \ldots } + \L^{(j)}_{\ldots\, , +^{m_i - 1} -^{k_i + 1}, \, \ldots }\right\} \\
= &  \sum_{i = 1}^N\, m_i \, \d \, \vf_{\ldots\, , +^{m_i} -^{k_i}, \, \ldots } \, = \, 0 \, .
\end{split}
\ee
We would like to show that the equations of motion, together with the gauge symmetry available after imposing \eqref{deltaMN}, allow to eliminate all unphysical components from $\vf$. 

The constraints \eqref{deltaMN} involve parameters possessing at least one ``$-$'' index; this means that those components of $\vf$ possessing at least one ``$+$'' index and no ``$-$'' indices at all can be eliminate via a gauge fixing of the form
\be
\vf_{+^{m_1}, \,  \ldots\, , +^{m_i}, \, \ldots\, , +^{m_N} } \, + \,\d \, \vf_{+^{m_1}, \,  \ldots\, , +^{m_i}, \, \ldots\, , +^{m_N}} \, = \, 0 \, .
\ee
The equations of motion for these components, in their turn, define systems for components of $\vf$ possessing one ``$-$'' index in the various possible positions; in particular, when only one ``$+$'' index is present they set to zero the corresponding component of $\vf$ with only transverse indices and just a single ``$-$'' index in the corresponding family:
\be
\begin{array}{lcl}
M_{+ i_{s_1 - 1},\, \ldots } \, = \, 0 &\qquad \rightarrow \qquad & \vf_{- i_{s_1 - 1},\, \ldots } \, = \, 0 \, , \\
M_{i_{s_1}, + i_{s_2 - 1},\, \ldots } \, = \, 0 &\qquad \rightarrow \qquad &\vf_{i_{s_1}, - i_{s_2 - 1},\, \ldots } \, = \, 0 \, , \\
\ldots &\qquad \rightarrow \qquad  & \ldots
\end{array}
\ee
while whenever two or more  ``$+$'' indices are involved the resulting equation bears the same form as \eqref{deltaMN}:
\be \label{MN+}
\begin{split}
M_{+^{m_1}, \, \ldots\, , +^{m_i}, \, \ldots } \, = \, 0  \qquad \rightarrow \qquad   \sum_{i = 1}^N\, m_i \,  \vf_{+^{m_1},\, \ldots\, , +^{m_i-1} -, \, \ldots } \, = \, 0 \, ,
\end{split}
\ee
with the implication that for all those subsystems in \eqref{MN+} that do not set to zero all the components of $\vf$ that they contain there is a corresponding subsystem of constraints in \eqref{deltaMN}; the latter will admit non-vanishing solutions for combinations of the parameters in one-to-one correspondence with the variables chosen as independent in the solution to \eqref{MN+}, so that one can use these solution to perform a gauge fixing eliminating the component of $\vf$ surviving from \eqref{MN+}. 

 While showing that all components with one ``$-$'' index eventually vanish, the previous argument also suggests the way to the general conclusion: if we assume that all components of $\vf$ with a given number of ``$-$'' indices altogether, say $t$, however displaced, vanish, then we are in the position of investigating the consequences of this assumption on the equations of motion:
\begin{itemize}
 \item for those components of $\vf$ that, on top of the $t$ ``$+$'' indices, also possess a single ``$+$'' index the equations of motion are easily seen to imply the vanishing of all components with $t + 1$ ``$-$'' indices placed in all possible ways and no ``$+$'' indices at all;
 \item whenever at least two ``$+$'' indices are present, the equations of motion for these components generate a system identical to that associated with the weak constraints, \eqref{deltaMN}. This implies that whenever a gauge fixing is obstructed because of the constraints then the corresponding component is set to zero by the equations of motion; vice-versa, in case the equations of motion are not strong enough to eliminate some of the unwanted components of $\vf$, the structure of \eqref{deltaMN} guarantees that the possibility of eliminating those components by means of an unconstrained gauge transformation.
\end{itemize} 
 Finally, let us observe that  our procedure automatically takes care of the reducibility of the gauge transformations: since all gauge fixings that we eventually perform involve combinations of the parameters in the form of variations $\d \vf$ of the corresponding components, then the possible linear dependence among the parameters themselves is automatically taken into account, and need not be discussed separately.

%%%%%%
\subsection{(A)dS backgrounds} \label{sec:dof_ads}
%%%%%%

 In this section we investigate the spectra described by the Lagrangians \eqref{lag_ads_symm} and \eqref{lag_mix_ads}. For the symmetric case our analysis mainly relies on one assumption: under ``smooth'' deformation of a Lagrangian gauge theory in Minkowski space to a Lagrangian gauge theory in (A)dS space the  number of degrees of freedom is unchanged. The deformation is termed ``smooth'' if it keeps the number of gauge symmetries. In Hamiltonian terms this statement is essentially equivalent to saying that a smooth deformation cannot introduce second-class constrains into the (A)dS system that were not already present in the flat one. We are not aware of any general proof of this otherwise reasonable\ft{Under such a deformation the number of primary constraints clearly does not change. However, it is a general result that whenever second-class constraints are present at least one of them should appear among primary constraints \cite{gitman}. Thus, assuming the flat theory to be free of second-class constraints, the possibility that they appear in the deformed (A)dS system would imply that some of the primary constraints changed their nature under the deformation, without the overall number of gauge generators being modified.} conclusion, and in the following we shall abide by the conventional wisdom of assuming its validity. Although our (A)dS Lagrangians for symmetric fields comply indeed with this criterion, we shall also provide a few additional independent arguments in support of their correctness, to be later adapted to the case of mixed-symmetry fields where our general solution does not preserve all the gauge symmetries of the corresponding flat theory. However, in the last part of this section we shall also provide some specific instances of smooth deformations of mixed-symmetry Lagrangians describing multiplet of fields in flat space, along the lines of \cite{bmv,zin-bose}.

\subsubsection{Symmetric tensors}\label{sec:dof_ads_symm}

For symmetric tensors in (A)dS background the Lagrangian \eqref{lag_ads_symm}  retains the same number of unbroken independent gauge symmetries as its Minkowskian counterpart \eqref{tlagr}, which is in fact true for both reducible and irreducible cases. Thus, the number of propagating polarisations is expected to coincide with that of the flat case. To provide further support to this conclusion let us also discuss a couple of independent arguments to the same effect.

For the irreducible case we have to analyse the content of the equations \eqref{ads_irr_eom},
\be \label{ads_irr_eom2}
M\, \vf \, - \, \fr{1}{L^{\, 2}} \, \left[\,(s - 2)(D + s - 3) \, - \, s\,\right] \, \vf \, + \, 
\fr{2}{D + 2 \, (s - 2)} \, g \, \nabla \cdot \nabla \cdot \, \vf \, =  \, 0 \, .
\ee
Since the number of first-class constraints is the same as for the Minkowskian case we can at least conclude that the degrees of freedom associated to the (A)dS equation cannot exceed those of the flat theory. The latter, on the other hand, also coincide with the propagating polarisations described by the Fierz system in (A)dS (see \cite{repr_ds} and \cite{repr_ads} for dS and AdS backgrounds, respectively):
\be \label{FSads}
\begin{split}
&\left\{\Box\, - \, \fr{1}{L^{\, 2}} \, \left[\,(s - 2)(D + s - 3) \, - \, s\,\right] \right\}\, \vf \, = \, 0 \, , \\
&\nabla \cdot \vf \, = \, 0 \, , \\
&\vf^{\, \pe} \, = \, 0 \, .
\end{split}
\ee
Thus, in order to prove that \eqref{ads_irr_eom} propagates the degrees of freedom of a single massless particle of spin $s$ it will be sufficient to show explicitly that it possesses all the solutions to \eqref{FSads}. Indeed in our framework $\vf^{\, \pe} = 0$ by assumption, while for fields in the kernel of the Klein-Gordon operator, i.e. for $\vf\, $ s.t.
\be
\left\{\Box\, - \, \fr{1}{L^{\, 2}} \, \left[\,(s - 2)(D + s - 3) \, - \, s\,\right]\right\} \, \vf \, = \, 0,
\ee
computing $n$ divergences of \eqref{ads_irr_eom2} we obtain
\be \label{div_ads}
\left\{\fr{n\, (n - 1)}{L^{\, 2}} \, \r_{n + 3} \, \r_{n + 4} \, (\nabla \cdot)^{\, n} \, + \, 
\r_{2n + 4} \, \nabla \, (\nabla \cdot)^{\, n + 1} \, - \, 2 \, g \,  (\nabla \cdot)^{\, n + 2} \right\} \, \vf \, = \, 0 \, ,
\ee
where we defined $\r_{\, n} \, = \, D \, + \, 2 s \, - \, n$. It is then possible to observe that \eqref{div_ads} recursively sets to zero all multiple divergences of $\vf$ in decreasing order, finally leading to $\nabla \cdot \vf \, = \, 0$. 

For the case of traceful tensors, described by the equations of motion \eqref{M_L_symm}
\be \label{M_L_symm2}
M_L\, \vf \,\equiv\, M\, \vf \, - \, \fr{1}{L^{\, 2}} \,\left\{\, \left[\,(s - 2)(D + s - 3) \, - \, s\,\right] \, \vf \, - \, 2 \, g \, \vf^{\, \pe} \,\right\},
\ee
while it is still true that the number of first class constraints is the same as the flat reducible theory, however it is not obvious what should be the proper ``Fierz system'' with which to compare our equations in order to prove that the degrees of freedom actually match those of the flat case \eqref{eomsymm}. The naive guess suggested by the flat-space example \eqref{FS} would be to reproduce the first two conditions in \eqref{FSads} while keeping the trace undetermined. However, it is simple to observe that, as a consequence of \eqref{M_L_symm2}, the first two conditions in \eqref{FSads} would anyway imply $\vf^{\, \pe} = 0$ thus leading to the contradictory conclusion that \eqref{M_L_symm2} describes the same degrees of freedom as the irreducible case. The reason behind this difference with respect to the case of flat background is that massless fields in (A)dS have mass-like terms depending on the spin, so that the various  propagating components  in $\vf$ actually satisfy different equations of motion.

However, the effective particle content associated to the Lagrangian \eqref{lag_ads_symm} can be identified comparing with the unconstrained Lagrangian for (A)dS triplets of \cite{st, FoTsu, FTads}. In that context the relevant equations after eliminating an auxiliary field are
\be \label{tripletAdS}
\begin{split}
& M_L \, \vf \, = \, - \, 2 \, \nabla^{\, 2} \, \cD \, + \, \fr{8}{L^{\, 2}} \, g \, \cD, \,    \\
& \hat{M}_L \, \cD \, = \, \nabla \cdot \nabla \cdot \, \vf \, - \, \fr{4}{L^{\, 2}} \, \vf^{\, \pe} \, ,
\end{split}
\ee
where $ \hat{M}_L$ is a deformation of the flat-space kinetic operator for $\cD$ \eqref{Mdeform},
\be
\hat{M}_L \, = \, 2 \, \Box \, + \, \nabla \, \nabla \cdot \, - \, \fr{2}{L^{\, 2}} \, \left[(s \, - \, 1) \, (D \, + \, s \, - \,3) \, + \, 3 \right]
\, + \, \fr{4}{L^{\, 2}} \, g \, T \, ,
\ee
while gauge invariance obtains choosing $\d \, \vf \, = \, \nabla \, \L$ and $\d \, \cD \, = \, \nabla \cdot \L$.
To make contact with our constrained theory, as already observed for the flat case, we remove the transversality constraint \eqref{gauge_ads} \`a la Stueckelberg, performing the gauge-invariant redefinition
\be
\vf \, \longrightarrow \, \vf \, - \, \nabla \, \th
\ee
where $\d \, \th \, = \, \L$. The resulting Lagrangian  
\be
\cL \, = \, \12 \, \vf \, M_L \, \vf \, + \, \vf \, \left(\nabla^{\, 2} \, - \, \fr{4}{L^{\, 2}} \, g\right)\, \nabla \cdot \th \, 
- \, 2 \, {s \choose 2} \, \nabla \cdot \th \, \hat{M}_L \, \nabla \cdot \th \, ,
\ee
actually coincides, upon renaming $\nabla \cdot \th \, \equiv \, \cD$, with the (A)dS triplet Lagrangian leading to \eqref{tripletAdS}, whose particle content was shown in \cite{FoTsu, FTads} to correspond to that of the flat space-time reducible system here computed in section \ref{sec:dof_flat}, thus completing our check. In section \ref{sec:ads_diag} we show how to decompose the field $\vf$ in order to identify in \eqref{lag_ads_symm} the propagating modes, each described by a single-particle Lagrangian leading to equations of the form  \eqref{ads_irr_eom}.

\subsubsection{Mixed-symmetry tensors}\label{sec:dof_ads_mixed}

 In this section we discuss the spectrum of the theory described by the Lagrangian \eqref{lag_mix_block} corresponding to the the AdS-unitary choice of keeping the gauge parameter lacking one box in the first rectangular block:
\be \label{lag_mix_ads_un}
\cL \,=\, \12 \ \vf \left\{\, \Box \,-\, \nabla^i\, \nabla_i\, - \, \frac{1}{L^2}\, \bigg[\, (s_{1}- t_1-1)(D+s_1- t_1-2) \,-\, \sum_{j\,=\,1}^p t_j s_j \,\bigg] \right\} \vf \, .
\ee
Here $\vf$ carries a representation of $O(D)$ described by  a diagram with $p$ rectangular blocks. The reduced amount of gauge invariance available for (A)dS tensors with mixed symmetry introduces additional complications if compared to more standard situations. For instance, for symmetric tensors on  flat space-time the variation of the divergence of the field in transverse-invariant theories is proportional to the D'Alembertian of the parameter:
\be
\d \, \prd \vf \, = \, \Box \, \L \, ,
\ee
thus implying that the transverse part of $\prd \vf$ can be removed upon partial gauge-fixing and need not be eliminated  manipulating the equations of motion. To appreciate the differences met in our present case it suffices to consider the simplest $O(D)$-hook field $\vf_{\, \m \n, \, \r}$, whose divergence varies according to
\be \label{example_div}
\d \, \nabla^{\, \a} \vf_{\, \a \, \n, \, \r} \, = \, \left(\Box  \, -\fr{D - 2}{L^2}\right)  \L_{\, \n, \, \r}\, ,
\ee 
where $\L_{\, \n, \, \r}$ is the antisymmetric parameter that, according to the general analysis of \cite{metsaev1, metsaev2}, must be kept in an Anti de Sitter background in order to preserve unitarity. It is then manifest that only the antisymmetric projection of $\nabla^{\, \a} \vf_{\, \a \, \n, \, \r}$ can be gauged away, while its symmetric projection, being gauge invariant, now has to be eliminated by other means. In general the variation under \eqref{gauge_expl_2} of the divergences of irreducible fields subject to \eqref{irr} can be cast in the form
\be
\d\, \nabla_i\, \vf \,=\, \left( \Box - \frac{1}{L^2} \left[ (s_n-n)(D+s_n-n-1)-\sum_{k\,=\,1}^N s_k +1 \right] \right) \L_{\,i} \, ,
\ee
where the left-hand side admits all Young projections labelled by available diagrams of the form $\{ s_1,\ldots,s_j-1,\ldots,s_N\}$ with $j>i$. If one keeps the parameter $\L^{(t_1)}$ all divergences but those taken in the first block are thus gauge invariant, similarly to the symmetric projection of \eqref{example_div}. This motivates the constraint \eqref{ultra-constr}, that we recall here for convenience,
\be \label{ultra-constr_2}
\nabla_{t_1+t_2}\, \vf \,=\, 0 \, ,
\ee
which sets to zero all gauge-invariant divergences and the corresponding components of $\nabla_i\,\vf$ with $i \leq t_1$. Note however that even after imposing \eqref{ultra-constr} not all gauge-invariant quantities automatically vanish. Indeed, in analogy with the flat-space case, the  irreducible components of the multiple divergence $(\nabla_{t_1})^n\, \vf$ with $n > 1$, denoted by $\{(s_1,t_1-1),(s_1-n,1),(s_2,t_2),\ldots,(s_p,t_p)\}$ in the notation where each pair $(s_i, t_i)$ indicates the dimensions of the i-th block, when available, is also gauge invariant and the field equations must dispose of it in order to avoid the propagation of spurious terms.

To show that this is indeed the case it is convenient to write explicitly the Lagrangian field equations. The traces of $M\vf$ are proportional to double divergences of the field,
\be
T_{ij}\, M\vf \,=\, -\, 2\,\nabla_{i}\nabla_{j}\,\vf \, ,
\ee
while all double traces vanish on account of \eqref{tracemix}. Therefore, when one imposes the constraint \eqref{ultra-constr_2} most of the terms in $g^{ij}$ that are needed to project $M\vf$ onto its traceless part are absent. In order to further simplify the discussion let us consider the case in which the first row of the diagram associated to the field is longer than the second, so that $s_1 = s_2 + k$. The general case can be discussed with minor modifications and with the proviso that the ensuing discussion applies to the last row of the first block. Under these conditions the constraint \eqref{ultra-constr_2} takes the form
\be \label{ultra2}
\nabla_2\,\vf \,=\, 0 \, ,
\ee
while the field equations following from the Lagrangian \eqref{lag_mix_ads_un} are
\be \label{eom-ultra}
E\,\vf \,=\, \left\{\, \Box \,-\, \nabla^1\, \nabla_1\, - \, m^2 \,\right\} \vf \,+\, \frac{2}{D+2(s_1-2)}\, g^{11}\, \nabla_1 \nabla_1\, \vf
\ee
with
\be \label{massE}
m^2 \,=\, \frac{1}{L^2}\, \bigg[\, (s_{1}- t_1-1)(D+s_1- t_1-2) \,-\, \sum_{j\,=\,1}^N s_j \,\bigg] \, .
\ee
All other terms that would enter the traceless projection in flat space here vanish on account of \eqref{ultra2}, in particular due to the fact that the commutator of two covariant divergences would be proportional to a trace, so that in our present setup with traceless fields there are no ordering issues for multiple divergences. Moreover, \eqref{eom-ultra} automatically satisfies the condition
\be \label{projE}
\nabla_2\, E\,\vf \,=\, 0 \, ,
\ee
and thus no additional projections are needed. In fact, exploiting the commutators collected in appendix \ref{app:form_mix} in combination with \eqref{ultra2} one finds
\be
\nabla_2\, E\,\vf \,=\, \frac{D-N+1}{L^2}\, S^1{}_2 \nabla_2\, \vf \,+\, \frac{1}{L^2}\, S^1{}_k S^k{}_2 \nabla_1\, \vf \, ,
\ee
where, in particular \eqref{ultra2} also implies that $\nabla_1\,\vf$ is irreducible; as such it is annihilated by all operators $S^i{}_j$ with $i < j$,
thus leading to \eqref{projE}.

Under our hypothesis the Lagrangian equations \eqref{eom-ultra} display essentially the same structure as their counterparts for the symmetric case  \eqref{ads_irr_eom2}, up to the form of the mass-like term \eqref{massE}. Here we would like to extend the argument we proposed in the previous section when discussing the vanishing on-shell of the divergence of the field, and to this end the difference between the mass term of the symmetric theory and \eqref{massE} will be crucial. In the first place one can notice that
\be
[\nabla_1, \nabla^1] \nabla_1 \vf \,=\, \left\{\, \Box \,-\, \frac{1}{L^2} \left[ s_1(D+s_1-2) - \sum_{k\,=\,1}^N s_k \right] \right\} \nabla_1 \vf \, ,
\ee
while all other commutators that appear in the computation of the  divergence of \eqref{eom-ultra} take exactly the same form as in the symmetric case. The point is that the sum over the length of the rows of the associated Young diagram cancels the identical contribution in the mass-like term \eqref{massE}, thus allowing to prove also in this context that on the solutions of Metsaev's equations all divergences vanish.  As we saw in the previous section, in the symmetric case the cancelation of all double divergences extends to fields off-shell and thus, since the structure of equations \eqref{eom-ultra} is identical to that of the symmetric case, we are led to conclude that this property holds in the mixed-symmetry case as well.

 Our conclusion is that the Lagrangian \eqref{lag_mix_ads_un}, supplemented with the constraints \eqref{ultra-constr_2}, describes the propagation of unitary irreducible representations of the AdS group. In the construction we took gauge invariance of the kinetic operator as our guiding principle, while the need for additional constraints on the fields with respect to the flat-space case was motivated by the reduced gauge symmetry that Maxwell-like Lagrangians unavoidably display on (A)dS backgrounds. It might be interesting to observe how the flat limit of the resulting equations of motion connects to our single-particle theory discussed in section \ref{sec:flat_mixed}. Setting to zero the cosmological constant indeed results in a partially gauge-fixed form of the flat-space equations for the $O(D)-$case, where the missing gauge  parameters associated to blocks other than the first can be interpreted in that context as due to the partial gauge fixing setting to zero the transverse parts of the corresponding divergences. In this sense, the two single-particle theories can be linked in a way similar to the case of symmetric tensors. However, while for symmetric tensors the limit $L \to \infty$ is smooth, in this case one experiences an abrupt change in the number of degrees of freedom, thus making it closer in spirit to the naive massless limit taken in Lagrangians for massive higher spins \cite{SH, dario07}.

Another possible approach to the problem would be to construct Stueckelberg Lagrangians for the degrees of freedom of interest, in the spirit of earlier proposals by Zinoviev \cite{zin-frame1, zin-fermi,zin-bose}.  They would provide ``smooth'' deformations of our Lagrangians in flat space, to be related upon a partial gauge fixing to our AdS Lagrangians \eqref{lag_mix_ads_un}, possibly supplemented by a set of Lagrange multipliers forcing on shell the condition \eqref{ultra-constr_2}. We are not going to fulfil this program in its full generality, but in the following we shall provide some explicit examples to this effect, together with some further comments on the structure of the problem. Besides those examples, an additional virtue of the Stueckelberg procedure, making it worth discussing it anyway, is that it allows to form an intuitive picture of the essential peculiarities of (A)dS gauge fields with mixed symmetry if compared to other more conventional classes of free fields. 
 
As we already discussed at length, for a gauge field $\vf$ in a given irrep of $O(D)$ or $GL(D)$ a natural road to its (A)dS deformation is to covariantise its flat gauge transformation,
\be \label{gauge_new}
\d \, \vf \, = \,  \nabla^{\, i} \, \L_{\, i}\, ,
\ee  
to then try and construct the corresponding gauge-invariant kinetic operator. To get a deeper insight into the reasons for the absence of general solutions to this program, here explicitly observed in section \ref{sec:ads_mixed}, one can appreciate a related difficulty whose clarification also bears the essence of its solution: \emph{in (A)dS backgrounds gauge-for-gauge invariance is generically broken}; indeed, the transformations of the parameters
\be \label{gpg}
\d \, \L_{\, i} \, = \, \nabla^{\, j} \, \L_{\, [ij]}\, , 
\ee 
that in flat space would leave $\vf$ unaltered,  now produce a variation of the field itself according to
\be \label{gpgbreak}
\d \, \vf \, = \, \12 \, [\nabla^{\, i},\, \nabla^{\, j}] \, \L_{\, [ij]} \, .
\ee  
Given that in a quantitative analysis of the consequences of \eqref{gpgbreak} one should take into account all generations of broken gauge-for-gauge transformations, it is anyway clear that in (A)dS the gauge-for-gauge parameters $\L_{\, [ij]}$, instead of providing a convenient bookkeeping for those combinations of the parameters $\L_{\, i}$ that do not affect the gauge field, encode instead true additional gauge redundancies  whose presence would eventually affect the counting of degrees of freedom of the resulting theory\ft{These considerations apply whenever the parameters of the gauge-for-gauge transformations do not sit in the kernel of the commutator in \eqref{gpgbreak}, which is the case relevant to the construction of the BMV multiplet. A simple counterexample is given by $p-$forms, for which clearly the deformation to (A)dS does not break gauge-for-gauge invariance.}.

 However, it might still be possible to propagate the polarisations of the $O(D-2)$ irrep associated with $\vf$ provided one ``neutralises'' the effect of those broken gauge-for-gauge transformations encoded in \eqref{gpgbreak} by \emph{promoting them to play the role of standard gauge parameters for new fields (and corresponding new degrees of freedom) to be introduced in the theory}. This is the basic idea underlying the Stueckelberg construction and, in our opinion, it also provides an interesting alternative insight into the mechanism encoded in the BMV pattern \cite{bmv}, showing how the degrees of freedom carried by individual (A)dS massless particles distribute over \emph{multiplets} of flat-space particles of zero mass.
 
 Considering for instance the case of $\{s, 1\}$ tensors of $O(D)$, their standard description as gauge fields in flat space would entail two gauge parameters with tableaux $\{s\}$ and $\{s-1,1\}$ respectively, and one gauge-for-gauge parameter given by a symmetric tensor of rank $s-1$:
\be \label{s1}
\vf:\, \hskip .3cm
\overbrace{\young(\hfil \hfil \cdots \hfil,\hfil)}^{s} \hskip .5cm \stackrel{G}{\longrightarrow} 
\begin{cases} {\small \overbrace{\young(\hfil \hfil \cdots \hfil,\hfil)}^{s-1} } \\
\\
{\small \overbrace{\young(\hfil \hfil \cdots \hfil)}^{s}}\\ 
\end{cases} \hskip .5cm \stackrel{G^2}{\Longrightarrow} \hskip .5cm
{\small \overbrace{\young(\hfil \hfil \cdots \hfil)}^{s-1} }  \, \\
\ee
 According to our previous discussion, in (A)dS  the latter has to play the role of a standard parameter for an additional field, that might be either of the form $\{s-1, 1\}$ or $\{s\}$, where the second option is the only one eventually resulting in a unitary theory in AdS. The corresponding Stueckelberg Lagrangian smoothly deforms the sum of the two flat Lagrangians for the $\{s, 1\}$ and for the $\{s\}$ representations, thus providing a description of the same $O(D-2)$ degrees of freedom. However, from the AdS vantage point, those degrees of freedom are to be viewed as corresponding to a single massless particle with the symmetries of the $\{s,\, 1\}$-tableau. 
 
 As an additional example of the general construction let us also discuss the case of an $O(D)$--tensor  with three families of indices and tableau structure $\{3, 2, 1\}$, that we denote $\vf^{(0)}$, which is instructive in particular due to the presence of one more generation of broken gauge-for-gauge invariances. 
  
  The first consequence of covariantising derivatives in the flat gauge transformation of $\vf^{(0)}$ is the appearance of three broken gauge-for-gauge parameters, with diagram structure $\{2, 1, 1\}$, $\{2, 2\}$ and $\{3, 1\}$ respectively; in addition, there is a third-generation gauge parameter with tableau $\{2, 1\}$ to be discussed later. The full pattern of gauge generations, including the first one, associated to $\vf^{(0)}$ is summarised in the following scheme:
\be \label{321}
\vf^{(0)}:\, \hskip .3cm
\yng(3,2,1) \hskip .5cm \stackrel{G}{\longrightarrow} 
\begin{cases} {\small \yng(2,2,1) } \\
\\
{\small \yng(3,1,1)}\\ 
\\
{\small \yng(3,2)}\,
\end{cases} \hskip .5cm \stackrel{G^2}{\Longrightarrow} \hskip .5cm
\begin{cases}
{\small \yng(2,1,1) } \\
\\
{\small \yng(2,2)}\\ 
\\
{\small \yng(3,1)}\,
\end{cases}
\hskip .5cm \stackrel{G^3}{\equiv} > \hskip .5cm
\yng(2,1)
\ee
As a first step, in order to deal with the additional gauge freedom emerging from the breaking of gauge-for-gauge invariance, we would include in our description two Stueckelberg fields, $\vf^{(1,\, 1)}$ and $\vf^{(1, \, 2)}$, with tableau structure $\{3, 1, 1\}$ and $\{3, 2\}$, while following \cite{metsaev1, metsaev2} we discard at this level the other possible choice of an additional Stueckelberg field $\vf^{(1, \, 3)}$ with structure $\{2, 2, 1\}$, assuming that it would lead to a non-unitary theory. Looking at \eqref{321} this choice is tantamount to promoting the two lost first-generation  parameters to play the role of Stueckelberg fields. A pictorial synopsis of the pattern of gauge generations for each of these two fields is provided in the following schemes:
\be \label{311}
\vf^{(1, 1)}:\, \hskip .3cm
\yng(3,1,1) \hskip .5cm \stackrel{G}{\longrightarrow} 
\begin{cases} {\small \yng(2,1,1) } \\
\\
{\small \yng(3,1)}\\ 
\end{cases} \hskip .5cm \stackrel{G^2}{\Longrightarrow} \hskip .5cm
\begin{cases}
{\small \yng(2,1) } \\
\\
{\small \yng(3)}\\ 
\end{cases}
\hskip .5cm \stackrel{G^3}{\equiv} > \hskip .5cm \yng(2)
\ee
\be \label{32}
\vf^{(1, 2)}:\, \hskip .3cm
\yng(3,2) \hskip .5cm \stackrel{G}{\longrightarrow} 
\begin{cases} {\small \yng(2,2) } \\
\\
{\small \yng(3,1)}\\ 
\end{cases} \hskip .5cm \stackrel{G^2}{\Longrightarrow} \hskip .5cm
{\small \yng(2,1) } \hskip 4.1 cm \, \\
\ee
Let us notice that the gauge variations of $\vf^{(1,\, 1)}$ and $\vf^{(1, \, 2)}$ comprise the three gauge-for-gauge parameters associated with $\vf^{(0)}$, together with an additional parameter with the tableau structure $\{3, 1\}$, needed to ensure that eventually both $\vf^{(1,\, 1)}$ and $\vf^{(1, \, 2)}$ propagate only their flat-space physical degrees of freedom, and not more. These Stueckelberg fields in their turn generate two broken gauge-for-gauge symmetries, both of the form $\{2, 1\}$, while in relation with the $\{3, 1, 1\}$-tensor one should also take into account the existence of a third-order transformation with a rank-two symmetric tensor parameter. 
  
  At this point an important novelty with respect to the two-family case manifests itself: as already observed, in the complete pattern of flat gauge transformations associated with  $\vf^{(0)}$  there is also a third-generation gauge parameter whose tableau structure is $\{2, 1\}$. In flat-space it would indicate the existence of combinations of the first-generation parameters apparently hampered by the existence of gauge-for-gauge transformations, but at a closer look effective on $\vf^{(0)}$. Thus, in order for the matching between field-components and gauge parameters to be exact, in our Stueckelberg construction we have to accommodate an additional gauge freedom with tableau structure $\{2, 1\}$. This means in practice that, of the two broken gauge-for-gauge parameters with structure $\{2, 1\}$ associated to  $\vf^{(1,\, 1)}$ and $\vf^{(1, \, 2)}$ one combination has to be left free, exactly to account for the part of the gauge symmetry of the initial field that has been removed after introducing the Stueckelberg fields $\vf^{(1,\, 1)}$ and $\vf^{(1, \, 2)}$. Thus, the presence of the pair of $\{2, 1\}$ parameters in \eqref{311} and \eqref{32} calls for the introduction of only one second-generation Stueckelberg  field $\vf^{(2, \, 1)}$, whose structure is again fixed by unitarity to be that of a $\{3,1\}$ tensor (while possible alternative options like tableaux $\{2,2\}$ or $\{2,1,1\}$ are discarded) and whose gauge pattern is given as follows:
\be \label{31}
\vf^{(2, 1)}:\, \hskip .3cm
\yng(3,1) \hskip .5cm \stackrel{G}{\longrightarrow} 
\begin{cases} {\small \yng(2,1) } \\
\\
{\small \yng(3)}\\ 
\end{cases} \hskip .5cm \stackrel{G^2}{\Longrightarrow} \hskip .5cm
{\small \yng(2) }  \, \\
\ee
Finally, let us notice that the additional gauge symmetry provided by the second-generation parameter in \eqref{31} is just what is needed to account for the third generation of broken gauge-for-gauge symmetry of $\vf^{(1,\, 1)}$, and thus the pattern of Stueckelberg fields that one needs to introduce does not include an additional rank-three symmetric tensor. The resulting system of $O(D-2)$ tableaux of the form $\{3, 2, 1\}$, $\{3, 1, 1\}$, $\{3, 2\}$ and $\{3,1\}$ matches the degrees of freedom of the massless AdS particle with the symmetries of the diagram $\{3, 2, 1\}$ as resulting from the BMV conjecture \cite{bmv, NCP, AGads}.

 These two examples should convey the general idea behind our interpretation of the BMV phenomenon while also suggesting the concrete procedure to build the Stueckelberg Lagrangian for  the degrees of freedom of a given (A)dS massless particle with mixed symmetry.
 
 In the general case $\vf^{(0)}$ can be an $O(D)$-tableau with $N$ rows (that for simplicity one might assume as being of different lengths) whose hierarchy  of flat gauge-transformations comprises $N$ gauge parameters, $({N \atop 2})$ gauge-for-gauge parameters, $({N \atop 3})$ third generation parameters and so on. To deal with the first instance of gauge-for-gauge breaking we would introduce $N-1$ first-generation Stueckelberg fields  $\vf^{(1, k)}, \, k = 1, \ldots, N-1$, (i.e. all possible Stueckelberg fields whose first row has the same length as that of $\vf^{(0)}$, effectively corresponding to all first-generation gauge parameters with first row of maximal length) to which one can associate an equivalent pattern of broken reducible gauge transformations. The generation of new fields will stop as soon as the overall gauge symmetry of the system will match that of its flat-space counterpart, accounting in particular for the full pattern of reducible gauge transformations for each mixed-symmetry field introduced in the spectrum. 
 
 Let us stress once more that insofar as gauge symmetry alone is concerned the pattern would not be uniquely determined: at each step different choices of Stueckelberg gauge fields would be indeed consistent with the additional gauge parameters emerging at the previous level.  In AdS all ambiguities are fixed performing at each step the unitary choice dictated by the analysis of \cite{metsaev1, metsaev2}, which amounts to choosing as allowed Stueckelberg fields only tableaux whose first row has the same length as that of $\vf^{(0)}$. We expect that, pursuing the construction of the corresponding Lagrangian for different choices, at least some of the Stueckelberg fields would eventually appear with kinetic terms of wrong signs.
 
 Notice that the previous discussion only rests upon properties of the gauge transformations \eqref{gauge_new}, and does not depend on the presence of possible constraints on the parameters. As a result, it applies both to our reducible and irreducible Lagrangians in flat space, thus suggesting a possible generalisation of the BMV mechanism to reducible multiplets of fields. We postpone a detailed investigation of this interesting possibility to future work, while here we would like to focus on the case of irreducible $O(D)$ fields. As we discussed in detail in section \ref{sec:flat_mixed}, in this case a safe way of defining the theory in flat space is to add some double-divergence constraints on the field. Obtaining a smooth deformation of the flat-space theory would entail also a deformation of these constraints. However, we were not able to build a completion of the constraints \eqref{mixtdiv} that is gauge invariant under the whole set of Stueckelberg transformations. 
 
 This does not imply that a Stueckelberg formulation of the theory is not possible. In the first place, although at the moment we are not able to produce a better solution for the general case, it is not at all obvious that the $O(D)$ theory really requires constraints of the form \eqref{mixtdiv}, that might very well be discarded if one were able to find the proper way to characterise the full amount of symmetry of the corresponding  kinetic operator. Actually, as we already noticed in section \ref{sec:flat_mixed} the constraints \eqref{mixtdiv} are manifestly too strong, since they set to zero also some components of the field that would be forced to vanish anyway by the equations of motion. Thus, even if some form of constraints might still be needed even in the pessimistic scenario, it is still possible that a covariant form of the ``minimal'' constraints would at least simplify the issue of the deformation. 
 
 While we shall defer the general analysis of these open issues to future work, in order to make the discussion more concrete in the remainder of this section we shall exhibit an explicit realisation of the mechanism previously discussed showing that, at least for two-family fields, it is possible to preserve the gauge-for-gauge symmetry by dealing with a suitable multiplet of fields. We shall conclude this section by displaying the full Stueckelberg Lagrangian for irreducible $\{s,1\}$ fields, both for our $O(D)$ theories and as deformations of the Labastida Lagrangian, taking advantage of the absence for these classes of fields of double-divergence constraints.

 According to the general discussion of the previous paragraphs, starting with an $O(D)$ tensor in the $\{s, k\}$ representation we expect our Lagrangian to involve a total of $k$ additional fields  with tableau structure $\{s, \, k-i\}$:
\be
\vf^{(i)} \, \, \sim \, \,  \{s, \, k-i\}; \hskip 1cm  i = 1, \ldots, k \, .
\ee
Each of the fields $\vf^{(i)} $ included in the system experiences gauge-for-gauge breaking phenomenon involving parameters having structure $\{s-1, \, k-i-1\}$, which is taken care of by the gauge transformation of the next field in the resulting hierarchy, $\vf^{(i-1)}$.

Working with two-family tensors allows for a more explicit notation already used in section \ref{sec:flat_symm}, according to which tensors of the form $\{s, k\}$ will be denoted by
\be
\vf_{\, \m_1 \, \cdots \, \m_s, \, \n_1 \, \cdots \, \n_k} \, \equiv \, \vf_{\, \m_s, \, \n_k} \, ,
\ee
while when computing products of tensors we will make use of the same symbols for indices that are meant to be totally symmetrised; e.g.
\be
\pr_{\, (\m_1} \, \vf_{\, \m_2 \, \cdots \, \m_{s+1}), \, \n_1 \, \cdots \, \n_k} \, \equiv \, \pr_{\, \m} \, \vf_{\, \m_s, \, \n_k}\, .
\ee
In this notation all rules for symmetric calculus collected in \eqref{A1} apply independently for the two sets of indices, while additional prescriptions for contracting indices belonging to different families are not difficult to derive case by case.

Taking into account the whole set of gauge parameters available within the system we can write the general form of the gauge transformation of each field as
\begin{align} \label{gauge-stueck-gen}
\d \vf^{(i)}{}_{\!\m_s,\,\n_{k-i}} & \,=\, \nabla_{\!\m}\, \L^{(i)}{}_{\!\m_{s-1},\,\n_{k-i}} \,+\, \nabla_{\!\n}\, \l^{(i)}{}_{\!\m_s,\,\n_{k-i-1}} - \frac{1}{s-k+i+1}\, \nabla_{\!\m}\, \l^{(i)}{}_{\!\m_{s-1}\,\n,\,\n_{k-i-1}} \nn \\
& + \frac{\a_i}{L} \l^{(i-1)}{}_{\!\m_s, \n_{k-i}} + \frac{\b_i}{L} \left\{2 g_{\m\m} \L^{(i+1)}{}_{\!\m_{s-2} \n, \n_{k-i-1}} - (s-k+i) g_{\m\n} \L^{(i+1)}{}_{\!\m_{s-1}, \n_{k-i-1}}\right\} \nn \\
& + \frac{\g_i}{L} \left\{\, 2\, g_{\m\m}\, \l^{(i+1)}{}_{\!\m_{s-2}\,\n\n ,\, \n_{k-i-2}} - (s-k+i+1)\, g_{\m\n}\, \l^{(i+1)}{}_{\!\m_{s-1}\,\n ,\, \n_{k-i-2}} \right. \nn\\
& \phantom{+ \frac{\g_i}{L}}\ \left. +\, (s-k+i+1)(s-k+i+2)\, g_{\n\n}\, \l^{(i+1)}{}_{\!\m_{s} ,\, \n_{k-i-2}} \,\right\} ,
\end{align}
where it is possible to appreciate that, besides  the parameters present in the flat gauge transformation, $\L^{(i)}{}_{\!\m_{s-1},\,\n_{k-i}}$ and $\l^{(i)}{}_{\!\m_s,\,\n_{k-i-1}}$, a number of additional contributions can also enter the  variation of $\vf^{(i)}{}_{\!\m_s,\,\n_{k-i}} $ (in the combinations needed to recover the corresponding $\{s,\, k-i\}$-projection, here collected in braces), exploiting parameters entering the system from the flat variation of other fields in the multiplet. It is due to this mixing of gauge transformations that gauge-for-gauge breaking at the level of a single field can be in principle reabsorbed in the whole system. Indeed, along with \eqref{gauge-stueck-gen} one can define the following transformations of the parameters:
\begin{subequations} \label{gfg}
\begin{align}
\d \L^{(i)}{}_{\! \m_{s-1},\, \n_{k-i}} \,=\, & \nabla_{\!\n}\, \Theta^{(i)}{}_{\!\m_{s-1},\,\n_{k-i-1}} - \frac{1}{s-k+i}\, \nabla_{\!\m}\, \Theta^{(i)}{}_{\!\m_{s-2}\,\n,\,\n_{k-i-1}} + \frac{b_i}{L}\, \Theta^{(i-1)}{}_{\! \m_{s-1},\, \n_{k-i}} \nn \\
& + \frac{d_i}{L} \left\{\, 2\, g_{\m\m}\, \Theta^{(i+1)}{}_{\!\m_{s-3}\,\n\n ,\, \n_{k-i-2}} - (s-k+i)\, g_{\m\n}\, \Theta^{(i+1)}{}_{\!\m_{s-2}\,\n ,\, \n_{k-i-2}} \right. \nn \\
& \phantom{+ \frac{\g_i}{L}}\ \left. +\, (s-k+i)(s-k+i+1)\, g_{\n\n}\, \Theta^{(i+1)}{}_{\!\m_{s-1} ,\, \n_{k-i-2}} \,\right\} \, ,
 \\[10pt]
\d \l^{(i)}{}_{\! \m_s,\, \n_{k-i-1}} \,=\, & -\, \frac{s-k+i+1}{s-k+i}\, \nabla_{\!\m}\, \Theta^{(i)}{}_{\!\m_{s-1},\,\n_{k-i-1}} \nn \\
& + \frac{e_i}{L} \left\{\, 2\, g_{\m\m}\, \Theta^{(i+1)}{}_{\!\m_{s-2}\,\n ,\, \n_{k-i-2}} - (s-k+i+1)\, g_{\m\n}\, \Theta^{(i+1)}{}_{\!\m_{s-1} ,\, \n_{k-i-2}} \,\right\} \, ,
\end{align}
\end{subequations}
smoothly deforming the gauge-for-gauge transformations of the flat theory. Imposing the invariance of \eqref{gauge-stueck-gen} under gauge-for-gauge transformations determines uniquely the coefficients appearing in the latter,
\begin{subequations}
\begin{eqnarray}
b_i &=& \frac{s - k + i}{s - k + i - 1} \, \a_i \, , \\[5pt]
d_i &=& \frac{2\, \b_i \,+\, (s - k + i + 2) (s - k + i + 3)\, \g_i}{(s - k + i + 1)^2} \, , \\[5pt]
e_i &=& \frac{(s - k + i + 2)\, \g_i \,-\, (s - k + i)\, \b_i}{s - k + i + 1} \, ,
\end{eqnarray}
\end{subequations}
while also imposing the following consistency conditions
\begin{subequations} \label{eq-gfg}
\begin{eqnarray}
\a_i \, e_{i - 1} \,+\, \b_i \, b_{i + 1} &=& \frac{s - k + i + 1}{s - k + i} \, , \\[5pt]
d_{i + 1} \, \b_i \,+\, e_{i + 1} \, \g_i &=& 0 \, .
\end{eqnarray}
\end{subequations}
One can thus conclude that, at least for two-family fields, it is possible to preserve all gauge-for-gauge transformations, but this requirement already fixes part of the structure of the gauge transformations. 

One should then check that a Stueckelberg Lagrangian invariant under these transformations exists, and we are now going to show this explicitly for 
$\{s,1\}$-projected $O(D)$ tensors. The frame-like Stueckelberg formulation of the $\{s,1\}$ case can be found in \cite{zin-frame2}. In this case the gauge transformations take the form
\begin{subequations} \label{gauge-stueck}
\begin{align}
\d \vf_{\m_s,\,\n} & =\, \nabla_{\!\m}\, \L_{\m_{s-1},\, \n} \,+\, \nabla_{\!\n}\, \l_{\m_s} -\, \frac{1}{s}\, \nabla_{\!\m}\, \l_{\m_{s-1}\n} \nn \\
& +\, \frac{\b}{L} \left(\, 2\, g_{\m\m}\, \x_{\m_{s-2}\n} \,-\, (s-1)\, g_{\m\n}\, \x_{\m_{s-1}} \,\right) \, , \label{gauge1} \\[10pt]
\d \c_{\m_s} & =\, \nabla_{\!\m}\, \x_{\m_{s-1}} \,+\, \frac{\a}{L} \, \l_{\m_s} \, , \label{gauge2}
\end{align}
\end{subequations}
where $\L_{\m_{s-1},\, \n}$ is a $\{s-1,1\}$-projected tensor. Moreover, we work with traceless gauge parameters as in the flat-space case. The gauge for gauge transformations are
\be
\begin{array}{lcl}
\d\, \L_{\m_{s-1},\, \n} \,&= &\, \nabla_{\!\n}\, \Theta_{\m_{s-1}}-\, \frac{1}{s-1}\, \nabla_{\!\m}\, \Theta_{\m_{s-2}\n} \, , \\
\d\, \l_{\m_s} \,&=&\, -\, \frac{s}{s-1}\, \nabla_{\!\m}\, \Theta_{\m_{s-1}} \, , \\
\d\, \x_{\m_s-1} \,&=&\, \frac{s\,\a}{(s-1)L}\, \Theta_{\m_{s-1}} \, ,
\end{array}
\ee
where we already fixed in $\d\, \x_{\m_s-1}$ the coefficient that guarantees the invariance of $\d\, \c_{\m_s}$ under gauge-for-gauge transformations. The consistency conditions \eqref{eq-gfg} then reduce in this case to $\b = \a^{-1}$. Moreover, working with traceless fields imposes the conditions
\begin{subequations} \label{constr-stueck}
\begin{align}
& \nabla\cdot\L_{\m_{s-2},\,\n} \,-\, \frac{1}{s}\, \nabla\cdot\l_{\m_{s-2}\n} \,+\, \frac{D+s-3}{\a\,L}\, \x_{\m_{s-2}\n} \,=\, 0\, , \\
& \nabla\cdot\x_{\m_{s-2}} \,=\, 0 \, ,
\end{align}
\end{subequations}
that deform the flat-space differential constraints. One can then check that the Stueckelberg Lagrangian
\be
\cL \,=\, \frac{1}{2}\,\vf^{\m_s,\,\n} \left\{ (M\vf)_{\m_s,\,\n} + m_0\, \vf_{\m_s,\,\n} \right\} + \frac{1}{2}\, \x^{\m_s} \left\{ (M\c)_{\m_s} +m_1\, \c_{\m_s} \right\} + \frac{c}{L}\, \c^{\m_s} \nabla^\l \,\vf_{\m_s,\,\l}
\ee
is left invariant by gauge transformations \eqref{gauge-stueck} subject to the constraints \eqref{constr-stueck} provided that
\begin{subequations}
\begin{align}
\a \, & =\, -\, c \,=\, \sqrt{(s+1)(d+s-5)} \, , \label{sol1-ex}\\
m_0 \, & =\, -\, \frac{1}{L^2} \left[ (s-2)(D+s-3)-(s+1) \right] \, , \\[2pt]
m_1 \, & =\,\frac{D+s-2}{L^2} \, .
\end{align}
\end{subequations}

We can also compare our result with the Stueckelberg deformation of the corresponding Labastida Lagrangians. The schematic form is again
\be \label{lag-stu-laba}
\cL \,=\, \cL_{\vf\vf} \,+\, \cL_{\vf\c} \,+\, \cL_{\c\c} \, .
\ee
The portion of $\cL$ that is quadratic in $\vf$ is the Labastida Lagrangian 
\be \label{lag_lab}
\cL_{\vf\vf} =\, \12\, \vf^{\,\m_s,\,\n} \left(\, \cF_{\m_s,\,\n} \,-\,\12\, g_{\m\m}\, \cF^{\,\pe}{}_{\!\!\!\m_{s-2},\,\n} \,+\,\frac{1}{4}\, g_{\m\n}\, \cF^{\,\pe}{}_{\!\!\!\m_{s-2},\,\m} \,\right) \, ,
\ee
where we used the irreducibility of $\vf$ to display only one trace of $\cF$. The Labastida tensor reads 
\be \label{labastida}
\begin{split}
\cF_{\m_s,\,\n} & =\, \Box\, \vf_{\m_s,\,\n} \,-\, \nabla_{\!\m} \nabla\cdot \vf_{\m_{s-1},\,\n} \,+\, \nabla_{\!\n} \nabla\cdot \vf_{\m_{s-1},\,\m} \,+\, \frac{1}{2}\, \nabla_{\!\m} \nabla_{\!\m}\, \vf^{\,\pe}{}_{\!\!\m_{s-2},\,\n} \\
& -\, \frac{1}{4}\, \{ \nabla_{\!\m}\, , \nabla_{\!\n} \} \, \vf^{\,\pe}{}_{\!\!\m_{s-2},\,\m} \,-\, \frac{1}{L^2} \left[ (s-2)(D+s-3) - (s+1) \right] \vf_{\m_s,\,\n} \\
& -\, \frac{1}{4L^2} \left(\, 10\, g_{\m\m}\, \vf^{\,\pe}{}_{\!\!\m_{s-2},\,\n} \,-\, (s+3)\, g_{\m\n}\, \vf^{\,\pe}{}_{\!\!\m_{s-2},\,\m} \,+\, 2\, g_{\m\m}\, \vf^{\,\pe}{}_{\!\!\m_{s-3}\,\n,\,\m} \,\right) ,
\end{split}
\ee
and the mass terms ensure its invariance under transformations generated by a traceless $\L_{\m_{s-1},\,\n}$. In a similar fashion, the portion of $\cL$ that is quadratic in $\c$ is the Fronsdal Lagrangian
\be \label{lag_fronsd}
\cL_{\c\c} =\, \12 \, \c^{\,\m_s} \left(\, F_{\m_s} \,-\,\12\, g_{\m\m}\, F^{\,\pe}{}_{\!\!\!\m_{s-2}} \,\right) ,
\ee
but in this case the mass terms that enter the Fronsdal tensor are unconventional since their variation has to compensate part of $\d \cL_{\vf\c}$:
\be
F_{\m_s} =\, \Box\, \c_{\m_s} -\, \nabla_{\!\m} \nabla\cdot \c_{\m_{s-1}} +\, \frac{1}{2}\, \nabla_{\!\m} \nabla_{\!\m}\, \c^{\,\pe}{}_{\!\!\m_{s-2}} +\, \frac{D+s-2}{L^2}\, \c_{\m_s} -\, \frac{1}{L^2}\, g_{\m\m}\, \c^{\,\pe}{}_{\!\!\m_{s-2}} \, .
\ee
The Lagrangian also contains a cross-coupling term that vanishes in the flat space limit and reads
\be
\cL_{\vf\c} =\, \frac{1}{\a\, L}\ \vf^{\,\m_s,\,\n} \left(\, \nabla_{\!\n}\, \c_{\m_s} \,-\, g_{\m\n} \nabla\cdot \c_{\m_{s-1}} \,-\, \12\, g_{\m\m} \nabla_{\!\n}\, \c^{\,\pe}{}_{\!\!\m_{s-2}} \,+\, \12\, g_{\m\n} \nabla_{\!\m}\, \c^{\,\pe}{}_{\!\!\m_{s-2}} \,\right) ,
\ee
where $\a$ is the constant \eqref{sol1-ex} that appears in the gauge transformations \eqref{gauge-stueck}, under which the Lagrangian is gauge invariant provided that the gauge parameters satisfy the Labastida constraint
\be \label{constr} 
\L_\r{}^\r{}_{\m_{s-3},\,\n} \,=\, \frac{1}{s}\, \l_\r{}^\r{}_{\m_{s-3}\n} \, .
\ee
Therefore, as expected, the invariance under gauge-for-gauge transformations is preserved also in the Labastida case, and our Lagrangians actually identify the couplings of \eqref{lag-stu-laba} that do not depend on the traces of the fields.

%%%%%%%%%%%%%%%%%%%%%%%%%%%%%%%%%%%%%%%%%%%%%%%%%%%%%%%%%%%%%%%%%%%%%

\section{Diagonalisation of reducible theories}\label{sec:diag}

%%%%%%%%%%%%%%%%%%%%%%%%%%%%%%%%%%%%%%%%%%%%%%%%%%%%%%%%%%%%%%%%%%%%%

Besides the analysis of the spectrum, we would also like to discern the proper combinations of the components of $\vf$ associated to each of the irreducible representations identified by our preceding analysis. Focussing on the case of symmetric tensors, in this section we present a systematic way to construct the field redefinitions needed to decompose the Lagrangians \eqref{tlagr} and \eqref{lag_ads_symm} in their block-diagonal form, where each block provides an action suitable for the description of the  irreducible polarisations of a given spin. In our opinion this latter approach retains some specific advantages: first, the resulting Lagrangians display at a glance both number and nature of the irreducible propagating degrees of freedom, including the relative signs among the various  kinetic terms making manifest the absence of ghosts; moreover, it allows in principle to interpret possible non-linear deformations of \eqref{tlagr} in terms  of couplings among single-particle fields. For the unconstrained versions of our Lagrangians provided by triplet systems the corresponding diagonalisation was discussed in \cite{ftcurrents,FTads}.

%%%%%%
\subsection{Symmetric tensors in flat backgrounds}\label{sec:flat_diag}
%%%%%%

Our starting point is a formal decomposition of  $\vf$ involving fields of decreasing spins: 
\be \label{phi}
\vf \, = \, \phi_s \, + \, O_{s-2} \, \phi_{s-2} \,  + \, O_{s-4} \, \phi_{s-4} \,+\, \cdots \, + \, O_{s-2k} \, \phi_{s-2k} \, + \cdots \, ,
\ee
where $\phi_{s-2k}$ is a  symmetric tensors of rank $(s - 2k)$, while the associated operators $O_{s-2k}$ are to be chosen so that when \eqref{phi} is inserted in \eqref{tlagr} the latter decomposes into a sum of decoupled Lagrangians.  Each of these Lagrangians will enjoy transverse gauge invariance and must ultimately describe irreducible, massless  spin-($s-2k$)  degrees of freedom, with $k = 0, 1,  \ldots, [\fr{s}{2}]$; as discussed in section \ref{sec:flat_symm}, this requires that the fields $\phi_{s-2k}$, and the corresponding gauge parameters, be \emph{traceless}. 

More explicitly, inserting \eqref{phi} into \eqref{tlagr} one gets
\be \label{tlagrO}
\cL \, = \,  \12 \, \sum_{k, l \,=\, 0}^{[\fr{s}{2}]} \, O_{s-2k} \, \phi_{s-2k} \, M \, O_{s-2l} \, \phi_{s-2l} \, ,
\ee
where $O_{\, s} \, = \, \mathbb{I}$ and where  contraction of indices is understood between $O_{s-2k} \, \phi_{s-2k}$ and $M \, O_{s-2l} \, \phi_{s-2l}$. From the previous expression one can see that the diagonalisation obtains if
\be  \label{diag}
O_{s-2k} \, \phi_{s-2k} \, M \, O_{s-2l} \, \phi_{s-2l} \, \sim  \, \d_{\, k, l} \, \phi_{s-2k} \, M \, \phi_{s-2l}\, ,
\ee
and we will show that the latter condition holds indeed if the operators $O_{s-2k}$ satisfy the equation
\be \label{MO}
M \, O_{s-2k} \, = \, \h^{\, k} \, M\, .
\ee
In general eq.~\eqref{MO} possesses several solutions, due to the invariance of the Maxwell-like operator
$M$ under the gauge transformation
\be \label{gauge}
\d \, O_{s-2k} \, = \, \pr \, \L_k \, ,
\ee
where $\L_k $ is itself an operator satisfying the transversality condition $\prd \L_k = 0$\ft{The solution 
would be unique if for some reasons there were no candidates for a divergenceless $\L_k$; while this is 
not the case in general, it happens indeed for a special subset of the operators $O_{s-2k}$, as we shall see in appendix \ref{app:diag}.}. Nonetheless, we shall see that whenever \eqref{MO} is satisfied the diagonalisation conditions \eqref{diag} holds as well, so that the explicit form of the operators $O_{s-2k}$ is not really needed for our present purposes. At any rate, it is possible to conclude on general grounds that all solutions to \eqref{MO} are to involve non-local operators, as we discuss in appendix \ref{app:diag} where we also exhibit an explicit solution.

Let us make use of  \eqref{MO} in \eqref{tlagrO}  assuming in addition, without loss of generality, $k \geq l$:
\be \label{diag_1}
\begin{split}
O_{s-2k} \, \phi_{s-2k} \, M \, O_{s-2l} \, \phi_{s-2l} \, & = \, O_{s-2k} \, \phi_{s-2k} \, \h^{\, l} M  \, \phi_{s-2l} \, \\
& = c_{\,l} \, \left\{M \, T^{\, l} \,  O_{s-2k} \, \phi_{s-2k}\right\} \, \phi_{s-2l} \, \\
& = c_{\, l} \,  \left\{[M, T^{\, l}] \,  O_{s-2k} \, \phi_{s-2k} \, + \,T^{\, l} \, \h^{\, k}\, M \, \phi_{s-2k} \right\} \phi_{s-2l} \, ,
\end{split}
\ee
where we exploited both the self-adjointness of $M$ (up to total derivatives) and \eqref{MO}, and where
\be \label{cl}
c_{\, l} \, = \, (2\, l \, - \, 1)! ! \, \binom{s}{2l} \, 
\ee
is a combinatorial factor coming from the contraction of the $l$ powers of $\h$, leading to the 
$l$ traces in the second line of \eqref{diag_1}, here denoted in operatorial notation as $T^{\, l}$.
Let us evaluate separately the two terms in the third line of \eqref{diag_1}.

In the first term, the commutator of $M$ and $T^{\, l}$ is proportional to a double divergence; 
more precisely:
\be \label{MT}
[M, T^{\, l}] \, = \, 2\,  l \, T^{\, l -1} \, \prd \pr\,\cdot \, ,
\ee
as can be verified recursively starting from $[M, T] \, = \, 2\, \prd \prd $\, and taking into account that
traces and divergences commute. In addition, the divergence of  \eqref{MO} gives
\be \label{ddO}
\prd \prd  O_{s-2k} \, = \, \h^{\, k} \, \prd \prd \, -\ \h^{\, k -1} \, M \, ,
\ee
where we factored out an overall gradient. Let us stress that \eqref{ddO} allows us to dispense with 
the detailed structure of the operators $O_{s-2k}$, which otherwise would make the general proof significantly more involved. All in all, we have to evaluate
\be
2\,  l \, T^{\, l -1}  \left\{\, \h^{\, k} \, \prd \prd \, - \ \h^{\, k -1} \, M \,\right\} \phi_{s-2k} \, \phi_{s-2l} \, ,
\ee
where, due to the tracelessness of  $\phi_{s-2l}$, for $k \geq l$ the first term never contributes\ft{More 
explicitly: $(T^{\, l -1} \, \h^{\, k} \, \prd \prd \phi_{s-2k} ) \phi_{s-2l} \, \sim \, 
\prd \prd \phi_{s-2k} (T^{\, k} \, \h^{\, l - 1} \, \phi_{s-2l}) = 0 $.} while the second term can be conveniently rewritten as
\be \label{1term}
- \, 2\,  l \, T^{\, l -1} \,\h^{\, k -1} \, M\, \phi_{s-2k} \, \phi_{s-2l}\, = 
- \, 2\,  l \, \tilde{c}_{\, l, k}\, M\, \phi_{s-2k} \, T^{\, k -1} \,\h^{\, l -1} \,\phi_{s-2l}\,  ,
\ee
up to an overall combinatorial coefficient $ \tilde{c}_{\, l, k}$, that we do not need to evaluate in general since
\eqref{1term} contributes only for $k = l$ when the coefficient itself  is trivial ($ \tilde{c}_{\, k, k} = 1$). 
For the same reason in the second term to be evaluated, 
\be \label{2term}
T^{\, l} \, \h^{\, k}\, M \, \phi_{s-2k} \, \phi_{s-2l} \, = \, \hat{c}_{\, l, k} \, M \, \phi_{s-2k} \,T^{\, k}  \, \h^{\, l} \phi_{s-2l} \, ,
\ee
the only contribution obtains for $k = l$; in both cases the relevant quantity to compute is
\be \label{TE}
T^{\, k}  \, \h^{\, l} \phi_{s-2l} \, = \, \d_{\, k, l} \, \prod_{i \,=\, 0}^{k - 1} \, [D + 2\,(s - 2k + i)]   \,  \phi_{s-2k} \, .
\ee
Substituting  \eqref{TE} in \eqref{1term} and \eqref {2term}, and then inserting the corresponding expressions 
in \eqref{diag_1}, we finally  obtain
\be \label{diag_2}
\cL \, = \, \12 \, \sum_{k, l \,=\, 0}^{[\fr{s}{2}]} \, 
O_{s-2k} \, \phi_{s-2k} \, M \, O_{s-2l} \, \phi_{s-2l} \, = \, \12 \, \sum_{k \,=\, 0}^{[\fr{s}{2}]}  \, 
c_{\, k} \, b_{\, k, s, D}\,  \phi_{s-2k} \, M \,  \phi_{s-2k}\, , 
\ee
where $c_{\, k}$ was given in \eqref{cl} and where we defined 
\be \label{b}
 b_{\, k, s, D}\,  = \, \prod_{i \,=\, 0}^{k - 1} \, [D + 2\,(s - 2k + i - 1)]  \, .
\ee
This proves that the redefinition \eqref{phi} in conjunction with the defining property \eqref{MO} of the operators
$O_{s-2k}$  actually diagonalise \eqref{tlagrO}. Each of the decoupled Lagrangians involves
traceless fields and displays transverse gauge invariance with traceless parameters, as required
for them to propagate each a single particle of a given spin.  The fact that all relative signs are equal confirms the absence of ghosts, 
while an additional rescaling would be needed in order to assign to the various fields their canonical normalisation.

%%%%%%
\subsection{Symmetric tensors in (A)dS backgrounds}\label{sec:ads_diag}
%%%%%%

The diagonalisation of the Lagrangian \eqref{lag_ads_symm} follows closely the corresponding procedure just presented for the flat case, and for this reason we shall limit ourselves to recalling its  main steps while stressing a few additional peculiar features of the (A)dS case.  We first introduce a set of traceless tensors of decreasing spins $ \phi_{s-2k}$  via
\be \label{phiL}
\vf \, = \, \phi_s \, + \, O^L_{s-2} \, \phi_{s-2} \,  + \, O^L_{s-4} \, \phi_{s-4} \,+\, \cdots \, + \, O^L_{s-2k} \, \phi_{s-2k} \, + \cdots \, ,
\ee
and then look for operators $O^L_{s-2k}$ implementing the diagonalisation condition for \eqref{lag_ads_symm}\ft{At the risk of being pedantic, here we add a label to specify the value of $s$ in the spin-dependent part 
of the kinetic operators; thus  $M_L^{(s)}$ corresponds to $M_L$ as defined in \eqref{M_L_symm}, 
while $M_L^{(s-2k)}$ can be obtained from \eqref{M_L_symm} by the substitution $s \rightarrow s - 2k$.
It might be also useful to stress that the operators $O^L_{s-2k}$ in \eqref{phiL} depend on the rank of $\vf$, 
so that if rank$(\vf) = s$ they are assumed to satisfy \eqref{MOs}  only under the action of $M_L^{(s)}$. }
\be  \label{diagL}
O^L_{s-2k} \, \phi_{s-2k} \, M_L^{(s)} \, O^L_{s-2l} \, \phi_{s-2l} \, \sim  \, \d_{\, k, l} \, \phi_{s-2k} \, M_L^{(s-2l)} \, \phi_{s-2l}\, .
\ee
The key to the whole procedure is to assume that the operators $O^L_{s-2k}$ satisfy the basic identity
\be \label{MOs}
M_L^{(s)}\, O^L_{s-2k} \,=\, g^k\, M_L^{(s-2k)} \, ,
\ee
which allows to write
\be \label{diag_L}
\begin{split}
O^L_{s-2k} \, \phi_{s-2k} \, M_L \, O^L_{s-2l} \, \phi_{s-2l} \, & = \, O^L_{s-2k} \, \phi_{s-2k} \,  g^{\, l}\, M_L^{(s-2l)} \, \phi_{s-2l}  \\
& = \, c_{\,l} \, M_L^{(s-2l)} \, T^{\, l} \,  O^L_{s-2k}  \, \phi_{s-2k} \, \phi_{s-2l} \, \\
& = c_{\, l}  \left\{\,[M_L^{(s-2l)}, T^{\, l}] \,  O^L_{s-2k} \, \phi_{s-2k} \right. \\
& + \,T^{\, l} \, (M_L^{(s-2l)}\, - \, M_L^{(s)}) \,O^L_{s-2k}\,  \phi_{s-2k}  \\
& \left. + \,T^{\, l} g^{\, k} \, M_L^{(s-2k)} \, \phi_{s-2k} \,\right\} \phi_{s-2l} \, ,
\end{split}
\ee
where the combinatorial coefficient $c_l$ is given in \eqref{cl}. 
Computing the commutator in \eqref{diag_L} gives
\be
[\, M_L^{(s-2k)}, T^{\, l} \,] \ = \, 2\, l \, T^{\, l - 1} \, \nabla \cdot \nabla \cdot \, - \, 
\frac{2}{L^2} \, l \, [D + 2\,(s - l - 1)]\, T^{\, l} \, ,
\ee
so that, after some manipulations, one finds that the term involving the commutator and the following one in \eqref{diag_L} sum up to
\be \label{inter}
2\, l \, T^{\, l - 1} \left(\, \nabla\cdot\nabla\cdot \, - \,  \frac{4}{L^2}\, T \,\right) O^L_{s-2k} \, \phi_{s-2k} \, \phi_{s-2l} \, .
\ee
To evaluate \eqref{inter} we make use of  the identity
\be
\left(\, \nabla\cdot\nabla\cdot \,-\, \frac{4}{L^2}\, T \,\right) O^L_{s-2k} \,=
\, -\, g^{k-1}\, M^{(s-2k)}_L \,+\, g^k\, \left( \nabla\cdot\nabla\cdot \, -\, \frac{4}{L^2}\, T \,\right)
\ee
which in itself is a consequence of the divergence of \eqref{MOs}. Assuming for simplicity $k \, \geq \, l$
and completing the computation as in section \ref{sec:flat_symm} it is then possible to conclude that the 
redefinition \eqref{phiL} decomposes the Maxwell-like Lagrangian \eqref{lag_ads_symm} on AdS as 
\be \label{diag_3}
\cL \,=\, \12\ \vf \, M_L^{(s)} \, \vf \,=\, \, \12 \, \sum_{k \,=\, 0}^{[\fr{s}{2}]}  \, 
c_{\, k} \, b_{\, k, s, D}\,  \phi_{s-2k} \, M_L^{(s-2k)} \,  \phi_{s-2k}\, , 
\ee
with the same combinatorial coefficients as for the flat case of \eqref{cl} and \eqref{b}, respectively.

%%%%%%%%%%%%%%%%%%%%%%%%%%%%%%%%%%%%%%%%%%%%%%%%%%%%%%%%%%%%%%%%%%%%%

\section{Discussion}\label{sec:discussion}

%%%%%%%%%%%%%%%%%%%%%%%%%%%%%%%%%%%%%%%%%%%%%%%%%%%%%%%%%%%%%%%%%%%%%
 
     In this work we performed a systematic exploration of theories describing massless bosons of arbitrary spin and symmetry under conditions of transversality on the corresponding gauge parameters, obtaining Lagrangians that are typically simpler than their more conventional counterparts.  

 Higher-spin free Lagrangians have been intensively studied  from several perspectives; in the metric-like approach, with second-order kinetic operators, the various options can be viewed as different solutions to the problem of dressing the D'Alembertian wave operator so that the resulting theory possesses a given amount of gauge invariance. As a necessary condition, the latter has to grant at least the elimination of field components whose presence would spoil the consistency of the theory. Aside from this requirement, however, stressing additional features can lead to different realisations of the same program, according to whether one aims to simplicity of the resulting action, to minimality --in terms of number of field components to be kept off-shell-- or to  the possibility of formulating the theory in terms of quantities amenable of a geometric interpretation, just to mention a few possible ancillary criteria. Clearly, the general goal lying on the background would be to prepare the stage for an investigation of interactions displaying in itself some advantages, either technical or conceptual, with respect to other known approaches.
 
 Without entering into a detailed illustration of the various directions explored so far, let us observe that at least some of them can be pictorially organised according to whether they refer more directly to the spin-two model of linearised gravity or to the spin-one example of Maxwell's theory. In both cases the corresponding higher-spin extension can be implemented with or without additional constraints, and both equations of motion and Lagrangians admit a formulation either in terms of suitably defined ``connexions'' or, when constraints are removed, more geometrically in terms of higher-spin curvatures \cite{dariokyoto}. 
 
 The Fronsdal-Labastida theory \cite{fronsdal,labastida}, together with its minimal unconstrained extensions \cite{fs3,cfms1}, can be safely placed in the first category due to the formal similarity of the corresponding kinetic operators  with the linearised Ricci tensor. The resulting equations, together with their non-local extensions formulated in terms of higher-spin curvatures \cite{fs1,fms1,bb}, naturally provide irreducible descriptions of free higher-spin propagation. Differently, as we discussed at length in the previous sections, the Maxwell-like theories that we explored in this work allow more naturally  for the description of reducible higher-spin spectra, insofar as trace constraints are not imposed. On the other hand, considering the same Lagrangians on the restricted space of traceless gauge potentials, possibly restricted by further additional constraints, like \eqref{mixtdiv}, leads to alternative formulations of irreducible theories that are somehow ``minimal'' with respect to their off-shell field content.  The unconstrained extensions of Maxwell-like theories are  attractive in their own right, since in their local form they bear a direct relation with free open strings in their tensionless regime \cite{triplets,fs2,st}, while their geometric incarnations display actions as simple as squares of curvatures, thus adding a piece of pictorial evidence to their formal relation with spin-one systems \cite{dariotripl}.
 
 The simplification obtained focussing  on the Maxwell operator allowed us to extend the scope of our construction to the case of mixed-symmetry fields in (A)dS backgrounds, providing a complete one-particle Lagrangian description of the corresponding representations in the general case, although it should be stressed that this is achieved imposing severe restrictions on the gauge potentials, eq. \eqref{ultra-constr_2}.  Moreover,  for this latter setting, far less explored in the literature if compared to the cases of flat backgrounds or symmetric tensors in (A)dS, its unconstrained extensions and their possible relation to tensionless strings are at present not known, and it is quite reasonable to expect that a necessary intermediate step would be to achieve a Lagrangian formulation of (A)dS fields describing multi-particle rather than single-particle spectra.
  
 The construction of a corresponding scheme for fermions appears to be less direct to implement. Indeed, if we were to follow closely the analogy with the bosonic case, starting from the Fang-Fronsdal equations for single massless fermions of spin $s + \12$ \cite{FF},
\be
\cS \, = \, i\, (\dsll \, \psi \, - \, \pr \, \psisl) \, = \, 0 \, , 
\ee
the simplest candidate to play the role of kinetic operator for a reducible theory in this case would seem to be the Dirac operator $\mbox{\boldmath{D}} \, = \, \not \!  \pr$. However, in order to allow for gauge invariance of the corresponding equation under $\d \psi = \pr \, \e$ one should also impose $\not \! \pr \,  \e = 0$, thus implying that only on-shell gauge invariance would be admissible. The counterpart of this observation from the point of view of fermionic triplets \cite{fs2, st} is that for those systems, differently from the bosonic ones, there are no fields satisfying purely algebraic equations of motion, so that the reduction to a simpler local system seemingly implies either to keep some auxiliary fields off-shell, or to impose constraints on the gauge parameters somehow stronger than the condition of transversality at the basis of our present construction. We leave to future work a more detailed analysis of the possible constrained theories for systems of reducible fermions.

In perspective, the main issue to investigate concerns the possibility that transverse-invariance might allow for a systematic study of higher-spin interactions while also retaining at least part of the advantages met for the free theory. To begin with, one might ask whether the simplicity of Maxwell-like Lagrangians survives in some forms when interactions are turned on. At the level of cubic vertices, and with the proviso that only  explicit calculations can really clarify the issue, one can expect the answer to be in the affirmative, given the minimal form of the completion needed in this case to promote the known, leading on-shell term in cubic interactions to a full off-shell form\ft{See e.g.  \cite{interactions}  for various approaches to the systematics of cubic vertices for higher-spin bosonic fields.}.  

After all, the  existence of non-linear theories for unimodular gravity  indicates that the transversality constraint should not represent an obstacle to this programme. On the other hand, one should anyway expect that the constraint \eqref{tdiff}, and generalisations thereof, be properly deformed at the non-linear level, and indeed uncovering the systematics of this deformation might represent one of the clues to the whole construction. 

 In addition, it would be interesting to investigate what would be at level of vertices the implications of the non-local redefinitions needed to diagonalise the reducible systems, described in appendix \ref{app:diag}. Indeed, given the existence of local interactions for single-particle couplings (at least to cubic order, insofar as flat space is considered), one would naturally expect the (cubic) couplings for reducible theories to reproduce the former, after diagonalising the quadratic part. However,  in order for the resulting vertices among single particles to stay local after the redefinitions, some non-trivial cancellations ought to occur whose systematics is yet to be explored. Let us mention that the issue does not appear to be related to the choice of flat background, given that in the field redefinition we found to diagonalise the (A)dS system the issue of non-locality appears even more severe than for its flat-space counterpart, and in this sense it can not be interpreted as a manifestation of yet another pathology of higher-spin interactions in Minkowski space-time. 

 One could also investigate directly the structure of couplings deforming single-particle Lagrangians, exploiting traceless fields. Once again, given the simplified kinematical setting at the level of fields involved, the possible complications are likely to come from the preservation or deformation of the constraints, and  it could well be that, at the end, the final balance would not especially favour transverse-invariance as a starting point for investigating interactions. However, an additional reason to explore this path is that, starting from Lagrangians \eqref{lag_mix_ads}, one has in principle the possibility to address  in a systematic and more direct fashion the interactions among bosonic gauge fields of mixed-symmetry on (Anti-)de Sitter backgrounds.

%%%
\subsection*{Acknowledgments}
%%%
We are grateful to K.~Alkalaev, G.~Barnich, N.~Boulanger, P.~A.~Grassi, M.~Grigoriev, M.~Henneaux, E.~Joung, E.~Latini, J.~J.~Lopez-Villarejo, K.~Mktrchyan, A.~Roura, E.~D.~Skvortsov, A.~Waldron, and especially to A.~Sagnotti for useful discussions and comments. We thank the Referee for comments on the AdS mixed-symmetry Lagrangians. We would like to thank APC-Paris VII, the Institute of Physics of the Academy of Sciences of the Czech Republic, the Scuola Normale Superiore, the INFN, the MPI-Albert Einstein Institute and the ULB Brussels for the kind hospitality extended to one or both of us during the preparation of this work. The work of A.C. was partially supported by the ERC Advanced Grant ``SyDuGraM", by IISN-Belgium (convention 4.4514.08) and by the ``Communaut\'e Francaise de Belgique" through the ARC program. The work of D.F. was supported in part by Scuola Normale Superiore, by INFN (I.S. TV12), by the MIUR-PRIN contract 2009-KHZKRX and by the EURYI grant EYI/07/E010 from EUROHORC and ESF.

\begin{appendix}

%%%%%%%%%%%%%%%%%%%%%%%%%%%%%%%%%%%%%%%%%%%%%%%%%%%%%%%%%%%%%%%%%%%%%

\section{Notation and useful formulae}\label{app:formulae}

%%%%%%%%%%%%%%%%%%%%%%%%%%%%%%%%%%%%%%%%%%%%%%%%%%%%%%%%%%%%%%%%%%%%%

%%%
\subsection{Symmetric tensors}\label{app:form_sym}
%%%

We work with mostly-positive metric in $D$ space-time dimensions. If not otherwise specified, symmetrised indices are left implicit, while symmetrisation is understood with no weight factors. Thus,  for instance, the symmetrised product $A\, B$ of two vectors $A_\mu$ and $B_\nu$ here stands for $A_\mu\, B_\nu + A_\nu\, B_\mu$, without additional factors of $1/2$. Traces can be denoted by ``primes'', by numbers in square brackets or even by means of the operator $T$:  $\vf^{\, \prime} \, \equiv \, T \, \vf$ is thus the trace of $\vf$, 
$\vf^{\, \prime\prime}$ is its double trace and  $\vf^{\, [n]} \, \equiv \, T^n \, \vf$ represents its $n-$th trace. Multiple gradients are denoted by symbols like $\pr^{\, k}$, while for divergences we use the symbol ``$\prd$''. The relevant combinatorics is summarised in the following rules  \cite{fs1}:
\begin{alignat}{4} \label{A1}
&\left( \pr^{\, p} \, \vf  \right)^{\, \pe} \ \, & = & \ \, \Box \,
 \pr^{\, p-2} \, \vf \ + \, 2 \, \pr^{\, p-1} \,  \prd \vf \ + \, \pr^{\, p} \,
\vf^{\, \pe} \,  , \nonumber \\
& \partial^{\, p} \, \partial^{\, q} \ \, & =  &\ \, {p+q \choose p} % \binom{p+q}{p} \ \
\partial^{\, p+q} \, , 
\nonumber \\
&\partial \cdot  \left( \partial^{\, p} \ \vf \right) \ \, & = & \ \, \Box \
\partial^{\, p-1} \ \vf \ + \
\partial^{\, p} \ \partial \cdot \vf \, ,  \nonumber \\
& \partial \cdot  (\eta^{\, k}\, \vf) \ \, & = & \ \, \partial \, \eta^{\, k-1} \, \vf \, + \, \h^{\, k} \, \prd \vf , \\
&\left( \eta^k \, \vf  \,  \right)^{\, \prime} \, \ & = & \ \, \left[ \, D
\, + \, 2\, (s+k-1) \,  \right]\, \eta^{\, k-1} \, \vf \ + \ \eta^k
\, \vf^{\, \prime} \, , \nonumber \\
&(\vf \, \psi)^{\, \pe} \ \, & =& \ \, \vf^{\, \pe} \, \psi \, + \, \vf \, \psi^{\, \pe} \, + \, 2
\, \vf \cdot \psi \, , \nonumber \\
& \eta \, \eta^{\, n-1} \ \, & = & \  \, n \, \eta^{\, n} \, . \nonumber 
%&\g \, \cdot \, (\g \, \psi) \ \, & = & \  \, (D \, + \, 2 \, s) \, \psi \, - \, \g \, \psisl \, , \nonumber
\end{alignat}
Switching to (A)dS backgrounds requires  the substitutions
\be \label{sub}
\pr \, \rightarrow \, \nabla\, ,  \hspace{2cm} \h \, \rightarrow \, g \, ,
\ee
where $g$ denotes the (A)dS metric, while also taking into account the following commutators,
\begin{align}
& [\, \nabla\cdot \,,\, \nabla \,]\, \vf \,=\, \Box\, \vf \, - \, \frac{1}{L^2}\, \left\{\, s(D+s-2)\, \vf \, - \, 2\,g\,\vf^\pe \,\right\} \label{comm_nn}\, , \\[5pt]
& [\, \Box \,,\, \nabla \,]\, \vf \,=\, - \, \frac{1}{L^2}\, \left\{\, (D+2s-1)\, \nabla \vf \,-\, 4\,g\,\nabla\cdot\vf \,\right\} \label{comm_bn}\, , \\[5pt]
& [\, \nabla\cdot \,,\, \Box \,]\, \vf \,=\, - \, \frac{1}{L^2}\, \left\{\, (D+2s-3)\, \nabla\cdot\vf \, - \, 2\,\nabla\vf^\pe \,\right\}\, . \label{comm_nb}
\end{align}
In several manipulations it is convenient to make use of the Lichnerowicz operator \cite{lichn}
\be
\Box_L\, \vf \,\equiv\, \Box\,\vf \,+\, \frac{1}{L^2}\, \left\{\, s\,(D+s-2)\, \vf \,-\, 2\, g\,\vf^{\,\pe} \,\right\} , 
\ee
defined so as to satisfy
\be \label{lich_property}
[\,\Box_L, \, \nabla \,] \, \vf \, = \, 0 \, .
\ee
%

%%%
\subsection{Mixed-symmetry tensors}\label{app:form_mix}
%%%

Unless otherwise specified we work with tensors $\varphi_{\mu_1 \cdots\, \mu_{s_1},\,\nu_1 \cdots\, \nu_{s_2},\,\cdots}$ often simply denoted by $\vf$ possessing several ``families'' of symmetric indices, with no additional symmetry properties relating different sets. In this sense they define reducible $GL(D)$ tensors, here often also referred  to as ``multi-symmetric'' tensors. In order to keep our formulas readable usually we do not display space-time indices, while we introduce family indices denoted by small-case Latin letters. We are thus able to identify tensors carrying a different number of  indices in some sets as compared to the basic field $\vf$, while also keeping track of index-reshuffling among different families. Thus, for instance, a gradient carrying a space-time index to be symmetrised with indices belonging to the $i-$th group is denoted by
\be \label{notation}
\pr^{\,i} \, \vf \, \equiv \, \pr_{\,(\,\m^i{}_1|} \, \vf_{\,\cdots \,,\, | \, \m^i{}_2 \, \cdots \, \m^i{}_{s_i+1} ) \,,\, \cdots} \, ,
\ee
with parentheses to signify symmetrisation with no additional overall factors, while for a divergence contracting an index in the $i-$th group we use the notation
\be
\pr_{\,i} \, \vf \, \equiv \, \pr^{\,\l} \, \vf_{\,\cdots \,,\, \l \, \m^i{}_1 \, \cdots \, \m^i{}_{s_i-1} \,,\, \cdots} \, .
\ee
Thus, as a basic rule, the position of the family indices carries information on their role, so that \emph{lower} family indices are associated to operators removing Lorentz indices, while \emph{upper} family indices are associated to operators adding Lorentz indices, to be symmetrised with their peers belonging to the group identified by the family label, as shown in \eqref{notation}. In a similar spirit, the gauge parameters are denoted by $\L_{\, i}$ to indicate that they carry one index less than the gauge field $\vf$ in the $i-$th family. The Einstein convention for summing over pairs of them is used throughout, although one should be careful not to confuse saturation in family indices with contraction between space-time indices. A notable example is the gauge transformation of $\vf$ 
\eqref{mixed_gauge},
\be
\d \, \vf \, = \, \pr^{\, i} \, \L_{\, i}\, ,
\ee
given by a sum of symmetrised gradients,  each for any of the families of $\vf$. Another important class of operators is defined by the following equations:
\begin{subequations} \label{def_S2app}
\begin{align} 
& S^i{}_i\, \vf \,\equiv\, s_i \ \vf_{\,\cdots\,,\,\m^i_1 \cdots\, \m^i_{s_i} ,\, \cdots} \, , \label{def_S1}\\
& S^i{}_j\, \vf \,\equiv\, \vf_{\,\cdots\,,\,(\m^i_1 \cdots\, \m^i_{s_i}| ,\, \cdots \,,\,|\m^i_{s_i+1})\,\m^j_1 \cdots\, \m^j_{s_j-1} ,\, \cdots} \quad \textrm{for}\ \ i \neq j \,   ,
\end{align}
\end{subequations}
whose effect for $i \neq j$ is thus to displace indices from one family to another, while also implementing the corresponding symmetrisation. For more general maximally symmetric backgrounds the flat metric $\h^{\, ij}$ gets replaced by  the (A)dS metric $g^{ij}$, while  $\d_k{}^{i\,}$ simply denotes a Kronecker $\d$-function in family space. In the following list we collect some useful (A)dS commutators, whose flat limit clearly obtains for $L^{\, 2} \to \, \infty$.
\begin{eqnarray}
& [\, S^i{}_j \,,\, \nabla^k \,] & = \, \nabla^{\,i}\, \d_j{}^k \, , \label{[S,d]} \\[5pt]
& [\, \nabla_k \,,\, S^i{}_j \,] & =\, \d_k{}^i\, \nabla_j \, , \label{[d,S]} \\[5pt]
& [\, S^i{}_j \,,\, S^k{}_l \,] & = \, \d_j{}^k S^i{}_l \, - \, \d_l{}^i S^k{}_j \, , \label{[S,S]} \\[5pt]
& [\, T_{ij} \,,\, \nabla^k \,] &=\, \nabla_{(i}\, \d_{\,j)}{}^k \, , \label{[T,d]} \\[5pt]
& [\, T_{ij} \,,\, g^{kl} \,] &= \, \frac{D}{2}\, \d_i{}^{(k} \d_j{}^{l)} \,+\, \frac{1}{2} \left(\, \d_i{}^{(k} S^{l)}{}_j \,+\, \d_j{}^{(k} S^{l)}{}_i \,\right) \, , \label{[T,g]} \\
& [\, \nabla_k \,,\, g^{ij} \,] & =\, \frac{1}{2}\, \d_k{}^{(i\,} \nabla^{\,j)} \, , \label{[d,g]} \\[5pt]
& [\, \nabla^i \,,\, \nabla^j \,] & = \, - \, \fr{2}{L^2} \, g^{k [i}\, S^{j]}{}_k \, ,\\[5pt]
& [\, \nabla_i \,,\, \nabla^j \,] & = \, \Box\, \delta_{i}{}^{\,j} -  \frac{1}{L^2} \,\left\{(D-N-1)\, S^j{}_i \,+\, S^j{}_k\, S^k{}_i \right\} + \frac{2}{L^2}\, g^{jk}\, T_{ik} \, , \label{[d,d]} \\[5pt]
& [\, \Box \,,\, \nabla^i \,] & = \, - \, \frac{1}{L^2} \left\{\, (D-1)\, \nabla^i \,+\, 2\, \nabla^j S^i{}_j \,\right\} \,+\, \frac{4}{L^2}\, g^{ij}\, \nabla_j \, , \label{[B,d]} \\[5pt]
& [\, \nabla_i \,,\, \Box \,] & = \, - \, \frac{1}{L^2} \left\{\, (D-1)\, \nabla_i \,+\, 2\, S^j{}_i \nabla_j \,\right\} \,+\, \frac{2}{L^2}\, \nabla^j\, T_{ij} \, . \label{[d,B]}  
\end{eqnarray}

%%%%%%%%%%%%%%%%%%%%%%%%%%%%%%%%%%%%%%%%%%%%%%%%%%%%%%%%%%%%%%%%%%%%%

\section{Variation of the Maxwell-like tensor in (A)dS}\label{app:proofs}

%%%%%%%%%%%%%%%%%%%%%%%%%%%%%%%%%%%%%%%%%%%%%%%%%%%%%%%%%%%%%%%%%%%%%

In section \ref{sec:ads_mixed} we argued that for \emph{irreducible} fields on (A)dS the gauge variation \eqref{var_mixed_ads} of the Maxwell-like tensor \eqref{maxwell_ads} should take the form
\be \label{todim}
M\, \d\vf \,=\, \sum_{n\,=\,1}^N \, \sum_{i\,=\,1}^n \, k_{\,n}\, \nabla^i \L^{(n)}_{\,i} \,+\, \textrm{divergences and traces}.
\ee
We are now going to prove that this is the correct structure of $M\d\vf$ and to compute the coefficients $k_{\,n}$ in order to prove eq.~\eqref{var_proj_k}. At the end of this appendix we also prove that the constraints \eqref{constrmix} imply the vanishing of all divergences of the single surviving irreducible gauge parameter.

The key of the proof is the possibility to treat independently contributions proportional to different irreducible components of the gauge parameters (labelled by $(n)$ in \eqref{todim}). A crucial ingredient is thus the decomposition of the reducible gauge parameters presented in eq.~\eqref{expansion},
\be \label{dec-app}
\L_k \,=\, \sum_{n\,=\,k}^N \left(1-\d_{s_n,\,s_{n+1}}\right) Y_{\{s_1,\ldots,\,s_n-1,\ldots,\,s_N\}}\, \L_k \,\equiv\, \sum_{n\,=\,k}^N \left(1-\d_{s_n,\,s_{n+1}}\right) \L^{(n)}_{\,k}\, ,
\ee
that determines the extrema of the sum over $i$ in \eqref{todim}. We therefore begin by showing how to derive the decomposition \eqref{dec-app} from the conditions \eqref{irr_par}, that we recall here for the reader convenience:
\be \label{irr-app}
S^i{}_j\, \Lambda_k \,+\, \delta^i{}_k\, \Lambda_j \,=\, 0 \, , \qquad \textrm{for}\ i < j \, .
\ee
In order to illustrate the meaning of eqs.\ \eqref{irr-app} it might be useful to first focus on the case of two families, where they take the explicit form
\be \label{2fam}
\begin{split}
&S^1{}_2\, \Lambda_1 \,+\,  \Lambda_2 \,=\, 0\, ,  \\
&S^1{}_2\, \Lambda_2  \,=\, 0 \, .
\end{split}
\ee
The second of \eqref{2fam} is the condition of irreducibility for $\L_2$, allowing to identify the latter with its homologous diagram:
\be
\L_2 \, = \, Y_{\{s_1, \, s_2 - 1\}}\, \L_2 \, \equiv \, \L^{(2)}_{\,2} \, ,
\ee
while from the first one we can now induce that, among all possible projections contained in $\L_1$, only two of them survive, namely $\L^{(1)}_{\,1} \sim \{s_1 - 1, \, s_2\}$, in the kernel of $S^1{}_2$, and $\L^{(2)}_{\,1} \sim \{s_1, \, s_2  - 1\}$, related to  $\L^{(2)}_{\,2}$ by
\be \label{L12}
S^1{}_2\, \L^{(2)}_{\,1} \,+\, \L^{(2)}_{\,2} \,=\, 0\, .
\ee
In the special case $s_1 \, = \, s_2$ there is no $\L^{(1)}_{\,1}$ projection, since the corresponding diagram does not exist, and the only independent parameter lives in the $\{s_1, \, s_2 - 1\}$ representation, with the corresponding components of $\L_1$ and $\L_2$ related as in \eqref{L12}. 

In the general case it is also convenient to analyse eqs.~\eqref{irr-app} starting from the highest value of the family label carried by the parameters: for $k = N$ these relations imply that $\L_N$ is irreducible since it is annihilated by all $S^i{}_j$ with $i<j$. As a result, it coincides with $\L^{(N)}_{\,N}$ in agreement with \eqref{dec-app}. On the other hand, if one decomposes the multi-symmetric parameter $\L_{N-1}$ in all its irreducible components the \eqref{irr-app} imply\footnote{The multi-symmetric tensor $\L_{N-1}$ carries additional components with respect to those that we labelled by the index $(n)$ in \eqref{dec-app}. However, the argument showing that those with $n<N-1$ are not compatible with \eqref{irr-app} applies also to those that we did not recall explicitly in eq.~\eqref{cond1n1} to simplify the presentation.}
\be \label{cond1n1}
S^i{}_j\, \L^{(n)}_{N-1} \,=\, 0 \, , \qquad \textrm{for}\ n < N \ \textrm{and} \ i<j \, ,
\ee
while for $n = N$ the parameter is annihilated only by the $S^i{}_j$ with $i<N-1$ and 
\be \label{cond2n1}
S^{N-1}{}_N\, \L^{(N)}_{N-1} \,=\, -\, \L^{(N)}_N \, .
\ee
To obtain these relations we used once more the fact that the operators $S^i{}_j$ commute with Young projectors, as discussed in section \ref{sec:flat_mixed}. The system of equations \eqref{cond1n1} is solved only by a tensor whose associated diagram has the same manifest symmetries, and we can thus conclude that $\L_{N-1}$ admits two irreducible components: the $\L^{(N-1)}_{N-1}$ and the $\L^{(N)}_{N-1}$ related to $\L^{(N)}_{N}$ via \eqref{cond2n1}. It should now be clear that one can show by induction that a generic $\L_k$ satisfies
\be
S^i{}_j\, \L^{(n)}_{k} \,=\, 0 \, , \qquad \textrm{for}\ n \leq k \ \textrm{and} \ i<j \, ,
\ee
while the components with $n > k$ are related to the $\L_i$ with $i>k$ via eqs.~\eqref{nonhom}, that generalise \eqref{cond2n1}.

We can now exploit \eqref{todim} in \eqref{var_mixed_ads}, focussing on the variation induced by a single irreducible component so as to obtain 
\begin{align}
M \d_{(n)}\vf \,=\, & -\, \frac{1}{L^2} \, \sum_{i\,=\,1}^n\, \nabla^i \bigg\{ (D + \sum_{l\,=\,1}^N s_l - 2)\, \L^{(n)}_{\,i} -\, \sum_{j\,=\,1}^{n} \bigg( (D-3)\,S^j{}_i  + \sum_{k\,=\,1}^{n} S^k{}_i S^j{}_k \bigg) \L^{(n)}_{\,j} \bigg\} \nn \\
& +\, \textrm{divergences and traces} , \label{app_var}
\end{align}
where we also used \eqref{[S,S]} to change the order of $S^i{}_j$ operators and we fixed the extrema of the sums according to \eqref{dec-app}. In order to proceed it is convenient to distinguish when the contracted indices are smaller, equal or greater than $i$. We shall thus treat separately
\be \label{brick1}
\a^{(n)}_{\,i} \,\equiv\, (D + \sum_{l\,=\,1}^N s_l - 2)\, \L^{(n)}_{\,i} -\, \sum_{j\,=\,1}^{i} \bigg( (D-3)\,S^j{}_i  + \sum_{k\,=\,1}^{i} S^k{}_i\, S^j{}_k \bigg) \L^{(n)}_{\,j} \, ,
\ee
that can be reduced to the form \eqref{todim} simply by exploiting \eqref{nonhom}, and
\begin{subequations} \label{bricks}
\begin{align}
& \b^{(n)}_{\,i} \,\equiv\, (D-3)\! \sum_{j\,=\,i+1}^{n} S^j{}_i\, \L^{(n)}_{\,j} \,+\, \sum_{j\,=\,1}^{i} \sum_{k\,=\,i+1}^{n} S^k{}_i\, S^j{}_k\, \L^{(n)}_{\,j} \, , \label{brick2} \\
& \g^{(n)}_{\,i} \,\equiv\, \sum_{j\,=\,i+1}^{n} \sum_{k\,=\,1}^{n} S^k{}_i\, S^j{}_k\, \L^{(n)}_{\,j} \label{brick3} \, ,
\end{align}
\end{subequations}
that require a more sophisticated discussion.

With the help of \eqref{def_S1} and \eqref{nonhom}, eq.~\eqref{brick1} can be cast in the form
\begin{align}
\a^{(n)}_{\,i} & =\, \left\{\, D + s_{tot} - (s_i-i)(D+s_i-4) - 2 \,\right\} \L^{(n)}_{\,i} - \sum_{k\,=\,1}^{i-1}\, [\, S^k{}_i \,, S^i{}_k \,]\, \L^{(n)}_{\,i} -\! \sum_{j,k\,=\,1}^{i-1}\! S^k{}_i S^j{}_k\, \L^{(n)}_{\,j} \nn \\
& =\, \left\{\, D + s_{tot} - (s_i-i)(D+s_i-4) + (i-1)(s_i-2) - 2\,\right\} \L^{(n)}_{\,i} \nn \\
& -\, \sum_{j\,=\,1}^{i-1}\, \left(\, \sum_{k\,=\,1}^{j-1}\, [\, S^k{}_i \,, S^j{}_k \,] \,+ \sum_{k\,=\,j+1}^{i-1} \! S^k{}_i S^j{}_k \,\right) \L^{(n)}_{\,j} \, ,
\end{align}
where we introduced the shorthand $s_{tot} = \sum_{l\,=\,1}^N s_l$.
Using again \eqref{nonhom} one can show
\be
\sum_{j\,=\,1}^{i-1}\, \left(\, \sum_{k\,=\,1}^{j-1}\, [\, S^k{}_i \,, S^j{}_k \,] \,+ \sum_{k\,=\,j+1}^{i-1} \! S^k{}_i S^j{}_k \,\right) \L^{(n)}_{\,j} \,=\, (i-1)(i-2)\, \L^{(n)}_{\,i} \, ,
\ee
and eventually conclude
\be \label{r1}
\a^{(n)}_{\,i} =\, - \, \left\{\, (s_i-i-1)(D+s_i-i-2) - s_{tot} \,\right\} \L^{(n)}_{\,i} \, .
\ee
If one supposes that \eqref{todim} holds this computation suffices to fix the coefficients $k_{\,n}$ since the term $\nabla^n \L^{(n)}_{\,n}$ cannot receive further corrections. At any rate, we shall proceed by evaluating also the remaining contributions collected in \eqref{bricks}.

Using \eqref{nonhom}, eq.~\eqref{brick2} can be cast in the form
\be \label{int1}
\b^{(n)}_{\,i} =\, (D-i-3)\! \sum_{j\,=\,i+1}^{n} S^j{}_i\, \L^{(n)}_{\,j} \, .
\ee
One cannot eliminate the remaining $S^j{}_i$ with \eqref{nonhom}, but one can use it to build a portion of the quadratic $gl(N)$ Casimir that was introduced in \eqref{casimir}: 
\be
\cC \,=\, \c \,+\, 2\, \sum_{i\,=\,1}^{N-1} \sum_{j\,=\,i+1}^N S^j{}_i S^i{}_j \, , \quad \textrm{with} \ \ \c \,=\, \sum_{i\,=\,1}^N\, S^i{}_i \left(\, S^i{}_i + N - 2i +1 \,\right) . \label{cas-app}
\ee
Using \eqref{nonhom} one can indeed add a $S^i{}_j$ operator in \eqref{int1} which becomes
\be
\b^{(n)}_{\,i} =\, -\, (D-i-3)\! \sum_{j\,=\,i+1}^{n} S^j{}_i\,S^i{}_j\, \L^{(n)}_{\,i} \,=\, -\, \12 \, (D-i-3) \left(\, \cC - \c \,\right) \L^{(n)}_{\,i} \, .
\ee
One can now observe that $\L^{(n)}_{\,n}$, like $\vf$, is a highest-weight state in a representation of the $gl(N)$ algebra generated by all $S^i{}_j$ (see \eqref{[S,S]}). This follows from its irreducibility that translates in
\be
S^i{}_j\, \L^{(n)}_{\,n} \,=\, 0\, , \quad \textrm{for}\ \ i < j \, ,
\ee
while \eqref{nonhom} implies that all $\L^{(n)}_{\,i}$ belong to the same representation. As a result $\cC$ takes the same value on all $\L^{(n)}_{\,i}$, and one can conveniently compute it on $\L^{(n)}_{\,n}$. On the other hand, $\c$ acts diagonally on any tensor and this leads to
\be \label{c-c}
\left(\, \cC - \c \,\right) \L^{(n)}_{\,i} \,=\, 2 \left(\, s_i - s_n + n - i \,\right) \L^{(n)}_{\,i} \, ,
\ee
and eventually to
\be \label{r2}
\b^{(n)}_{\,i} =\, - \ (D-i-3)( s_i - s_n + n - i )\, \L^{(n)}_{\,i} \, .
\ee

The leftover term \eqref{brick3} can be simplified with a similar strategy: we shall build again $(\cC-\c)$ with the help of \eqref{nonhom}. To this end, one can begin by distinguishing various contributions in the sum over $k$:
\be \label{interm}
\begin{split}
\g^{(n)}_{\,i} = \sum_{j\,=\,i+1}^n \bigg\{\,& \sum_{k\,=\,1}^{i-1}\,[\, S^k{}_i \,,\, S^j{}_k \,]\, \L^{(n)}_{\,j} \,+\, (s_i+s_j-2)\, S^j{}_i\, \L^{(n)}_{\,j} \\
& - \sum_{k\,=\,j+1}^n \!S^k{}_i\, \L^{(n)}_{\,k} \,- \sum_{k\,=\,i+1}^{j-1} \!S^k{}_i S^j{}_k S^k_j\, \L^{(n)}_{\,k} \,\bigg\} \, .
\end{split}
\ee
In \eqref{interm} we already used \eqref{nonhom} to manipulate the two sums in the second line. There the extrema of the sums over $k$ depend on $j$, but they can be reorganised such that
\be
\g^{(n)}_{\,i} = \sum_{j\,=\,i+1}^n (s_i+s_j-j)\, S^j{}_i\, \L^{(n)}_{\,j} \,-\, \sum_{k\,=\,i+1}^n S^k{}_i \left( \sum_{j\,=\,k+1}^n S^j{}_k S^k{}_j\, \L^{(n)}_{\,k} \right) \, .
\ee
While one cannot built $(\cC-\c)$ in the first sum due to the $j-$dependent coefficients, the terms between parentheses in the second one can be substituted by $(\cC-\c) \L^{(n)}_{\,i}$. The result is
\be \label{r3}
\begin{split}
\g^{(n)}_{\,i} & =  \sum_{j\,=\,i+1}^n \{ (s_i+s_j-j) - (s_j-s_n+n-j) \}\, S^j{}_i\, \L^{(n)}_{\,j} \\
& =\, -\ (s_i+s_n-n)(s_i-s_n+n-i)\, \L^{(n)}_{\,i} .
\end{split}
\ee
In conclusion, summing \eqref{r1}, \eqref{r2} and \eqref{r3} one obtains
\be
\begin{split}
M\, \d_{(n)}\vf \,& =\, -\, \frac{1}{L^2}\, \sum_{i\,=\,1}^n\, \nabla^i  \left( \a^{(n)}_{\,i} - \b^{(n)}_{\,i} - \g^{(n)}_{\,i} \right) \\
& =\, \frac{1}{L^2} \, \left\{\, (s_n-n-1)(D+s_n-n-2) - s_{tot} \,\right\}\,  \sum_{i\,=\,1}^n\, \nabla^i \L^{(n)}_{\,i} \, .
\end{split} 
\ee
As expected we obtained an overall coefficient that depends on the chosen irreducible component and coincides with the one appearing in \eqref{r1} for $i=n$. 

The previous discussion suffices to conclude that the Lagrangian \eqref{lag_mix_ads} is invariant under transformations generated by a single irreducible and fully divergenceless parameter. In section \ref{sec:dof_ads_mixed} we also checked that this amount of gauge symmetry suffices to remove the unphysical components, at least in the two-family case. However, here we also would like to show that the vanishing of all divergences of the residual parameter is already forced by the apparently weaker condition \eqref{constrmix}, that in this case reads
\begin{subequations} \label{divcomp}
\begin{align}
& \nabla_i\, \L^{(n)}_{\,i} \,=\, 0 \, , \label{diagdiv} \\
& \nabla_i\, \L^{(n)}_{\,j} \,+\, \nabla_j\, \L^{(n)}_{\,i}  \, = \, 0 \, ,\quad \textrm{for} \ i < j \, , \label{offdiv}
\end{align}
\end{subequations}
while if different irreducible components were present they would mix in \eqref{divcomp}. Eqs.~\eqref{nonhom} and \eqref{[d,S]} then imply
\be
\nabla_j\, \L^{(n)}_{\,i} \,=\, -\ S^k{}_j\, \nabla_i\, \L^{(n)}_{\,k} \, , \qquad \textrm{for fixed}\ \ k < j \,,
\ee
where we recalled that no summation over $k$ is implied. Thus, one can choose the value $k = i$ and then exploit \eqref{diagdiv} to conclude that 
\be
\nabla_j\, \L^{(n)}_{\,i} \,=\, 0 \, , \qquad \textrm{for fixed}\ \ i \leq n<j \, .
\ee
The remaining divergences can be shown to vanish with a recursive argument that relies again on \eqref{c-c}. In fact, combining this result with \eqref{nonhom} enables one to obtain\footnote{This relation also allows to rewrite the gauge variation \eqref{todim} only in terms of the irreducible $\L^{(n)}_{\,n}$, that have the right structure to be identified with the parameters of \cite{labastida-morris}.}
\be
\L^{(n)}_{\,i} \,=\, -\ \frac{1}{s_i - s_n + n - i} \sum_{j\,=\,i+1}^n S^j{}_i\, \L^{(n)}_{\,j} \,  
\ee
which in its turn, upon substitution in eq.~\eqref{offdiv} and with the help of \eqref{[d,S]}, 
gives
\be \label{syst_div}
(s_i-s_n+n-i-1)\, \nabla_i\, \L^{(n)}_{\,n} \, - \sum_{j\,=\,i+1}^n S^j{}_i\, \nabla_n\, \L^{(n)}_{\,j} \, .
\ee
For $i = n-1$ eq.~\eqref{syst_div} implies $\nabla_{n-1}\, \L^{(n)}_{\,n} = 0$ and, a posteriori, also $\nabla_n\, \L^{(n)}_{\,n-1} = 0$. Increasing the value of $i$ taking into account the previous outcomes eventually implies the vanishing of all divergences of all $\L^{(n)}_{\,i}$.

%%%%%%%%%%%%%%%%%%%%%%%%%%%%%%%%%%%%%%%%%%%%%%%%%%%%%%%%%%%%%%%%%%%%%

\section{Explicit forms of diagonal Lagrangians}\label{app:diag}

%%%%%%%%%%%%%%%%%%%%%%%%%%%%%%%%%%%%%%%%%%%%%%%%%%%%%%%%%%%%%%%%%%%%%

In this section we discuss explicit solutions to \eqref{MO}, that we report here for the sake of clarity:
\be \label{maxwell}
M \, O_{s-2k} \,  = \, \h^{\, k} \,  M \, ,
\ee
where the dimensionless operators $O_{s-2k}$ appearing in the redefinition of $\vf$ \eqref{phi}
\be \label{phiapp}
\vf \, = \, \phi_s \, + \, O_{s-2} \, \phi_{s-2} \,  + \, O_{s-4} \, \phi_{s-4} \,+\, \cdots \, + \, O_{s-2k} \, \phi_{s-2k} \, + \cdots \, ,
\ee
consist of linear combinations of monomials involving the metric tensor $\h$, suitable powers of gradients and divergences, together with the appropriate inverse powers of the D'Alembertian operator.

 We see from \eqref{maxwell} that the operators $O_{s-2k}$ satisfy a Maxwell equation sourced by $\h^{\, k}   M$; the general solution is thus expected to be of the form 
\be 
O_{s-2k} \,  = \, O^{\, o}{}_{s-2k} \,  + \, O^{\, *}{}_{s-2k} \, ,
\ee
with $O^{\, *}{}_{s-2k}$ a particular solution to \eqref{maxwell}, while among the solutions of the homogeneous equation there should be pure-gauge operators of the form
\be
O^{\, o}{}_{s-2k} \, = \, \pr \, \L^{o} \, ,
\ee
with $\L^{o}$ satisfying the transversality condition 
\be \label{optdiff}
\prd \L^{o} \, = \, 0 \, .
\ee
While these observations imply that the general solution to \eqref{maxwell} is not unique, in special circumstances it might happen that it is not possible to construct an operator with the properties of $\L^{o}$. Let us consider for instance the spin$-2$ case,
\be
\vf \, = \, \phi_2 \, + \, O_{0} \, \phi_{0}\, ,
\ee
and let us construct a solution $O_{0}{}^{*}$ to \eqref{maxwell} by iteration:
\begin{align}
& O_{0}{}^{(0)}&&=& &\h & & \longrightarrow &  &M \, O_{0}{}^{(0)} \, = \, [M, \, \h] \, + \, \h \, M \, = \, -\,2 \, \pr^{\, 2} \, + \, \h \, M \, ,& \\ 
& O_{0}{}^{(1)}&& =& & \h \, + \, a \, \fr{\pr^{\, 2}}{\Box} & &\longrightarrow & 
&M \, O_{0}{}^{(1)} \, = \, -\,(a \, + \, 2)  \, \pr^{\, 2} \, + \, \h \, M \, ,& \\
& O_{0}{}^{*} &&=& &\h \, - \, 2 \, \fr{\pr^{\, 2}}{\Box} \, .& 
\end{align}
In this particular case the solution is thus completely fixed, since in the various steps of the construction there was never the possibility to choose among alternative options. In view of the previous observations we can interpret this result as due to the impossibility of building the gauge parameter $\L^{o}$ for this special case. Indeed $\L^{o}$ should be a rank$-1$ operator acting on scalars, thus implying that neither $\h$ nor $\prd$ can appear in its definition, while a pure gradient is also excluded in force of the transversality condition \eqref{optdiff}. Actually this is a manifestation of a general phenomenon valid for all even spins, when it comes to solving the equation \eqref{maxwell} for the rank$-s$ operator $O_{0}$, since in all those cases it is impossible to build the corresponding gauge parameter. However, for the general case parameters $\L^{o}$ can be constructed, leading to solutions depending on a number of arbitrary coefficients. In the following we will exhibit a particular solution to \eqref{maxwell}, which is tantamount to choosing  a specific gauge.

 To begin with, we would like to expand the operators $ O_{s-2k}$ in their monomial components, so as to translate \eqref{maxwell} into an explicit system for the coefficients of those terms. Each coefficient can be identified by means of three labels:
$
a^{\, (m, k)}_{\, i}\, , 
$
where
\be
\begin{split}
& m \, \rightarrow \, \mbox{denotes the power of} \ \h \, ; \\
& k \, \rightarrow \, \mbox{is related to the rank of the operator} \ O_{\, s - 2k} \, : rk \{O_{\, s - 2k}\} = 2k\, ;\\
&  i \, \rightarrow \, \mbox{denotes the number of divergences} \, ,
\end{split}
\ee
so that in general  $O_{\, s - 2k}$ can be cast in the form
\be \label{O}
\begin{split}
O_{\, s - 2k} \, & = \, \sum_{i = 0}^{s - 2k} \, a^{\, (0, k)}_{\, i}\, \fr{\pr^{\, 2k + i}}{\Box^{\, k + i}} \, {\pr \cdot}^i \, + \, \h \, \, \sum_{i = 0}^{s - 2k} \, a^{\, (1, k)}_{\, i}\, \fr{\pr^{\, 2(k-1) + i}}{\Box^{\, k - 1 + i}} \, {\pr \cdot}^i \, + \cdots \\
& +  \h^{\, m}\,  \sum_{i = 0}^{s - 2k} \, a^{\, (m, k)}_{\, i}\, \fr{\pr^{\, 2(k-m) + i}}{\Box^{\, k - m + i}} \, {\pr \cdot}^i \, + \, \cdots \, \\
& = \, \sum_{i = 0}^{s - 2k}\ \sum_{m = 0}^{k + \left[\frac{i}{2}\right]} a^{\, (m, k)}_{\, i}\, \h^m\, \fr{\pr^{\,2(k-m) + i}}{\Box^{\, k - m + i}} \, {\pr \cdot}^i \, .
\end{split}
\ee
Moreover, it is understood that
\be
2\, (k - m) \, + \, i \, \geq 0 \, ,
\ee
otherwise the corresponding coefficients are simply not present. We fix the set of initial data
\be
a^{\, (k, k)}_{\, 0}\, = \, 1 \, ,
\ee
corresponding to a choice for the normalisation of the $\phi_{s - 2k}$'s convenient for our 
manipulations.  In terms of these definitions \eqref{maxwell} translates into the system
\be \label{MO=eO}
[2(k-m)+i-1] \left\{\, a_i^{(m,k)} + [2(k-m)+i]\, a_i^{(m+1,k)} \,\right\} + [2(k-m)+i]\, a_{i-1}^{(m,k)} \,=\, 0 \, ,
\ee
that for  $m \neq k$  provides a set of conditions $\forall \, i$, while it applies only for  $i > 1$ for $m = k$. In particular for $i = 0$ we have the initial datum  $a^{\, (k, k)}_{\, 0}\, = \, 1$, and  for $i = 1$ we just get
\be \label{ini_kk}
a^{\, (k, k)}_{\, 1}\, - \, a^{\, (k, k)}_{\, 1}\, - a^{\, (k, k)}_{\, 0}\, - \, 0 \cdot a^{\, (k+ 1, k)}_{\, 1}\, = \, - \, 1 \, ,
\ee
thus ensuring that the two terms with zero divergences and one divergence respectively correctly recombine to give $M$.

Since the coefficients depend on $m$ and $k$ only through the combination $k-m$, it is convenient to define
\be
n \,=\, k - m \, ,
\ee
and to introduce the shorthand
\be
a_{i,n} \,=\, a_i^{(k-n,k)} \, ,
\ee
so that \eqref{MO=eO} simplifies to
\be \label{recursion}
(2n+i-1) \left\{\, a_{i,n} + (2n+i)\, a_{i,n-1} \,\right\} + (2n+i)\, a_{i-1,n} \,=\, 0
\ee
with $n \leq k$ (corresponding to $m \geq 0$), with the proviso that for $n < 0$ one has $i \geq -\,2n$, while for $n > 0$ one has $i \geq 0$. 

Eq.~\eqref{recursion} simplifies for the minimum values of $i$ admitted for a given $n$. For $n \geq 0$ and $i = 0$ it becomes
\be
a_{0,n} +\, 2n\, a_{0,n} \,=\, 0 \, .
\ee
With the initial condition $a_{0,0} = 1$ this recursion relation is solved by
\be \label{ini0}
a_{0,n} \,=\, (-1)^n\, (2n)!! \, .
\ee
For $n < 0$ and $i = -\,2n$ eq.~\eqref{recursion} implies
\be \label{inimin}
a_{-2n,n} \, = \, 0 \, .
\ee
For generic values of $i$ the structure of \eqref{recursion} and of the conditions \eqref{ini0} and \eqref{inimin} suggest to consider the ansatz
\be
a_{i,n} \,=\, (-1)^{n+i}\, k_i\, (2n+i)\,(2(n+i-1))!! \, .
\ee
It manifestly satisfies the condition \eqref{inimin} due to the factor $(2n+i)$ and it reduces to \eqref{ini0} for $i = 0$. Moreover, it enables one to factor out various terms so that \eqref{recursion} becomes
\be
(-1)^{n+i} (2n+i-1)(2n+i) (2(n+i-2))!! \left\{\, 2(n+i-1)k_i - (2n+i-2)k_i - k_{i-1} \,\right\} \,=\, 0
\ee
and reduces to
\be
i\,k_i \,-\, k_{i-1} \,=\, 0 \quad \Rightarrow \quad k_i \,=\, \frac{1}{i!} \ .
\ee
Notice that the structure of the double factorial was chosen in order to let $k_{i-1}$ contribute only through a constant term. In conclusion, a particular solution of eq.~\eqref{maxwell} is provided by \eqref{O} with the coefficients
\be \label{fullsol}
a_i^{(k-n,k)} \,=\, (-1)^{n+i}\, 2^{n+i-1}\, (2n+i)\, \frac{(n+i-1)!}{i!} \ .
\ee

For the (A)dS case we expect to be able to find solutions for the operators $O_{\, s - 2k}^L$ as deformations of any flat solution
by terms of $\cO\, (\fr{1}{L^2})$. It is interesting that, at least  for spin $2$, the operator $O_{\, 0}^L$ satisfying
\be \label{MOs'}
M_L^{(2)}\, O^L_{0} \,=\, g\, M_L^{(0)} \, ,
\ee
actually coincides with its flat counterpart, up to covariantisation of the derivatives:
\be
O^L_{0} \, = \, g \, - \, 2 \, \frac{\nabla^2}{\Box_L} \, ,
\ee
where in particular in the construction of the corresponding projector we make use of the Lichnerowicz operator.

However, for tensors of higher ranks the naive covariantisation of the flat-space $O_{s-2k}$ does not solve \eqref{MOs}, that is the (A)dS counterpart of \eqref{maxwell}. The correct deformation involves infinite series of terms with growing powers of the inverse Lichnerowicz operator. This phenomenon can be conveniently illustrated in the simplest example given by the $O_1$ operator, that suffices to complete the decomposition of a rank-$3$ field. In this case the general solution of \eqref{maxwell} contains a free parameter and reads
\be \label{O1}
O_1 =\, \h \,-\, 2\ \frac{\pr^2}{\Box} \,+\, a\ \h\, \frac{\pr}{\Box}\,\prd \,+\ 3(1-a)\, \frac{\pr^3}{\Box^2}\,\prd \, .
\ee
It coincides with \eqref{fullsol} for $a = -1$, but for any value of the parameter \eqref{MOs} can be solved by deforming \eqref{O1} with an infinite number of terms that are proportional to negative powers of $L^2\, \Box_L$:
\be \label{O1L}
O^L_1 =\, g \,- \sum_{k\,=\,0}^\infty \frac{1}{L^{2k}\Box_L^k} \left\{\, 2 \left[\,2(s-2)(D+s-4)\right]^k \frac{\nabla^2}{\Box_L} \,+\, a_k\, g\, \frac{\nabla}{\Box_L}\, \nabla\!\cdot \,+\ b_k\, \frac{\nabla^3}{\Box_L^2}\, \nabla\cdot \,\right\}
\ee
where the coefficients $a_k$ and $b_k$ satisfy
\be
\begin{split}
& 3 a_k \,+\, b_k \,-\, \left[ (s-2)(D+s-4)+(s-3)(D+s-5) \right] b_{k-1} \\
&=\, -\, 3 \left[\, 2(s-2)(D+s-4)\right]^k  .
\end{split}
\ee
The free parameters thus reside only in the divergence terms, as in flat space, while the infinite tower of contributions in \eqref{O1L} appears unavoidable. However, before drawing a definite conclusion, it would be advisable to explore alternative deformations of the inverse D'Alembertian other then the inverse of the Lichnerowicz operator, here used to avoid order ambiguities.

\end{appendix}

\end{document}